\documentclass[review]{elsarticle}
 \biboptions{comma,sort&compress}
\usepackage{graphicx}
\usepackage{amsmath}
\usepackage{here}
\usepackage{cuted}
\def\beq{\begin{equation}}
\def\eeq{\end{equation}}
\def\beqn{\begin{equation*}}
\def\eeqn{\end{equation*}}
\def\bea{\begin{eqnarray}}
\def\eea{\end{eqnarray}}
\def\bq{\begin{quote}}
\def\eq{\end{quote}}

\def\nnb{\nonumber}
\def\ga{\left(}
\def\dr{\right)}

\def\rar{\rightarrow}

\def\nnb{\nonumber}

\def\la{\langle}
\def\ra{\rangle}

\def\ba{\vspace*{-0.2cm}\begin{array}}
\def\ea{\end{array}\vspace*{-0.2cm}}

\def\b{$\bullet~$}

\def\als{\alpha_s}

\def\gg2{ \la\alpha_s G^2 \ra}
\def\gg3{g^3f_{abc}\la G^aG^bG^c \ra}
\def\ggg4{\la\als^2G^4\ra}

\def\qq{\la\bar{q}q\ra}

\usepackage{amsmath}
\usepackage{slashed}
\usepackage{color}

\begin{document}

\begin{frontmatter}

\title{ \large $X_{0,1}(2900)$ and 
 $(D^-K^+)$ invariant mass
from QCD Laplace sum rules at NLO}
\author{R. Albuquerque}
\address{Faculty of Technology,Rio de Janeiro State University (FAT,UERJ), Brazil}
\ead{raphael.albuquerque@uerj.br}
\author{S. Narison\corref{cor1}
}
\address{Laboratoire
Univers et Particules de Montpellier (LUPM), CNRS-IN2P3, \\
Case 070, Place Eug\`ene
Bataillon, 34095 - Montpellier, France\\
and\\
Institute of High-Energy Physics of Madagascar (iHEPMAD)\\
University of Ankatso, Antananarivo 101, Madagascar}
\ead{snarison@yahoo.fr}
\author{D. Rabetiarivony}
\ead{rd.bidds@gmail.com}
\author{G. Randriamanatrika}
\ead{artesgaetan@gmail.com}

\address{Institute of High-Energy Physics of Madagascar (iHEPMAD)\\
University of Ankatso, Antananarivo 101, Madagascar}



\begin{abstract}
\noindent

We revisit, improve and complete some recent estimates of the $0^{+}$ and $1^-$ open charm $(\bar c \bar d)(us)$   tetraquarks and the corresponding molecules masses and decay constants from QCD spectral sum rules (QSSR) by using QCD Laplace sum rule (LSR) within stability criteria where the factorised perturbative NLO corrections   and the contributions of quark and gluon condensates up to dimension-6 in the OPE are included. We confront our results with the  $D^-K^+$ invariant mass recently reported by LHCb from $B^+\to D^+(D^-K^+)$ decays. We expect that the bump near the $D^-K^+$ threshold can be originated from  the $0^{++}(D^-K^+)$ molecule and/or $D^-K^+$ scattering. The  prominent $X_{0}$(2900) scalar peak  and the bump $X_J(3150)$ (if $J=0$) can emerge from a {\it minimal mixing model}, with a tiny mixing angle $\theta_0\simeq (5.2\pm 1.9)^0$, between a scalar {\it Tetramole} (${\cal T_M}_0$) (superposition of nearly degenerated hypothetical molecules and compact tetraquarks states with the same quantum numbers) having a mass $M_{{\cal T_M}_0}$=2743(18) MeV and the first radial excitation of the $D^-K^+$ molecule with mass $M_{(DK)_1}=3678(310)$ MeV. In an analogous way, the $X_1$(2900) and the $X_J(3350)$ (if $J=1$) could be a mixture between the vector {\it Tetramole} $({\cal T_M}_1)$ with a mass $M_{{\cal T_M}_1}=2656(20)$ MeV and its first radial excitation having a mass $M_{({\cal T_M}_1)_1}=4592(141)$ MeV with an angle $\theta_1\simeq (9.1\pm 0.6)^0$. A (non)-confirmation of the previous {\it minimal mixing models} requires an experimental identification of the quantum numbers of the bumps at 3150 and 3350 MeV.  

\end{abstract}

\end{frontmatter}
\pagestyle{plain}
 \section{Introduction}
QCD spectral sum rules (QSSR)\`a la SVZ\,\cite{SVZa,SVZb,ZAKA} have been applied 
since 41 years\,\footnote{For revieews, see e.g\,\cite{SNB1,SNB2,SNB3,SNREV15,SNREV10,SNB4,SNB5,IOFFEb,RRY,DERAF,BERTa,YNDB,PASC,DOSCH}.}  to study successfully the hadron properties (masses, couplings and widths) and to extract some fundamental QCD parameters ($\alpha_s$, quark masses, quark and gluon condensates,...).

Beyond the successful quark model of Gell-Mann\,\cite{GELL}  and Zweig\,\cite{ZWEIG}, Jaffe\,\cite{JAFFE1,JAFFE2} has introduced the four-quark states within the framework of the bag models for an attempt to explain the complex structure of the $I=1,0$ light scalar mesons (see also\,\cite{ISGUR,ACHASOV,THOOFT}). 

In earlier papers, QSSR has been used to estimate  the $I=0$ light scalar mesons ($\sigma, f_0,$) masses and widths\,\cite{LATORRE,SN4} assumed to be four-quark states.  However, the true nature of these states remains still an open question as they can be well interpreted as glueballs / gluonia\,\cite{VENEZIA, BRAMON,SNG,OCHS,MENES3,MENES2}. 

More recently, after the recent discovery of many exotic states beyond the quark model found in different accelerator experiments\,\footnote{For recent reviews, see e.g. \,\cite{MOLEREV,MAIANI,RICHARD} and references quoted therein.}, there was a renewed interest on the four-quarks and molecule states for explaining the properties of these new states. 

In previous papers\,\cite{X5568,SU3,QCD16,MOLE16}, we have systematically studied the masses and couplings of the open-charm and -beauty molecules and tetraquark states using QSSR with the inclusion of factorised contributions at next-to-next-to leading order (N2LO) of perturbation theory (PT) and of the quark and gluon condensates up to dimension 5-7 using the inverse Laplace transform (LSR)\,\cite{BELLa,BELLb,BECCHI,SNR}  of  QCD spectral sum rules (QSSR). More recently, we have extended the analysis to the fully hidden scalar molecules and tetraquark states\,\cite{4Q}.  We have emphasized the importance of these PT corrections for giving a meaning on the input heavy quark mass which plays an important role in the heavy quark sector analysis.  However, these corrections are numerically small in the $\overline{MS}$-scheme as there is a partial compensation of the radiative corrections in the ratio of sum rules used to extract these masses. This property (a posteriori) justifies the uses of the  $\overline{MS}$ running masses in different channels at lowest order (LO)\,\cite{MOLEREV}.
\begin{figure}[hbt]
\begin{center}
\includegraphics[width=7.cm]{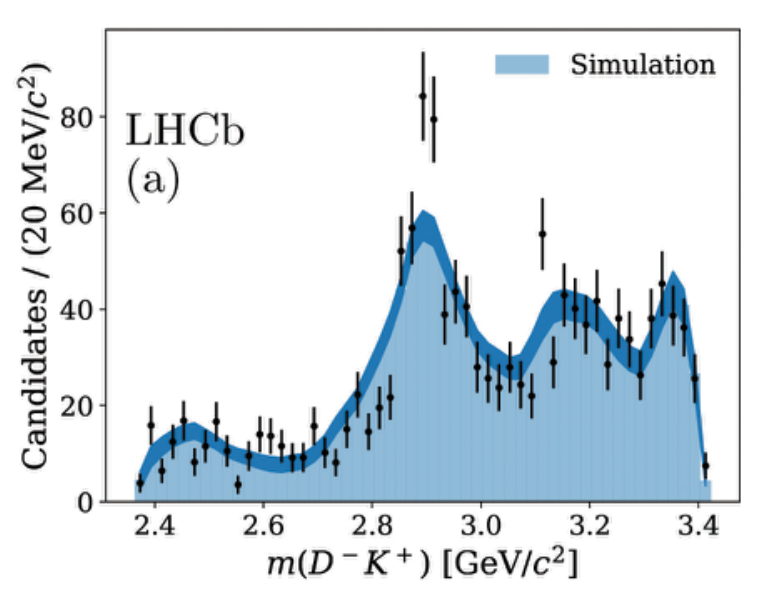}
\includegraphics[width=8.2cm]{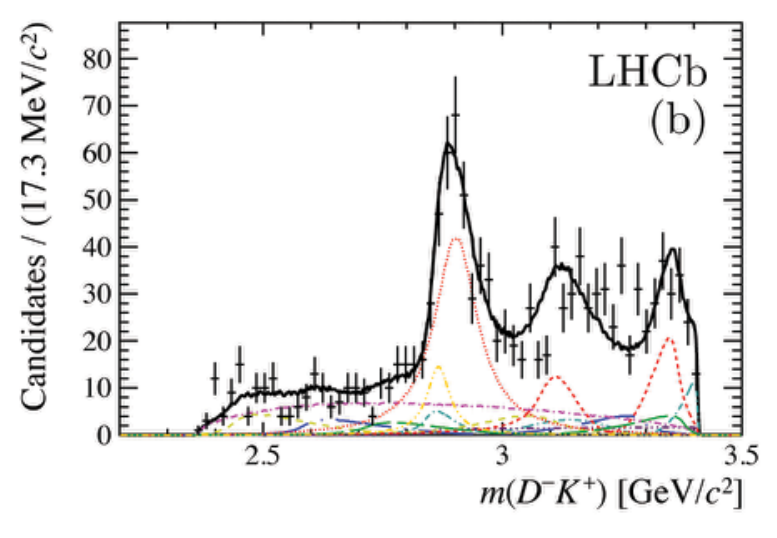}

\vspace*{-0.5cm}
\caption{\footnotesize  LHCb preliminary results for the $D^-K^+$-invariant mass from $B\to D^+D^-K^+$-decays.} 
\label{fig:LHCb}
\end{center}
\vspace*{-0.25cm}
\end{figure}
In this paper, we attempt to estimate, from LSR, the masses and couplings of the $0^{++}$ and $1^-$ molecules and compact tetraquarks states for interpreting the recent LHCb data from  $B\to D^+(D^-K^+)$ decays  where in the $D^-K^+$ invariant mass shown in Fig.\ref{fig:LHCb}\,\cite{LHCb1,LHCb2}
, one finds two prominent peaks (units of MeV):
\bea
\hspace*{-0.6cm} M_{X_0}(0^+)&=&(2866.3\pm 6.5\pm 2.0),~~\Gamma_{X_0}=(57.2\pm 12.9), \nnb\\
\hspace*{-0.6cm} M_{X_1}(1^-)&=&(2904.1\pm4.8\pm 1.3), ~~\Gamma_{X_1}=(110.3\pm 11.5). 
\eea
We have studied in Ref.\,\cite{X5568} the masses and couplings of the $D^0K^0 (0^{++})$ molecule and of the corresponding tetraquark states decaying into $D^0K^0$ but not into $D^-K^+$ and found the lowest ground state masses :
\bea
M_{DK}&= &2402(42)~{\rm MeV},~~~~f_{DK}=254(48)~{\rm keV},\nnb\\
M_{\bar c\bar d{us}}&=&2395(68)~{\rm MeV},~~~~f_{\bar c\bar d{us}}=221(47)~{\rm keV}.
\label{eq:lowest-dk}
\eea
where the LSR parameters  at which one obtains the previous optimal results are:
\beq
\tau\simeq 0.7~{\rm GeV}^{-2},~~~t_c\simeq (12\sim 18)~{\rm GeV}^2.
\eeq
We have used this result to interpret the nature of the $D^*_{s0}$(2317) compiled by PDG\,\cite{PDG} where the existence of a $DK$ pole at this energy has been recently confirmed from lattice calculations of scattering amplitudes\,\cite{LATT}.

For the molecule state, we can interchange the $u$ and $d$ quarks in the interpolating current and deduce from $SU(2)$ symmetry that the $D^-K^+(0^{++})$ molecule mass is degenerated with the $D^0K^0$ one. Compared with the LHCb data, one may invoke that this charged molecule can be responsible of the bump near the DK threshold around 2.4 GeV but is too light to explain the $X_{0,1}$ peaks.

For the tetraquark state, one may not use a simple $SU(2)$ symmetry (rotation of $u$ and $d$ quarks) to deduce the ones decaying into $D^-K^+$ due to our present ignorance of the diquark dynamics (for some attempts see\,\cite{NEU1,NEU2}). 

Therefore, recent analysis based on QSSR at lowest order (LO) of perturbation theory (PT) using some specific tetraquarks and / or molecules configurations appear in the literature\,\cite{ZHANG,CHEN,WANG,STEELE} (see also\,\cite{AGAEV,TURC} which appeared after the completion of this work).

However, due to the complexity of the QCD calculations, to some other possible configurations and to the different ways for extracting these predictions, we think that it is important to revisit and to improve these LO results by adding the NLO perturbative contributions and by using an optimization procedure where the values of the external LSR parameters [sum rule inverse energy variable ($\tau$), QCD continuum threshold ($t_c$) and subtraction scale ($\mu$)] are left as free parameters.  This is the aim of the present paper.

\section{The Laplace sum rule (LSR)}
We shall work with the  Finite Energy version of the QCD Inverse Laplace sum rules (LSR) and their ratios :
\beq
\hspace*{-0.2cm} {\cal L}^c_n(\tau,\mu)=\int_{(M_c+m_s)^2}^{t_c}\hspace*{-0.5cm}dt~t^n~e^{-t\tau}\frac{1}{\pi} \mbox{Im}~\Pi_{\cal M,T}(t,\mu)~,~~~~~~~~~~~
 {\cal R}^c_n(\tau)=\frac{{\cal L}^c_{n+1}} {{\cal L}^c_n},
\label{eq:lsr}
\eeq
 where $M_c$ and $m_s$ (we shall neglect $u,d$ quark masses) are the on-shell / pole charm and running strange quark masses, $\tau$ is the LSR variable, $n=0,1$ is the degree of moments, $t_c$ is  the threshold of the ``QCD continuum" which parametrizes, from the discontinuity of the Feynman diagrams, the spectral function  ${\rm Im}\,\Pi_{\cal M,T}(t,m_c^2,m_s^2,\mu^2)$   where  $\Pi_{\cal M,T}(t,m_Q^2,\mu^2)$ is the  scalar correlator defined as :
 \bea
\hspace*{-0.6cm} \Pi_{\cal M,T}(q^2)\hspace*{-0.2cm}&=&\hspace*{-0.25cm}i \hspace*{-0.15cm}\int \hspace*{-0.15cm}d^4x ~e^{-iqx}\la 0\vert {\cal T} {\cal O}^J_{\cal M,T}(x)\ga {\cal O}^J_{\cal M,T}(0)\dr^\dagger \vert 0\ra~,
 \label{eq:2-pseudo}
 \eea
 where ${\cal O}^{J}_{\cal M,T}(x)$ are the interpolating currents for the tetraquarks ${\cal T}$ and molecules ${\cal M}$ states. The  superscript $J$ refers to the spin of the particles.  
\section{The interpolating operators}
We shall be concerned with the interpolating given in Table\,\ref{tab:current}.   
\vspace*{-0.5cm} 
   {\scriptsize
\begin{table}[hbt]
\setlength{\tabcolsep}{2.2pc}
    {\small
  \begin{tabular}{ll}
&\\
\hline
\hline
Scalar states ($0^+$) & Vector states ($1^-$) \\
\hline
  Tetraquarks \\
 $ {\cal O}^0_{SS} = \epsilon_{i j k} \:\epsilon_{m n k} \left(
  u_i^T\, C \gamma_5 \,d_j \right) \left( \bar{c}_m\, 
  \gamma_5 C \,\bar{s}_n^T\right) $ &  ${\cal O}^1_{AP} = \epsilon_{m n k}\: \epsilon_{i j k}\left( \bar{c}_m\, \gamma_\mu C \,
  \bar{s}_n^T\right) \left(
  u_i^T\, C \,d_j \right) $ \\
  
$   {\cal O}^0_{PP} = \epsilon_{i j k} \:\epsilon_{m n k} \left(
  u_i^T\, C \,d_j \right) \left( \bar{c}_m\, 
  C \,\bar{s}_n^T\right) $ & ${\cal O}^1_{PA} = \epsilon_{m n k}\: \epsilon_{i j k}\left( \bar{c}_m\,  C \,
  \bar{s}_n^T\right) \left(
  u_i^T\, C\gamma_\mu \,d_j \right) $  \\
  
 $  {\cal O}^0_{VV} = \epsilon_{i j k} \:\epsilon_{m n k} \left(
  u_i^T\, C \gamma_5 \gamma_\mu \,d_j \right) \left( \bar{c}_m\, 
  \gamma^\mu \gamma_5 C \,\bar{s}_n^T\right) $ & $ {\cal O}^1_{SV} = \epsilon_{i j k} \:\epsilon_{m n k} \left(
  u_i^T\, C \gamma_5 \,d_j \right) \left( \bar{c}_m\, 
  \gamma_\mu \gamma_5 C \,
  \bar{s}_n^T\right) $ \\
  
 $  {\cal O}^0_{AA} = \epsilon_{i j k} \:\epsilon_{m n k} \left(
  u_i^T\, C \gamma_\mu \,d_j \right) \left( \bar{c}_m\, 
  \gamma^\mu C \,\bar{s}_n^T\right) $ & $ {\cal O}^1_{VS} =\epsilon_{i j k} \:\epsilon_{m n k} \left(
  u_i^T\, C \gamma_5 \gamma_\mu \,d_j \right) \left( \bar{c}_m\, 
  \gamma_5 C \,
  \bar{s}_n^T\right) $\\
  Molecules \\
$  {\cal O}^0_{DK}=(\bar c\gamma_5 d)(\bar s\gamma_5 u)$& ${\cal O}^1_{D_1 K} =\left(\bar{c} \gamma_\mu \gamma_5 d \right) 
  \left( \bar{s} \gamma_5\,u\right) $  \\
${\cal O}^0_{D^*K^*}=(\bar c\gamma^\mu d)(\bar s\gamma_\mu u)$&$ {\cal O}^1_{D K_1} = \left(\bar{c} \gamma_5 d \right) 
  \left( \bar{s}\, \gamma_\mu \gamma_5 \,u\right) $\\
$ {\cal O}^0_{D_1K_1}=(\bar c\gamma^\mu\gamma_5 d)(\bar s\gamma_
  \mu\gamma_5 u)$ & $ {\cal O}^1_{D^* K^*_0} = \left(\bar{c} \gamma_\mu  d \right) 
  \left( \bar{s}\,u\right) $ \\
 $  {\cal O}^0_{D^*_0K^*_0}=(\bar c d)(\bar s u)$ & 
$ {\cal O}^1_{D^*_0 K^*} =   \left(\bar{c} \,d \right) 
  \left( \bar{s} \,\gamma_\mu \,u\right)$ \\
  \hline\hline
  \vspace*{-0.5cm}
\end{tabular}}
 \caption{Interpolating operators describing the scalar $(0^+)$ and vector $(1^-)$ molecules and teraquark states.}  

\label{tab:current}
\end{table}
} 
The lowest order (LO) perturbative (PT) QCD expressions  including the quark and gluon condensates contributions up to dimension-six condensates of the corresponding two-point spectral functions are given in the Appendix.  
\subsection*{\b Higher Orders PT corrections to the Spectral functions}
We extract the NLO PT corrections by considering that the molecule /tetraquark two-point spectral function is the convolution of the two ones built from 
two quark bilinear currents (factorization) which is justified because we have seen for the LO that the non-factorized part of the QCD diagrams gives negligible contribution and behaves like 1/$N_c$ where $N_c$ is the number of colours (see some explicit examples in\,\cite{MOLE16,4Q}), while at order $\alpha_s$, this feature has been shown from the analysis of the four-quark correlator governing the $B^0-\bar B^0$ mixing\,\cite{SNPIVO,HAGIWARA}.
\bea\label{eq:qqcurrent}
J^{P,S}(x)\equiv\bar c [i\gamma_5,1] c ~~\rar ~~ \frac{1}{\pi}{\rm Im}\,\psi^{P,S}(t)\sim\frac{3}{8\pi^2}~, ~~~~~~
J^{V,A}(x)\equiv\bar c [\gamma_\mu,\gamma_\mu\gamma_5] c ~~ \rar ~~ \frac{1}{\pi}{\rm Im}\,\psi^{V,A}(t)\sim\frac{1}{4\pi^2} ~,
\eea
where the spectral functions behave as a constant in the limit $m^2_c\ll t$ in order to be consistent with the $t$-behaviour of the one from the four-quark current given in the Appendix. 
In this way, we obtain the convolution integral\,\cite{SNPIVO,PICH}:
\\
-- {\it Molecules} :
\bea
\hspace*{-0.85cm}\frac{1}{ \pi}{\rm Im}\, \Pi_{\cal M,T}(t)= \theta (t-(M_c+m_s+m_d)^2)\times \ga  \frac{k}{4\pi}\dr^2\hspace*{-0.15cm} t^2 \hspace*{-0.15cm}\int_{(M_c+m_d)^2}^{(\sqrt{t}-m_s)^2}\hspace*{-0.7cm}dt_1\int_{m_s^2}^{(\sqrt{t}-\sqrt{t_1})^2}  \hspace*{-1cm}dt_2~\lambda^{1/2}{\cal K}^H~ ,
\label{eq:mole-int}
\eea
Here : 
\bea
\hspace*{-0.7cm}{\cal K}^{SS,PP}\hspace*{-0.25cm}&\equiv&\hspace*{-0.25cm}\ga \frac{t_1}{ t}+ \frac{t_2}{ t}-1\dr^2
\times ~ \frac{1}{\pi}{\rm Im} \,\psi^{S,P}(t_1) \frac{1}{\pi} {\rm Im}\, \psi^{S,P}(t_2)~,\nnb\\
\hspace*{-0.7cm}{\cal K}^{VV,AA}\hspace*{-0.25cm}&\equiv&\hspace*{-0.25cm}\Bigg{[}\ga \frac{t_1}{ t}+ \frac{t_2}{t}-1\dr^2 \hspace*{-0.15cm}+8\frac{t_1t_2}{ t^2}\Bigg{]}
\times \frac{1}{\pi}{\rm Im} \,\psi^{V,A}(t_1)  \frac{1}{\pi} {\rm Im}\, \psi^{V,A}(t_2)~,
\eea
for spin zero scalar state and :
\beq
\hspace*{-0.7cm}{\cal K}^{VS,AP}\equiv
2\lambda\times ~ \frac{1}{\pi}{\rm Im} \,\psi^{V,A}(t_1) \frac{1}{\pi} {\rm Im}\, \psi^{S,P}(t_2)~,
\eeq
for spin one vector state, with the phase space factor:
\beq
\lambda=\ga 1-\frac{\ga \sqrt{t_1}- \sqrt{t_2}\dr^2}{ t}\dr \ga 1-\frac{\ga \sqrt{t_1}+ \sqrt{t_2}\dr^2}{ t}\dr~.
\eeq
$M_c$ is the on-shell / pole perturbative heavy quark mass while $k$ is an appropriate normalization factor for matching the spectral function with the one from a direct calculation of the four-quark correlator given in the Appendix. 

-- {\it Tetraquarks} :  
\\
One interchanges $s$ and $d$ in the integrals of Eq.\,\ref{eq:mole-int}. We have taken $m_u=0$ for simplfying the expression but we shall also neglect $m_d$ in the numerical analysis. 

-- The NLO perturbative expressions of the spectral functions built from bilinear quark - antiquark currents are known in the literature\,\cite{SNB1,SNB2,RRY,BECCHI,BROAD,CHET1,PIVOSU3}. 

-- We estimate the N2LO contributions assuming a geometric growth of the numerical coefficients\,\cite{SZ,CNZa,CNZb,ZAKa,ZAKb}. We consider this contribution as an estimate of the error due to the truncation of the PT series. 

\subsection*{\b QCD input parameters}
{\scriptsize
\begin{table}[hbt]
\setlength{\tabcolsep}{1.9pc}
    {\small
  \begin{tabular}{llll}
&\\
\hline
\hline
Parameters&Values&Hadron sources& Ref.    \\
\hline
$\alpha_s(M_Z)$& $0.1181(16)(3)$&$M_{\chi_{0c,b}-M_{\eta_{c,b}}}$&
 \, \cite{SNparam,SNparam2} \\
$\overline{m}_c(m_c)$ [MeV]&$1266(6)$ &$D, B_c\oplus {J/\psi}, \chi_{c1},\eta_c$&
\cite{SNm20,SNparam,SNbc20,SNmom18,SNFB15}\\
$\hat \mu_q$ [MeV]&$253(6)$ &Light  &\,\cite{SNB1,SNp15,SNLIGHT} \\
$\hat m_s$ [MeV]&$114(6)$ &Light &\,\cite{SNB1,SNp15,SNLIGHT} \\
$\kappa\equiv\la \bar ss\ra/\la\bar dd\ra$& $0.74\pm 0.06$&Light \& heavy&\cite{SNB1,SNp15,HBARYON1,HBARYON2}\\
$M_0^2$ [GeV$^2$]&$0.8 \pm 0.2$ &Light \& Heavy&\,\cite{SNB1,DOSCH,HEIDa,HEIDb,JAMI2a,JAMI2c,HEIDc,SNhl} \\
$\la\alpha_s G^2\ra$  [GeV$^4$]& $(6.35\pm 0.35)\times 10^{-2}$&Light \& Heavy &
 \cite{SNparam}\\
${\la g^3  G^3\ra}/{\la\alpha_s G^2\ra}$& $(8.2\pm 1.0)$ GeV$^2$&${J/\psi}$&
\cite{SNH10,SNH11,SNH12}\\
$\rho \alpha_s\la \bar qq\ra^2$ [GeV$^6$]&$(5.8\pm 0.9)\times 10^{-4}$ &Light, $\tau$-decays &\cite{DOSCH,SNTAU,LNT,JAMI2c,LAUNERb}\\
\hline\hline
\end{tabular}}
 \caption{QCD input parameters estimated from QSSR (Moments, LSR and ratios of sum rules). 
 }  
\label{tab:param}
\end{table}
} 
We shall use the QCD inputs in Table\,\ref{tab:param}. The Renormalization Group Invariant parameters are defined as\,\cite{SNB1,SNB2}: 
\beq
{\bar m}_{s}(\tau)=
{\hat m}_{s}  \ga-\beta_1a_s\dr^{-2/{
\beta_1}},~~~
{\la\bar qq\ra}(\tau)=-{\hat \mu_q^3  \ga-\beta_1a_s\dr^{2/{
\beta_1}}},~~~
{\la\bar q Gq\ra}(\tau)=-{M_0^2{\hat \mu_q^3} \ga-\beta_1a_s\dr^{1/{3\beta_1}}}~,
\eeq
where $\beta_1=-(1/2)(11-2n_f/3)$ is the first coefficient of the $\beta$ function 
for $n_f$ flavours; $a_s\equiv \alpha_s(\tau)/\pi$; 
$\hat\mu_q$ is the spontaneous RGI light quark condensate \cite{FNR}. The running charm mass $ \overline{m}_c$ is related to the on-shell (pole) mass $M_c$ used to compute the two-point correlator from the NLO relation\,\cite{TAR,COQUEa,COQUEb,SNPOLEa,SNPOLEb} :
\beq
M_c(\mu) = \overline{m}_c(\mu)\Bigg{[}
1+\frac{4}{3} a_s(\mu)+\ln{\ga\frac{\mu}{M_c}\dr^2} a_s(\mu)+{\cal O}(a_s^2)\Bigg{]}
\eeq
The QCD condensates entering in the analysis are the light quark condensate $\qq$,  the gluon condensates $ \la
\alpha_sG^2\ra
\equiv \la \alpha_s G^a_{\mu\nu}G_a^{\mu\nu}\ra$ 
and $ \la g^3G^3\ra
\equiv \la g^3f_{abc}G^a_{\mu\nu}G^b_{\nu\rho}G^c_{\rho\mu}\ra$, 
the mixed quark-gluon condensate $\la\bar qGq\ra
\equiv {\la\bar qg\sigma^{\mu\nu} (\lambda_a/2) G^a_{\mu\nu}q\ra}=M_0^2\la \bar qq\ra$ 
and the four-quark 
 condensate $\rho\alpha_s\la\bar qq\ra^2$, where
 $\rho\simeq (3\sim 4)$ indicates the deviation from the four-quark vacuum 
saturation. 

\section{Extracting the lowest ground state mass and coupling}
In Ref.\,\cite{X5568}, we have extracted the lowest ground state mass by using the minimal duality
ansatz:
 \beq
\hspace*{-0.65cm} \frac{1}{\pi}{\rm Im} \Pi_{\cal M/T}\simeq f_{\cal M/T}^2 M_{\cal M/T}^{8} \delta(t-M_{\cal M/T}^2) +\Theta(t-t_c) ``{\rm  Continuum}",
 \eeq
 where the decay constant $f_{\cal M}$ (analogue of $f_\pi$) is defined as :
 \beq
\la 0\vert  {\cal O}_{\bar DK}\vert {\bar DK}\ra = f_{\bar DK} M_{\bar DK}^{4}, ~~~~~~~~~~~~~
\la 0\vert  {\cal O}^\mu_{\bar D^*K}\vert {\bar D^*K}\ra = \epsilon^\mu f_{ D^*K} M_{\bar D^*K}^{5},
\eeq
and analogously for the one $f_{\cal T}$ of tetrquark state. Interpolating currents constructed from bilinear (pseudo)scalar currents are not renormalization group invariants such that the corresponding  decay constants possess anomalous dimension:
\beq
f_{\bar DK} (\mu)=\hat f_{\bar DK} \ga -\beta_1a_s\dr^{4/\beta_1}(1-k_fa_s) ,~~~~~~~~~~
f_{ D^*K} =\hat f_{\bar D^*K} \ga -\beta_1a_s\dr^{2/\beta_1}(1-k_fa_s/2) ~,
\eeq
where : $\hat f_{\cal M}$ is the renormalization group invariant coupling and $-\beta_1=(1/2)(11-2n_f/3)$ is the first coefficient of the QCD $\beta$-function for $n_f$ flavours. $a_s\equiv (\alpha_s/\pi)$ is the QCD coupling and $k_f=2.028 (2.352)$ for $n_f=4(5)$ flavours. 

Within a such parametrization, one  obtains: 
 \beq
  {\cal R}^{c}_0\equiv {\cal R}\simeq M_{\cal M}^2~,
  \label{eq:mass}
  \eeq
 indicating that the ratio of moments appears to be a useful tool for extracting the mass of the hadron ground state as shown in the original SVZ papers\,\cite{SVZa,SVZb}, different books, reviews and papers\,\cite{SNB1,SNB2,SNB3,SNB4,IOFFEb,RRY,DERAF,BERTa,YNDB,PASC,DOSCH}. 
 
 As $\tau,~ t_c$ and $\mu$ are free external parameters , we shall use stability criteria (minimum senstivity on the variation of these parameters) to extract the lowest ground state mass and coupling (see more details discussions in the previous books and reviews). 

Within the approach, one has obtained the masses of the lowest ground state $\bar D^0K^0$ molecule and of its $\bar c\bar u ds$ tetraquark states analogue quoted in Eq.\,\ref{eq:lowest-dk}. 
\section{The $0^{++}\bar SS$ and $\bar AA $ tetraquarks}

The two channels present similar features. Then, we show only explicitly the analysis of the $SS$ channel for a better understanding on the extraction of our numbers.
\subsection*{\b $\tau$- and $t_c$-stabilities}
\begin{figure}[H]
\begin{center}
\centerline {\hspace*{-7.5cm} \bf a)\hspace*{8cm} \bf b) }
\vspace{0.25cm}
\includegraphics[width=8cm]{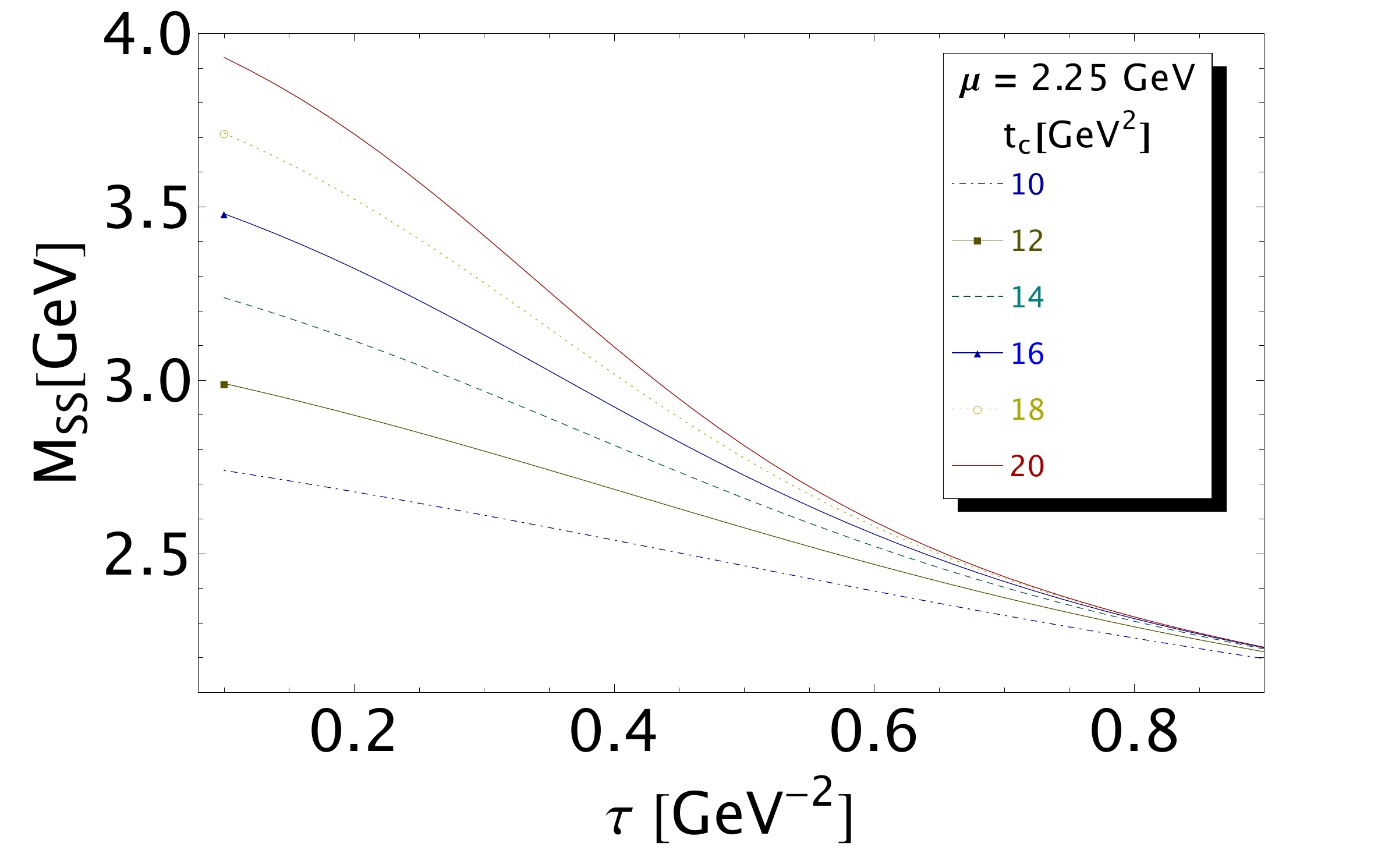}
\includegraphics[width=8cm]{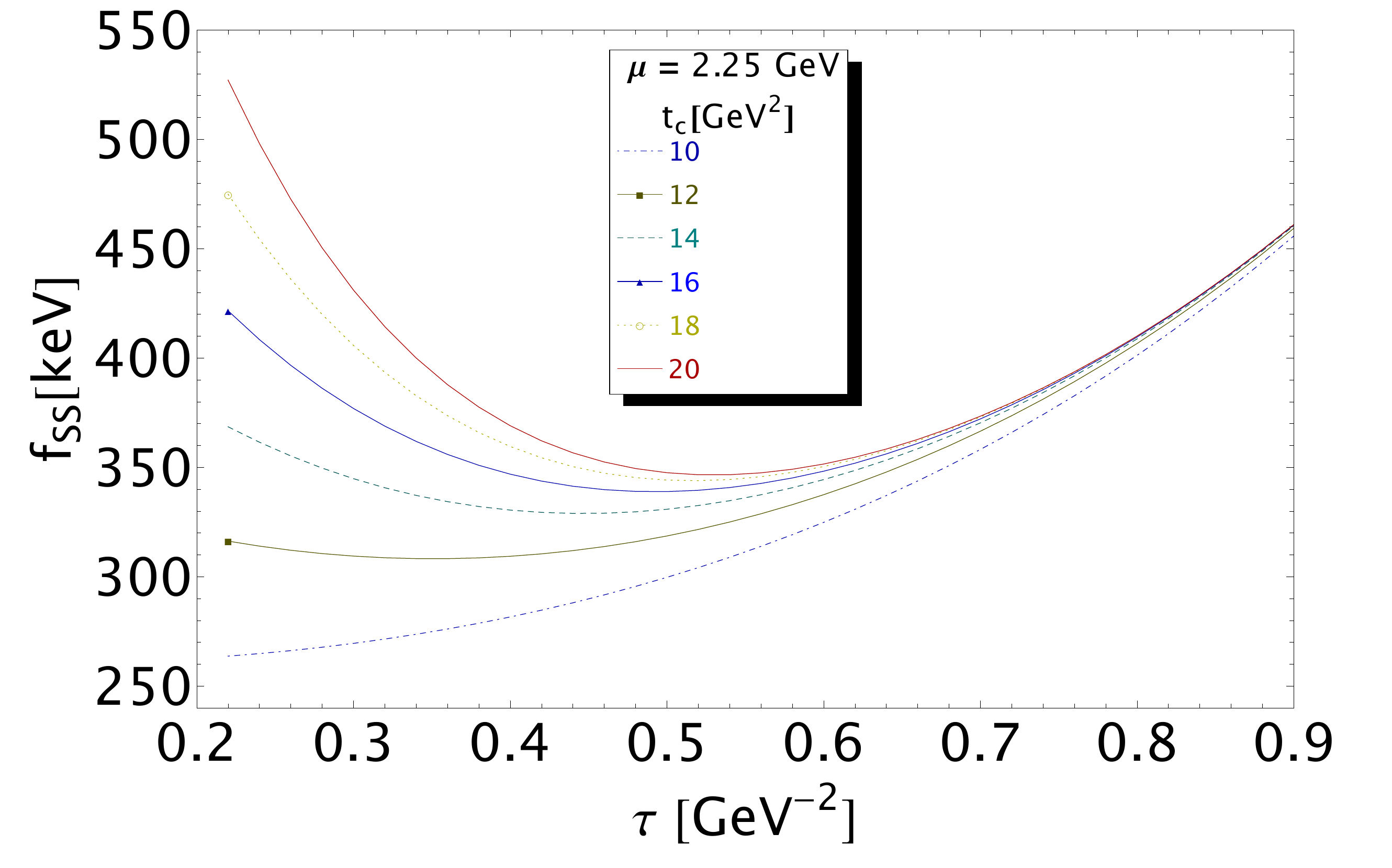}
\vspace*{-0.5cm}
\caption{\footnotesize  $f_{SS}$ and $M_{SS}$ as function of $\tau$ at NLO for different values of $t_c$, for $\mu$=2.25 GeV and for values of the QCD parameters given in Table\,\ref{tab:param}.} 
\label{fig:sstau}
\end{center}
\end{figure}

-- We show in Fig.\ref{fig:sstau}a) the $\tau$- and $t_c$- dependence of the mass obtained from ratio of moments ${\cal R}_0$.  We have used 
$\mu$=2.25~{\rm GeV}
obtained in\,\cite{X5568} which we shall check later on.
The analysis of the coupling from the moment ${\cal L}^c_0$ is shown in Fig.\,\ref{fig:sstau}b). The results stabilize at $\tau\simeq 0.5 $ GeV$^{-2}$(inflexion point for the mass and minimum for the coupling). 

-- To extract our numbers, we proceed iteratively. From Fig.\ref{fig:sstau}a), we extract the mass as the mean value of the one for $t_c\simeq$ 12 GeV$^2$ (beginning of the inflexion point) and of the one at beginning of $t_c$-stability of about 18 GeV$^2$. 

-- We use this (physical) mass value in ${\cal L}^c_0$ to draw Fig.\,\ref{fig:sstau}b). We check the range of $t_c$-values where the above-mentioned stabilities have been obtained by confronting Figs.\,\ref{fig:sstau}a) and b). Here, one can easily check that this range of $t_c$-values is the same for the mass and coupling. If the range does not co\"\i ncide, we take the common range of $t_c$ and redo the extraction of the mass. 

-- One can also see that the range of $\tau$-stabilities co\"\i ncide in Fig.\,\ref{fig:sstau}a) (inflexion points) and in Fig.\,\ref{fig:sstau}b) (minimas). It is obvious that the value of $\tau$ from the minimum is more precise which we re-use to fix the final value of the mass. 
\subsection*{\b $\mu$-stability}
\begin{figure}[hbt]
\vspace*{-0.25cm}
\begin{center}
\centerline {\hspace*{-7.cm} \bf a)\hspace*{8cm} \bf b) }
\vspace{0.25cm}
\includegraphics[width=8cm]{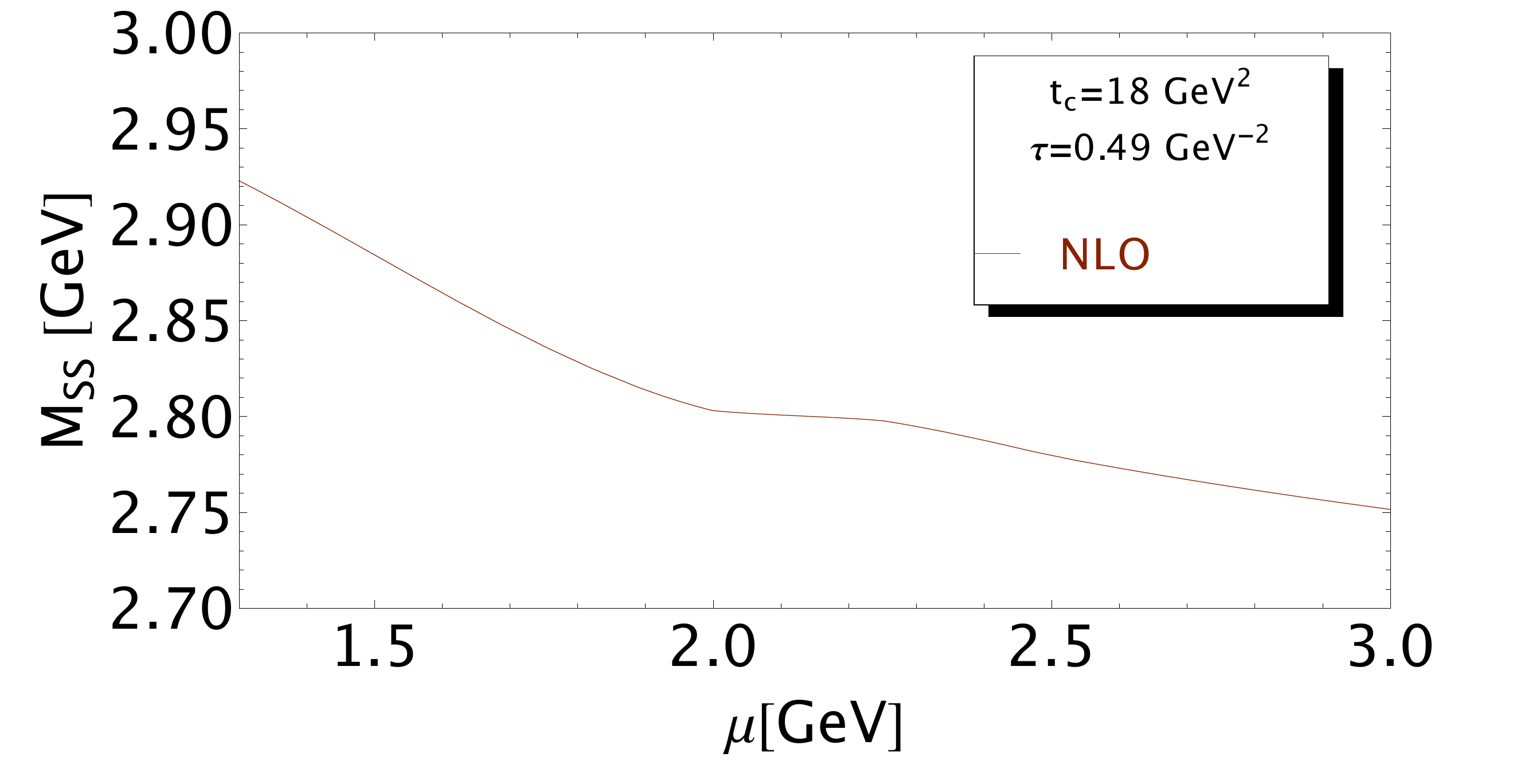}
\includegraphics[width=8cm]{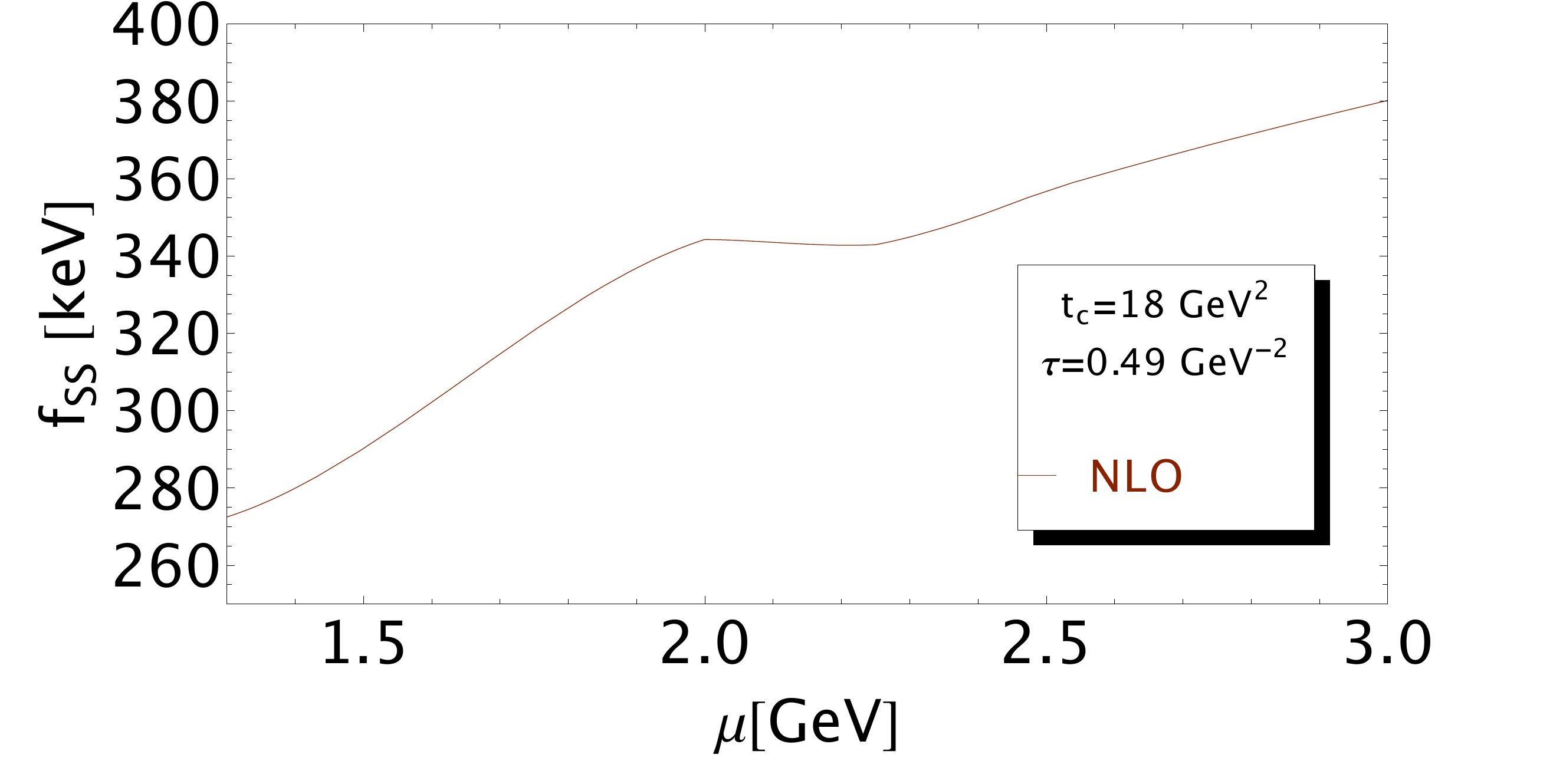}
\vspace*{-0.5cm}
\caption{\footnotesize  $M_{SS}$ and $f_{SS}$  as function of $\mu$ at NLO for fixed values of $t_c$ and $\tau$ and for the values of the QCD parameters given in Table\,\ref{tab:param}.} 
\label{fig:ssmu}
\end{center}
\end{figure}

We show in Fig.\,\ref{fig:ssmu} the $\mu$-dependence of the results for given $t_c$=18 GeV$^2$ and $\tau$=0.49 GeV$^{-2}$. One finds a common stability for :
\beq
\mu=(2.25\pm 0.25)~{\rm GeV},
\label{eq:mu}
\eeq
which confirms the result in Ref.\,\cite{X5568}. 
\subsection*{\b Final results}
Our final results are obtained at the stabilities of the set of parameters $(\tau, t_c,\mu)$. They  are compiled in Table\,\ref{tab:res0} together with the different sources of errors which we shall comment later on in Section 13.  

\begin{figure}[H]
\vspace*{-0.25cm}
\begin{center}
\centerline {\hspace*{-7.cm} \bf a)\hspace*{8cm} \bf b) }
\vspace{0.25cm}
\includegraphics[width=8.cm]{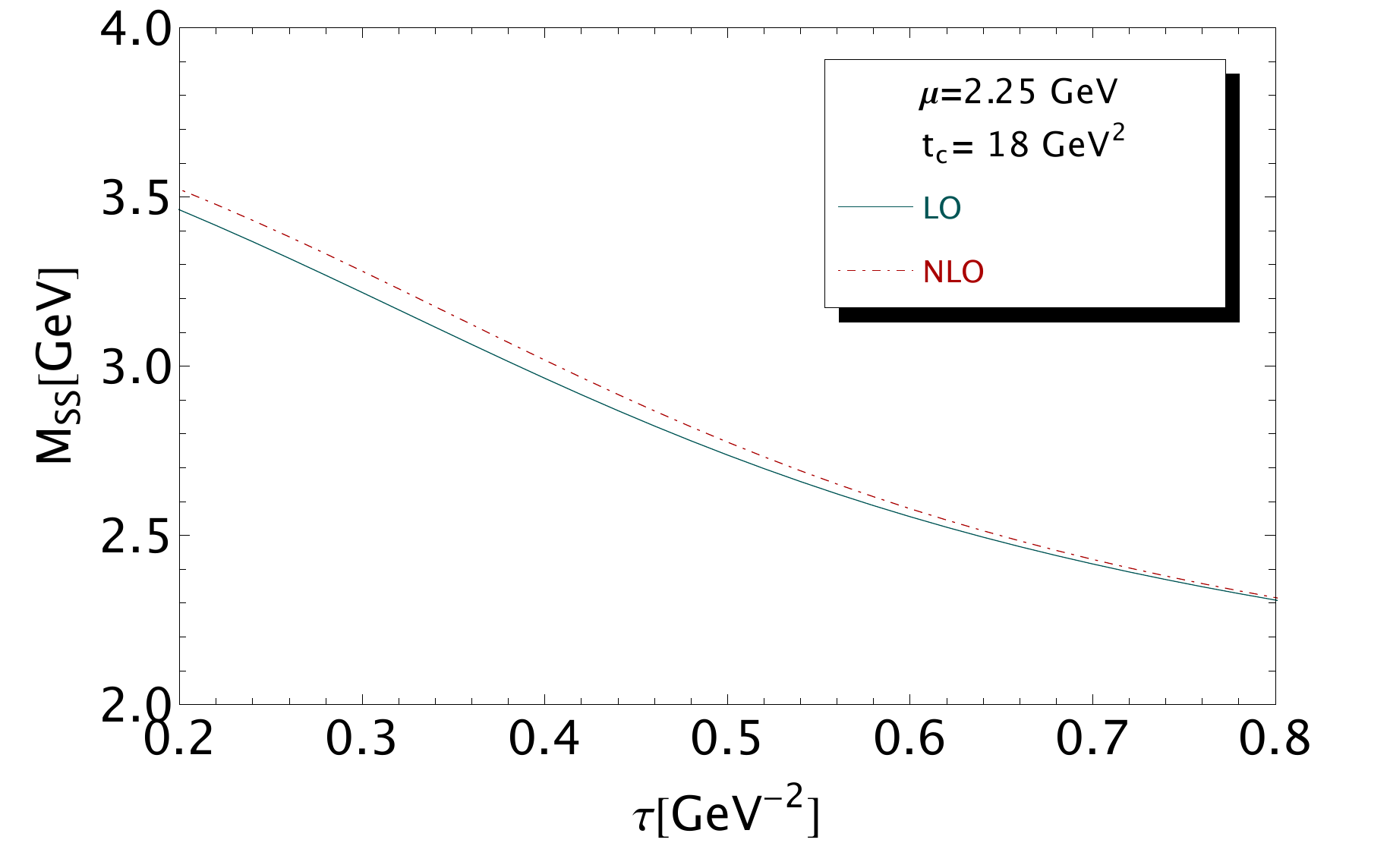}
\includegraphics[width=8.cm]{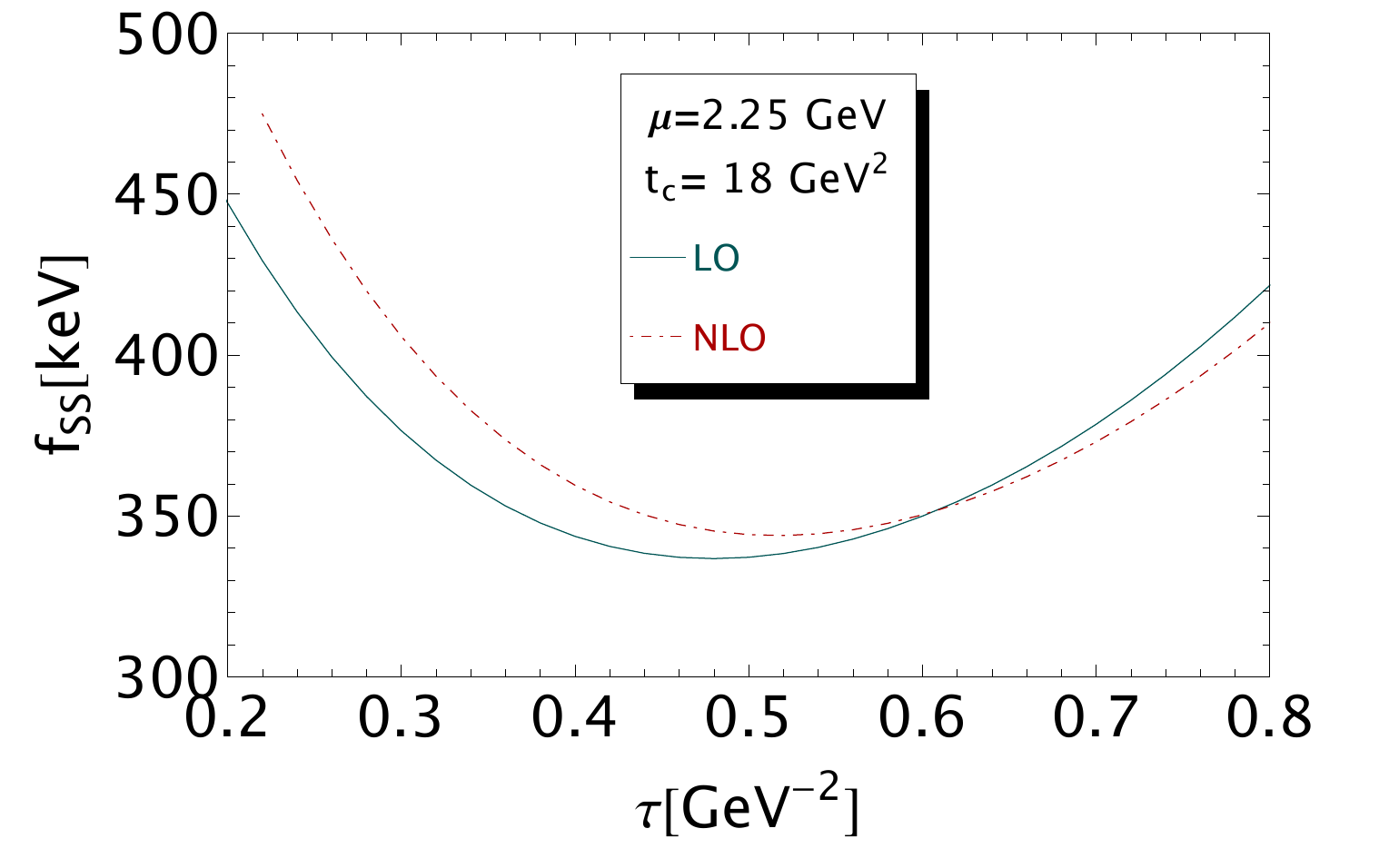}
\vspace*{-0.5cm}
\caption{\footnotesize  $M_{SS}$ and $f_{SS}$  as function of $\tau$ at LO and NLO for fixed values of $t_c$ and $\mu$ and for the values of the QCD parameters given in Table\,\ref{tab:param}.} 
\label{fig:ss-lo-nlo}
\end{center}
\end{figure} 
\subsection*{\b LO versus NLO results}
We compare in Fig.\,\ref{fig:ss-lo-nlo} the $\tau$-behaviour of the mass and coupling for fixed $\tau$ and $\mu$ 
at LO and NLO of perturbative QCD in the $\overline{MS}$-scheme. One can notice that  the NLO corrections are relatively small.  At the stability point, the radiative corrections decreases the $SS$ (rep. $AA$) mass by 46 (resp. 22) MeV and increases the coupling  by 7 (resp. 8) keV. 
\section{The $0^{++}~ \bar PP$ and $VV$ tetraquarks}
The two channels present similar features. Then, it suffices to show explicitly the analysis for the $PP$ channel.
\subsection*{\b $\tau$- and $t_c$-stabilities}
\begin{figure}[hbt]
\begin{center}
\centerline {\hspace*{-7.cm} \bf a)\hspace*{8cm} \bf b) }
\vspace{0.25cm}
\includegraphics[width=7.7cm]{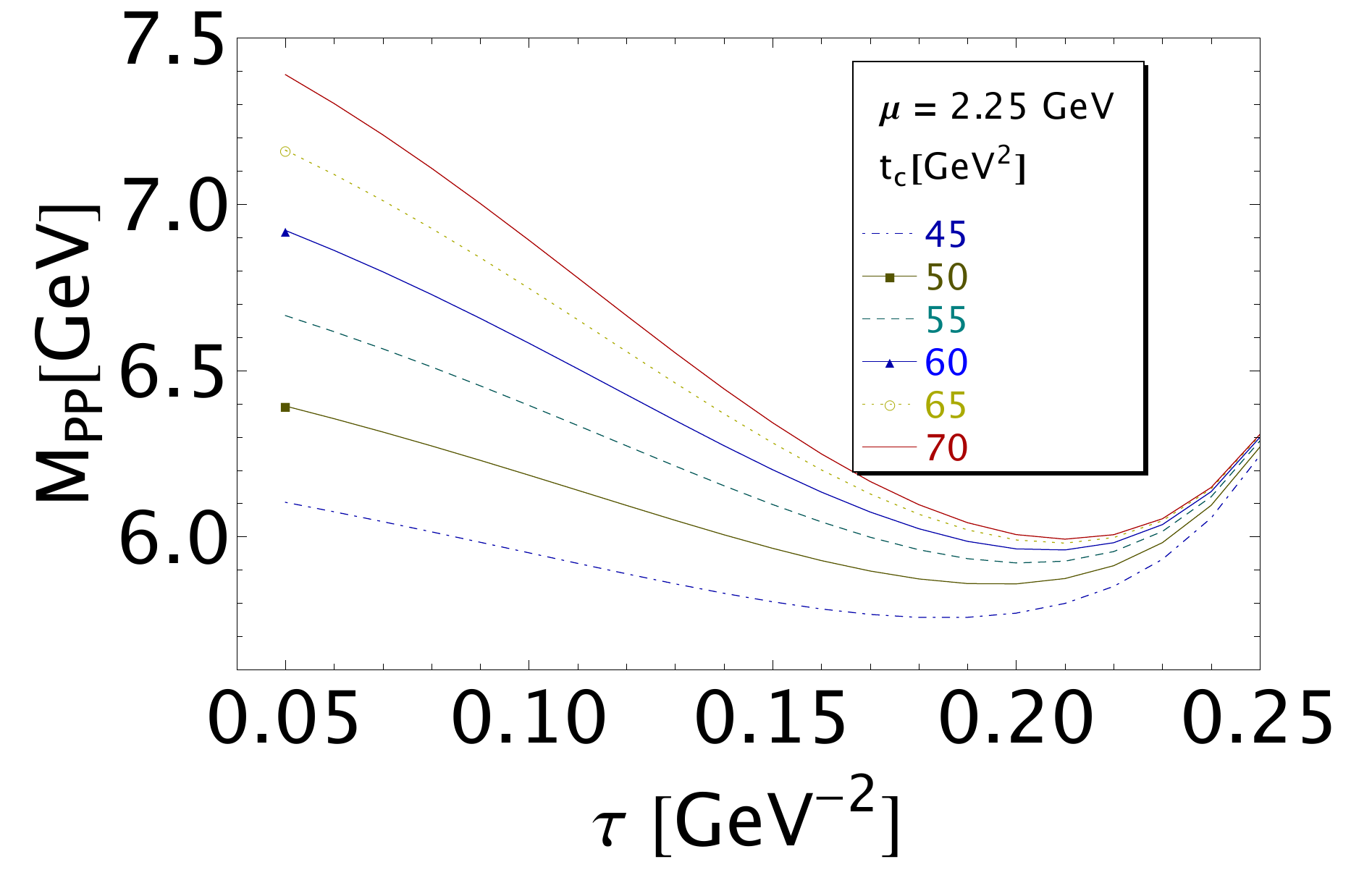}
\includegraphics[width=8cm]{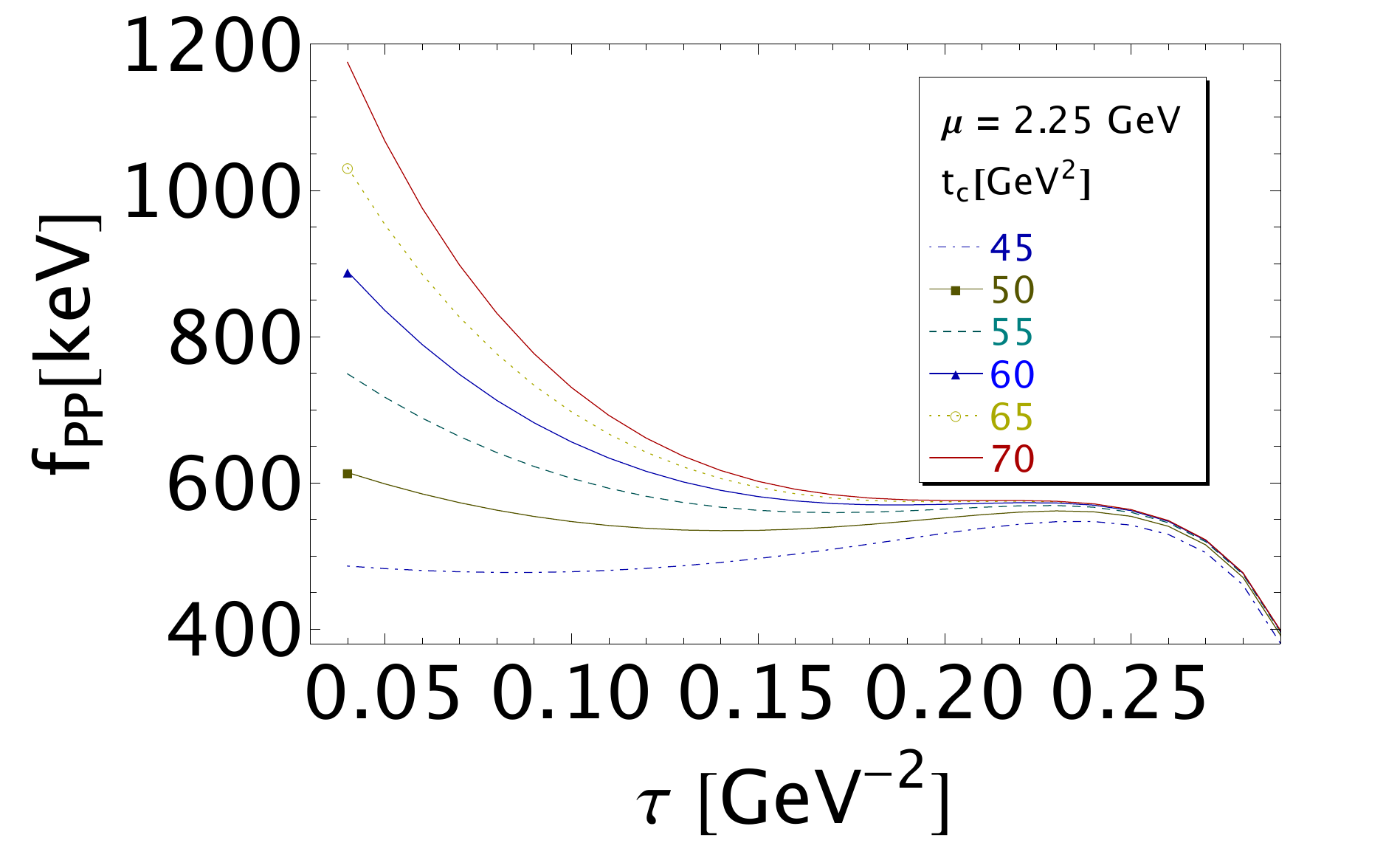}
\vspace*{-0.5cm}
\caption{\footnotesize  $f_{PP}$ and $M_{PP}$ as function of $\tau$ at NLO for different values of $t_c$, for $\mu$=2.25 GeV and for values of the QCD parameters given in Table\,\ref{tab:param}.} 
\label{fig:pptau}
\end{center}
\end{figure} 
The analysis is shown in Fig.\,\ref{fig:pptau}. Compared to the previous cases of $SS$ and $AA$ configurations, one can notice that the stabilities are reached for smaller values of $\tau\simeq (0.15\sim 0.20)$ GeV$^{-2}$ and for larger values of $t_c\geq 45$ GeV$^2$. This peculiar feature can be understood from the QCD expression of the corresponding correlators, where the $\la\bar\psi\psi\ra$ and $\la\bar\psi\psi\ra^2$ contribute largely and in a negative way which necessites to work at higher energies for having a positive QCD expression of the spectral function and a convergence of the OPE. As a consequence of the duality between the QCD and experimental sides, the resulting value of the lowest resonance mass becomes relatively high (see Table\,\ref{tab:res0}). Notice that working only with the ratio of moments ${\cal R}_0$ to extract the meson mass without inspecting the moment ${\cal L}^c_0$ leads to misleading results as one can obtain a lower mass at larger values of $\tau$ but one does not find that this low mass comes from the ratio of imaginary decay constants from ${\cal L}^c_0$. 
\subsection*{\b $\mu$-stability}
\begin{figure}[hbt]
\vspace*{-0.25cm}
\begin{center}
\centerline {\hspace*{-7.cm} \bf a)\hspace*{8cm} \bf b) }
\vspace{0.25cm}
\includegraphics[width=8.cm]{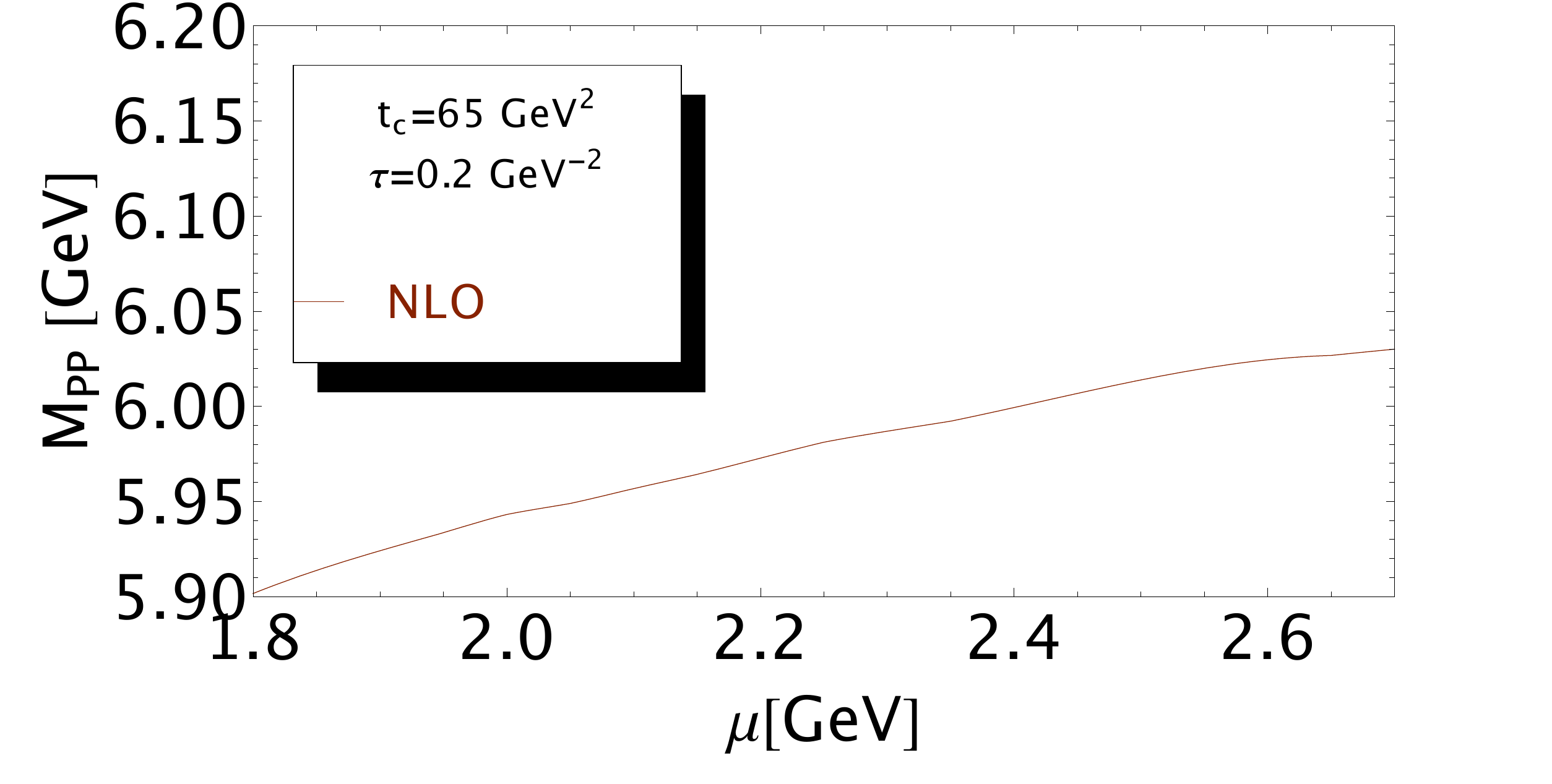}
\includegraphics[width=8.cm]{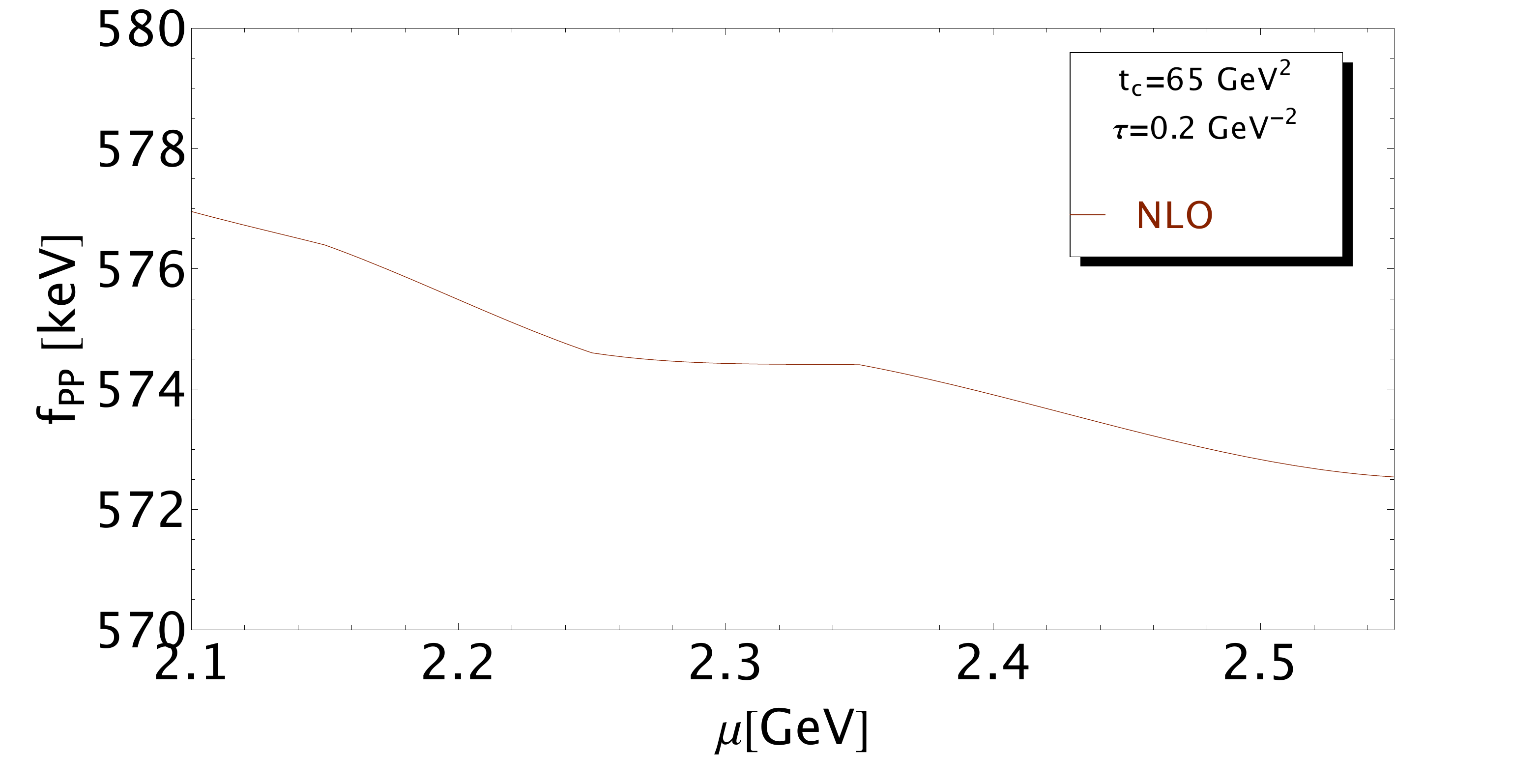}
\vspace*{-0.5cm}
\caption{\footnotesize  $M_{PP}$ and $f_{PP}$  as function of $\mu$ at NLO for fixed values of $t_c$ and $\tau$ and for the values of the QCD parameters given in Table\,\ref{tab:param}.} 
\label{fig:ppmu}
\end{center}
\end{figure}

The $\mu$-behaviour of the mass and coupling is shown in Fig.\,\ref{fig:ppmu} where one can see inflexion points at $\mu\simeq (2.25\sim 2.35)$ GeV which are consistent with the one for the $SS$ and $AA$ discussed previously. 
\subsection*{\b LO versus NLO results}
\begin{figure}[H]
\vspace*{-0.25cm}
\begin{center}
\centerline {\hspace*{-7.cm} \bf a)\hspace*{8cm} \bf b) }
\vspace{0.25cm}
\includegraphics[width=8.cm]{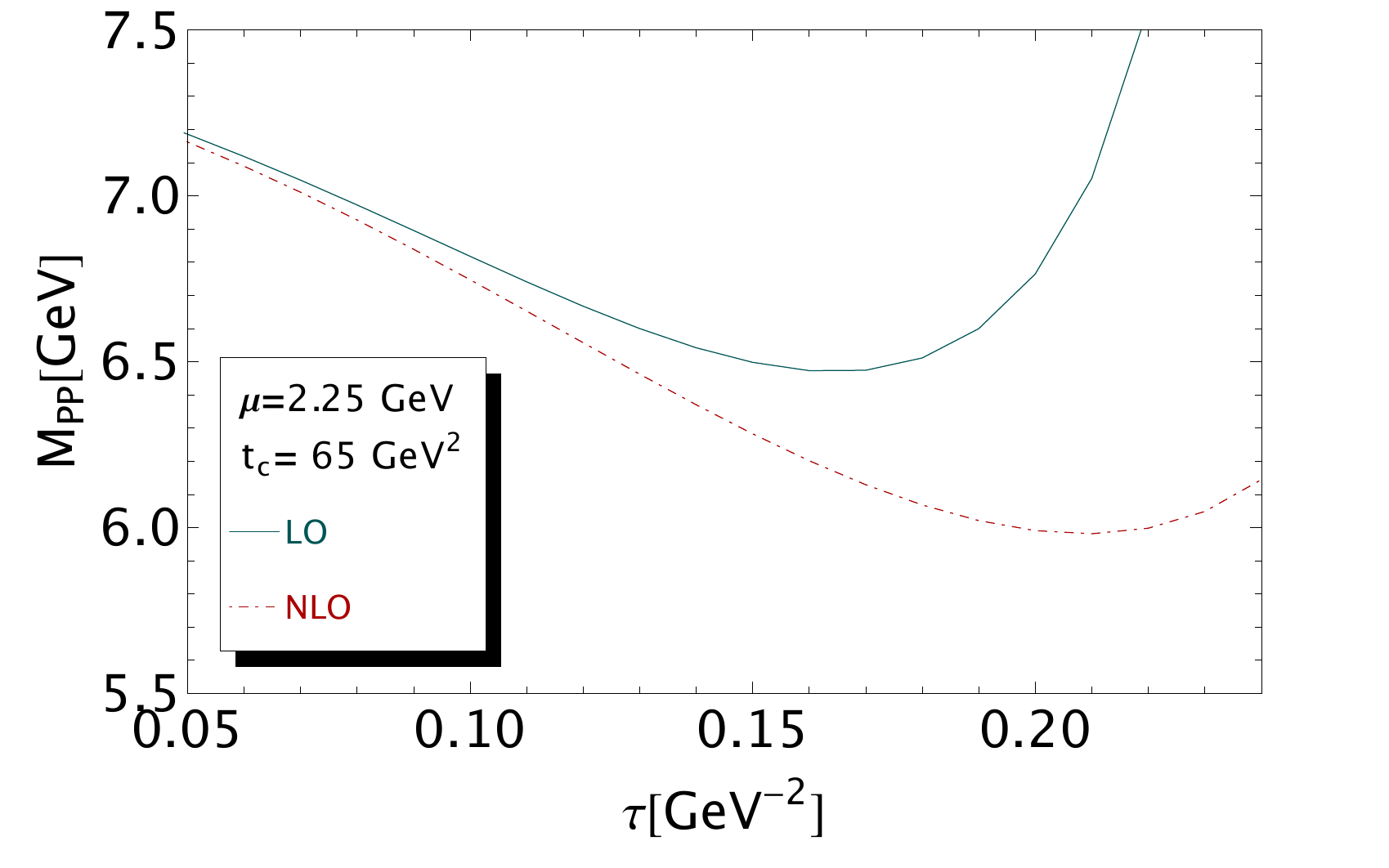}
\includegraphics[width=8.cm]{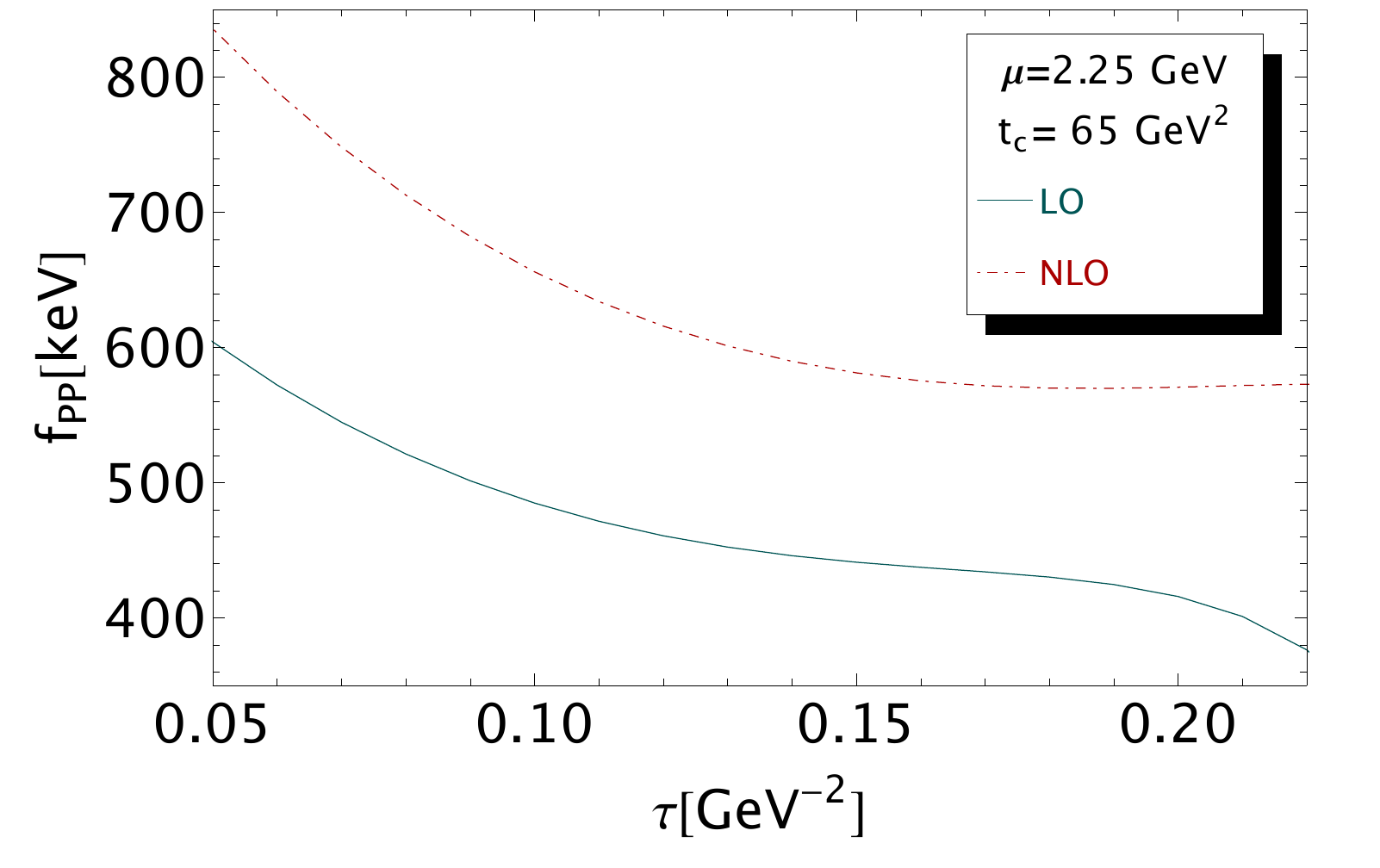}
\vspace*{-0.5cm}
\caption{\footnotesize  $M_{PP}$ and $f_{PP}$  as function of $\tau$ at LO and NLO for fixed values of $t_c$ and $\mu$ and for the values of the QCD parameters given in Table\,\ref{tab:param}.} 
\label{fig:pp-lo-nlo}
\end{center}
\end{figure}

We compare in Fig.\,\ref{fig:pp-lo-nlo} the $\tau$-behaviour of the mass and coupling for fixed $t_c$ and $\mu$ 
at LO and NLO of perturbative QCD in the $\overline{MS}$-scheme. One can notice that the $\alpha_s$ corrections are large for $PP$ which decrease the  mass by 495 MeV while  increase the coupling by 137 keV.  On the contrary, the NLO corrections for $VV$ are relatively small which decrease the mass by 20 MeV and increase the coupling by  52 MeV. 
\section{The $0^{++}$ molecules}
The behaviours of the different curves are similar to the previous cases.
\subsection*{\b $D^*K^*$}
The curves for the $D^*K^*$ molecule are similar to the cases of $SS$ and $AA$ tetraquarks. Here the NLO corrections are $- 50$ MeV for the mass and $+16$ keV for the coupling. 
\subsection*{\b $D_1K_1,~D^*_0K^*_0$}
The curves for the $D_1K_1,~D^*_0K^*_0$ molecules are similar to the cases of the $PP$ and $VV$ tetraquarks.
The NLO corrections are  $-394$ (resp. $+36$) MeV for the mass  and +46 (resp. $-106$) keV for the coupling of the $D_1K_1$ (resp. $D^*_0K^*_0$) molecules. 
\begin{table*}[hbt]
\setlength{\tabcolsep}{0.1pc}
\catcode`?=\active \def?{\kern\digitwidth}
{\scriptsize
\begin{tabular*}{\textwidth}{@{}l@{\extracolsep{\fill}}|cccccc  cc | cccc cccc}
\hline
\hline
      States         & \multicolumn{8}{c |}{Scalars ($0^+$)} 
                  & \multicolumn{8}{ c }{Vectors ($1^-$) } \\
\cline{1-9} \cline{10-17}
Parameters          & \multicolumn{1}{c}{$SS$} 
       & \multicolumn{1}{c}{$AA$} 
                 & \multicolumn{1}{c}{{$PP$}} 
                 & \multicolumn{1}{c}{{$VV$ }} 
              & \multicolumn{1}{c}{$DK$} 
       & \multicolumn{1}{c}{$D^*K^*$} 
                 & \multicolumn{1}{c}{{$D_1K_1$}} 
                 & \multicolumn{1}{c |}{{$D^*_0K^*_0$}} 
                 
                  & \multicolumn{1}{ c}{$AP$} 
       & \multicolumn{1}{c}{$PA$}
            & \multicolumn{1}{c}{{$SV$}} 
                 & \multicolumn{1}{c}{{$VS$}} 
                       & \multicolumn{1}{c}{$D_1K$} 
       & \multicolumn{1}{c}{$DK_1$}
               & \multicolumn{1}{c}{{$D^*_0K^*$}} 
                 & \multicolumn{1}{c}{{$D^*K^*_0$}}
                                \\
\cline{1-5} \cline{6-9} \cline{10-13} \cline{14-17}
$t_c$ [GeV$^2$]&14-18&14-18&50-65&42-55&12-18&14-18&40-55&50-65&40-55&12-18&12-18&40-55&12-18&40-55&14-18&40-55\\
$\tau$ [GeV]$^{-2} 10^2$&45-51&47-53&17-21&17-18&73-77&45-53&26-28&13-15&20-22&36-50&41-53&17-20&32-47&22-24&41-49&20-22\\

\hline\hline
\end{tabular*}
}
 \caption{Values of the LSR parameters $t_c$ and the corresponding $\tau$ at the optimization region for the PT series up to NLO and for the OPE truncated at $\la g^{3}_{s} G^3\ra$.}
\label{tab:lsr-param}
\end{table*}
\subsection*{\b Results}
We show in Table\,\ref{tab:lsr-param} the different values of the LSR parameters ($t_c,\tau$) used to deduce the optimal results given in Table\,\ref{tab:res0}. The results will be commented later on.

\begin{table}[H]
\setlength{\tabcolsep}{0.15pc}
\catcode`?=\active \def?{\kern\digitwidth}
    {\small
  \begin{tabular*}{\textwidth}{@{}l@{\extracolsep{\fill}}|ccccccccc   c ccc  l}
\hline
\hline
 Observables\,&$ \Delta t_c$&$\Delta\tau$&$ \Delta\mu $ &$\Delta \alpha_s$& $\Delta m_s$ & $\Delta m_c$&$\Delta \bar\psi\psi$&$\Delta \kappa$&$\Delta \alpha_s G^2 $&$\Delta M_0^2$&$\Delta\bar\psi\psi^2$&$\Delta G^3 $&$\Delta M_G$& Values\\
\hline
$\bf 0^+$ {\bf States}\\
\cline{0-0} 
$f_{G}$ [keV] \\
\cline{0-0} 
 {\it Tetraquark}\\
$SS$&7.00&0.48&14.55&3.95&0.66&1.85&1.30&0.85&0.15&0.95&22.15&0.05&1.50&345(28)\\
$PP$&10.00&1.70&1.10&0.75&0.97&2.79&3.32&3.80&0.29&0.55&13.09&0.05&37&538(41)\\
$VV$&27.00&4.50&7.05&1.84&0.01&3.75&3.86&4.40&0.25&0.90&14.13&0.06&58&713(66)\\

$AA$&8.00&0.15&21.22&6.25&1.22&2.70&1.35&1.03&0.12&1.10&35.11&0.07&2.10&498(43) \\
 {\it Molecule}\\
 $DK$&&&&&&&&&&&&&&254(48)\,(Ref.\,\cite{X5568})\\
$D^*K^*$ &12.95&0.40&16.6&4.4&1.1&2.25&1.15&9.4&0.0&3&22.25&0.0&0.4&405(33)\\
$D_1K_1$ &10.00&0.54&1.43&0.48&0.44&3.56&5.55&9.9&0.21&4.10&15.75&0.0&52&664(57)\\
$ D^*_0K^*_0$&11.32&0.30&2.02&0.48&6.96&0.41&4.73&2.78&0.65&0.65&5.0&0.06&10&249(18)\\
\cline{0-0} 
$M_{G}$ [MeV]\\
\cline{0-0}
 {\it Tetraquark} \\
 $SS$&7.00&19&1.2&3.50&1.15&1.90&2.45&2.83&0.3&0.45&1.5&0.02&--&2736(21) \\
 
$PP$&36.00&19.5&29.3&7.70&2.30&5.25&14.8&16.8&1.41&4.20&80.8&0.32&--&5917(98)\\
$VV$&96.00&26.0&34.4&2.4&1.13&4.12&10.9&12.4&0.8&3.67&62.9&0.22&--&5704(149) \\

$AA$&17.00&62&1.86&2.80&2.52&2.05&2.17&2.47&0.21&0.40&1.81&0.03&--&2675(65)\\
 {\it Molecule}\\
$DK$&&&&&&&&&&&&&&2402(42)\,(Ref.\,\cite{X5568})\\
$D^*K^*$ &2.91&38.45&5.25&4.01&2.14&1.39&3.07&9.3&0.1&0.9&7.0&0.0&--& 2808(41)\\
$D_1K_1$ &50.50&22.70&29.4&8.0&3.0&4.34&16.8&49.0&0.65&18.3&75.6&0.1&--&5258(113)\\
$ D^*_0K^*_0$&123.0&37.50&33.3&9.18&14.9&2.82 &39.1&39.1&5.01&4.04&67.8&0.7&--&6270(160)\\
\\
\hline\hline
\end{tabular*}
{\scriptsize
 \caption{Sources of errors and predictions from LSR at NLO and  for the decay constants and masses of the ($0^+$) scalar molecules and tetraquark states. The errors from the QCD input parameters are from Table\,\ref{tab:param}. $\Delta \mu$ is given in Eq.\,\ref{eq:mu}. We take $|\Delta \tau|= 0.02$ GeV$^{-2}$.
 }
 \label{tab:res0}
}
}
\end{table}

\begin{table}[hbt]
\setlength{\tabcolsep}{0.15pc}
\catcode`?=\active \def?{\kern\digitwidth}
    {\small
  \begin{tabular*}{\textwidth}{@{}l@{\extracolsep{\fill}}|ccccccccc   c ccc  l}
\hline
\hline
 Observables\,&$ \Delta t_c$&$\Delta\tau$&$ \Delta\mu $ &$\Delta \alpha_s$& $\Delta m_s$ & $\Delta m_c$&$\Delta \bar\psi\psi$&$\Delta \kappa$&$\Delta \alpha_s G^2 $&$\Delta M_0^2$&$\Delta\bar\psi\psi^2$&$\Delta G^3 $&$\Delta M_G$& Values\\
\hline

$\bf 1^-$ {\bf States}\\
\cline{0-0} 
$f_{G}$ [keV] \\
\cline{0-0} 
  {\it Tetraquark}\\
 $AP$&18.7&5.4&5.64&0.70&0.24&2.02&2.53&2.88&0.13&1.48&11.62&0.03&30&416(38)\\
  $PA$&10.4&0.22&13.0&3.62&0.71&1.46&1.04&1.19&0.07&0.97&23.0&0.05&2.5&285(29)\\
 $SV$&7.1&0.19&11.0&3.22&0.83&1.30&0.97&1.10&0.11&2.37&20.4&0.04&2.2&259(25)\\
 $VS$&27.8&7.3&2.35&0.41&0.32&1.98&2.10&2.39&0.10&0.29&11.45&0.03&29&412(43)\\
   {\it Molecule}\\
  $D_1K$&8.0&0.20&8.2&2.90&0.1&0.9&0.8&8.3&0.3&2.45&14.5&0.1&2.9&191(21)\\
  
  $DK_1$&12&1.6&1.0&0.15&0.50&1.50&2.65&3.95&0.13&0.75&10&0.08&26&351(31)\\
  
  $D^*_0K^*$&5.42&0.23&9.26&2.47&0.26&1.11&1.38&9.52&0.13&0.13&16.6&0.1&1.7& 216(22)\\
 $ D^*K^*_0$&7.54&0.16&3.63&0.94&2.91&0.97&3.39&4.38&0.33&2.08&8.66&0.05&18&255(23)\\
 \cline{0-0} 
$M_{G}$ [MeV] \\
\cline{0-0} 
 {\it Tetraquark}\\
 $AP$&112&27.1&28.1&1.82&0.78&4.21&10.9&12.5&0.84&10.2&87.9&0.18&--&5542(139)\\
 $PA$&1.68&29.0&1.7&3.83&1.05&2.3&4.17&4.75&0.14&0.30&3.11&0.0&--&2666(32)\\
 $SV$&1.17&27.6&5.0&4.08&1.41&2.48&5.62&6.40&0.24&2.59&1.43&0.0&--&2593(31)\\
 $VS$&149&28.2&17.6&3.36&0.33&4.06&9.69&11.1&0.57&2.73&84.6&0.23&--&5698(175)\\
  {\it Molecule}\\
  $D_1K$&6&38.5&22&5.0&5.0&2.5&6&5.5&2.0&5.0&3.5&0.0&--& 2676(47)\\
  $DK_1$&80&28&16&4.0&4.0&4.0&15&55&1&4&130&0.65&--&5377(166)\\
  $D^*_0K^*$&3.6&38.5&1.8&5.3&2.4&1.9&6.5&8.8&0.2&2.8&6.9&0.05&--&2744(41)\\
 $ D^*K^*_0$&80.4&31.2&9.5&2.9&1.2&4.2&20.9&53.4&3.1&23.4&109&0.58&--&5358(153)\\

\hline\hline
\end{tabular*}
{\scriptsize
 \caption{Same as in Table\,\ref{tab:res0} but for the ($1^-$) vector states.}
 \label{tab:res1}
}
}
\end{table}

\section{The $1^{-}$ Vector states}
\subsection*{\b $AP,~VS$ tetraquarks and $DK_1,~D^*K^*_0$ molecules}
Their corresponding curves behave like the ones of the $PP,~VV$ $(0^{++})$ tetraquarks and of the $D_1K_1,~D^*_0K^*_0$ $(0^{++})$ molecules.    Including the NLO corrections, the $AP$ (resp. $VS$) mass decreases by 164 (resp.117) MeV while the coupling increases by 29 (resp. 63) keV. For the $DK_1$ (resp. $D^*K^*_0$) molecules, the mass decreases by 351 (resp. 48) MeV while the coupling increases by 17 (resp. decreases by  50) keV. 
\subsection*{\b $PA,~SV$ tetraquarks and $D_1K,~D^*_0K^*$ molecules}
Their corresponding curves behave like the ones of $SS,~AA$ $(0^{++})$ tetraquarks and of the $DK,~D^*K^*$ $(0^{++})$ molecules.  Including the NLO corrections, the $PA$ (resp. $SV$) mass changes by $-3$ (resp. +26) MeV while the coupling changes by (+2 (resp. $-2$) keV. For the $D_1K$ (resp. $D^*_0K^*$) molecules, the mass increases by 56 (resp. 46) MeV while the coupling decreases by 1 (resp. increases by 9) keV. 
\subsection*{\b Results}
The results are shown in Table\,\ref{tab:res1} and will be commented later on.
The different values of the LSR parameters ($t_c,\tau$) used to deduce the optimal results are shown in Table\,\ref{tab:lsr-param}.
\section{The first radial excitation $(\bar DK)_1$ of the $0^{++}(\bar D^-K^+)$ molecule}
For this purpose, we extend the analysis in Ref.\,\cite{X5568} by using a ``Two resonances" + $\theta(t-t_c)$``QCD continuum" parametrization of the spectral function. To enhance the contribtuion of the 1st radial excitation [hereafter called $(DK)_1$], we shall also work with the ratio of moments ${\cal R}_1$ in addition to ${\cal R}_0$ for getting the mass of $(DK)_1$. 
\subsection*{\b $\tau$- and $t_c$-stabilities}
\begin{figure}[hbt]
\begin{center}
\vspace{0.25cm}
\includegraphics[width=9cm]{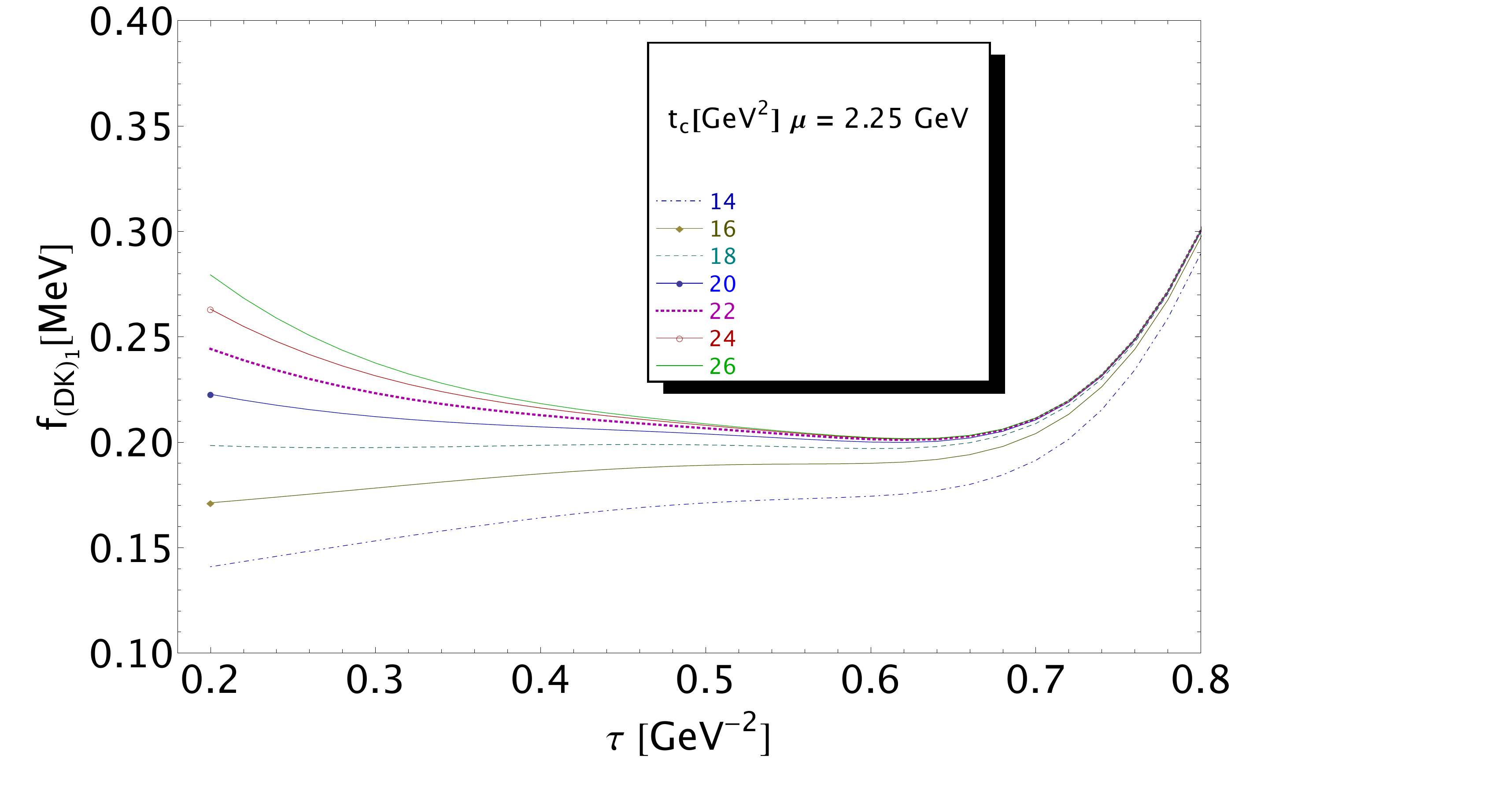}
\vspace*{-0.5cm}
\caption{\footnotesize  $f_{(DK)_1}$  from the first moment ${\cal L}^c_0$ as function of $\tau$ at NLO for different values of $t_c$, for $\mu$=2.25 GeV and for values of the QCD parameters given in Table\,\ref{tab:param}.} 
\label{fig:f-dk1}
\end{center}
\end{figure}

\begin{figure}[hbt]
\vspace*{-0.5cm}
\begin{center}
\centerline {\hspace*{-7.cm} \bf a)\hspace*{8cm} \bf b) }
\vspace{0.25cm}
\includegraphics[width=8.5cm]{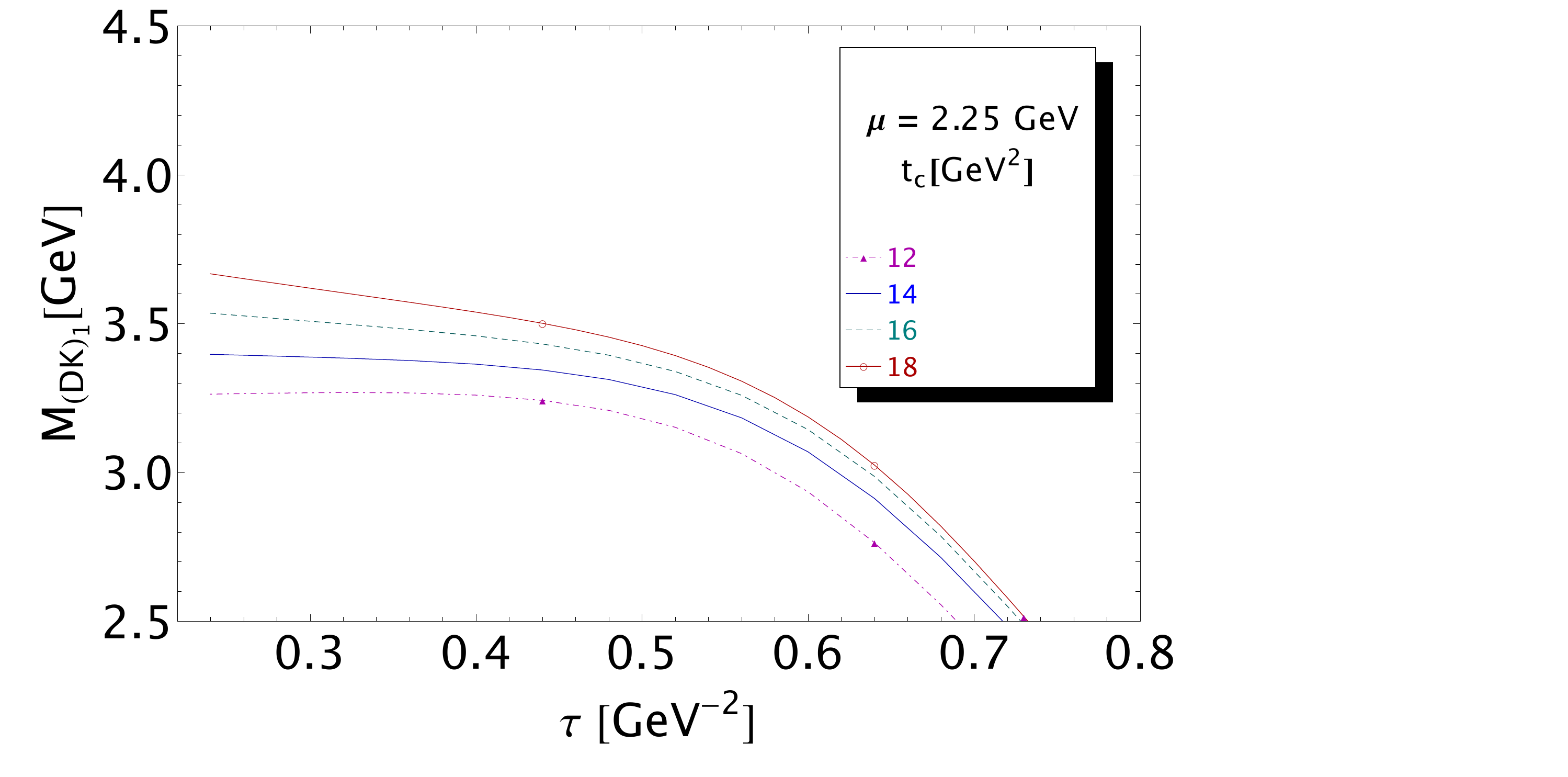}
\includegraphics[width=7.8cm]{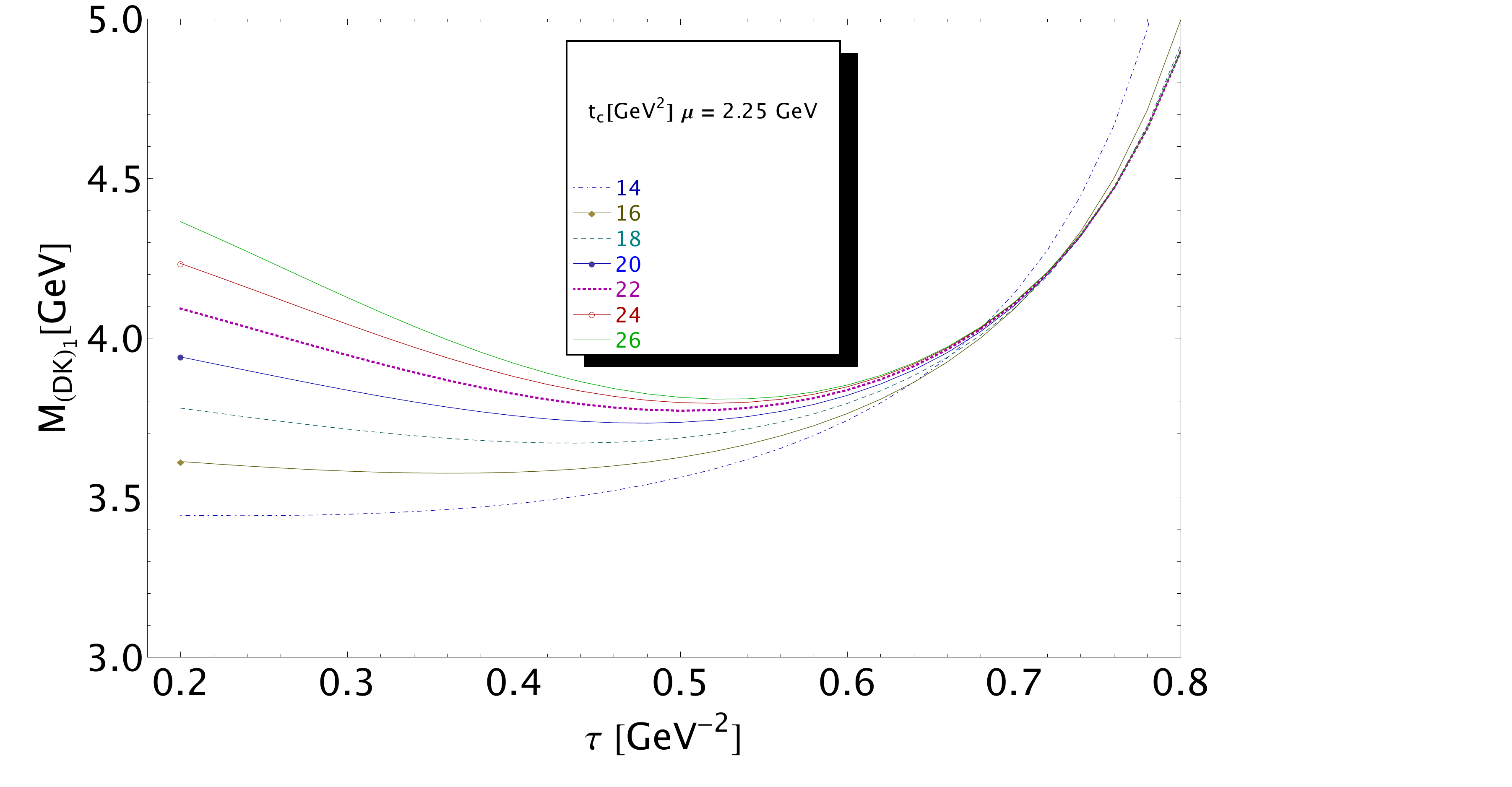}
\vspace*{-0.5cm}
\caption{\footnotesize   $M_{(DK)_1}$ as function of $\tau$ at NLO for different values of $t_c$, for $\mu$=2.25 GeV and for values of the QCD parameters given in Table\,\ref{tab:param} : {\bf a)} from the lowest ratio of moments ${\cal R}_0$;  {\bf b)} from the 2nd ratio of moments ${\cal R}_1$.} 
\label{fig:m-dk1}
\end{center}
\end{figure}

We show in Fig.\,\ref{fig:f-dk1} the $\tau$- and $t_c$-behaviours of the coupling from ${\cal L}^c_0$ and in Fig\,\ref{fig:m-dk1} the ones of the mass from ${\cal R}_0$ and ${\cal R}_1$ using as input the values of the lowest ground state mass and coupling obtained in Eq.\,\ref{eq:lowest-dk}. 
-- One can notice that the coupling  from ${\cal L}^c_0$ stabilizes for $\tau\simeq (0.55\sim 0.65)$ GeV$^{-2}$ which is slightly lower than the value $\tau=0.7$ GeV$^{-2}$ corresponding to the one-resonance parametrization. The corresponding values of $t_c$ are 18 to 24 GeV$^2$ compared to 12 to 18 GeV$^2$ for the one resonance case.  The result is given in Table\,\ref{tab:resdk0} where one can notice that the largest error comes from the coupling of lowest ground state.

-- The analysis of the mass from ratio of moments ${\cal R}_0$ and  ${\cal R}_1$ is shown in Figs.\,\ref{fig:m-dk1} a) and b). One can notice that the prediction from ${\cal R}_1$ is more precise due to its more sensitivity on the contribution of $(DK)_1$ in the high-energy region ($\tau\simeq 0.5$ GeV$^{-2}$ and $t_c\simeq (18-24)$ GeV$^2$) from which we extract the final result compiled in Table\,\ref{tab:resdk0}.

-- One can notice that the mass of the radial excitation is in the range of $t_c\simeq (12\sim 18)$ GeV$^2$ where the mass of the lowest ground has been obtained indicating that the value of the QCD continuum threshold $t_c$ in the ``One resonance" parametrization gives an approximate value of the 1st radial excitation. 

-- The set of $(\tau, t_c)$-values where the optimal results have been obtained are compiled in Table\,\ref{tab:tc-radial}. 
\begin{table*}[hbt]
\setlength{\tabcolsep}{0.01pc}
\catcode`?=\active \def?{\kern\digitwidth}
{\scriptsize
\begin{tabular*}{\textwidth}{@{}l@{\extracolsep{\fill}}|cccc | cccc}
\hline
\hline
   States            & \multicolumn{4}{c}{Scalars ($0^+$)} 
                  & \multicolumn{4}{ c }{Vectors ($1^-$)} \\
         \cline{1-2}         
\cline{2-5} \cline{5-9}
    Parameters     & \multicolumn{1}{c}{$(SS)_1$} 
       & \multicolumn{1}{c}{$(AA)_1$} 
              & \multicolumn{1}{c}{$(DK)_1$} 
       & \multicolumn{1}{c |}{$(D^*K^*)_1$} 
                 
       & \multicolumn{1}{c}{$(PA)_1$}
                   & \multicolumn{1}{c}{{$(SV)_1$}} 
                   & \multicolumn{1}{c}{$(D_1K)_1$} 
                   & \multicolumn{1}{c}{{$(D^*_0K^*)_1$}} 
                              \\
\cline{1-5} \cline{5-9} 
$t_c$ [GeV$^2$]&28-36&28-36&18-24&32-40&28-36&28-36&28-36&28-36\\
\cline{1-1}
$\tau$ [GeV]$^{-2}10^2$&&&&&&&&\\
\cline{1-1}
$f_{(G)_1}$&44-46&48-50&55-65&35-45&24-36&26-40&28-36&36-42\\
$M_{(G)_1}$&36-40&36-40&50&40&36-40&38-42&34-38&34-38\\
\hline\hline
\end{tabular*}
}
 \caption{Values of the LSR parameters $(t_c, \tau)$ at the otpimization region where the masses and couplings of the 1st radial excitations are obtained for the PT series up to NLO and for the OPE truncated at $\la g^{3}_{s} G^3\ra$.}
\label{tab:tc-radial}
\end{table*}

\subsection*{\b $\mu$-stability}
\begin{figure}[hbt]
\vspace*{-0.25cm}
\begin{center}
\centerline {\hspace*{-7.cm} \bf a)\hspace*{8cm} \bf b) }
\vspace{0.25cm}
\includegraphics[width=8.cm]{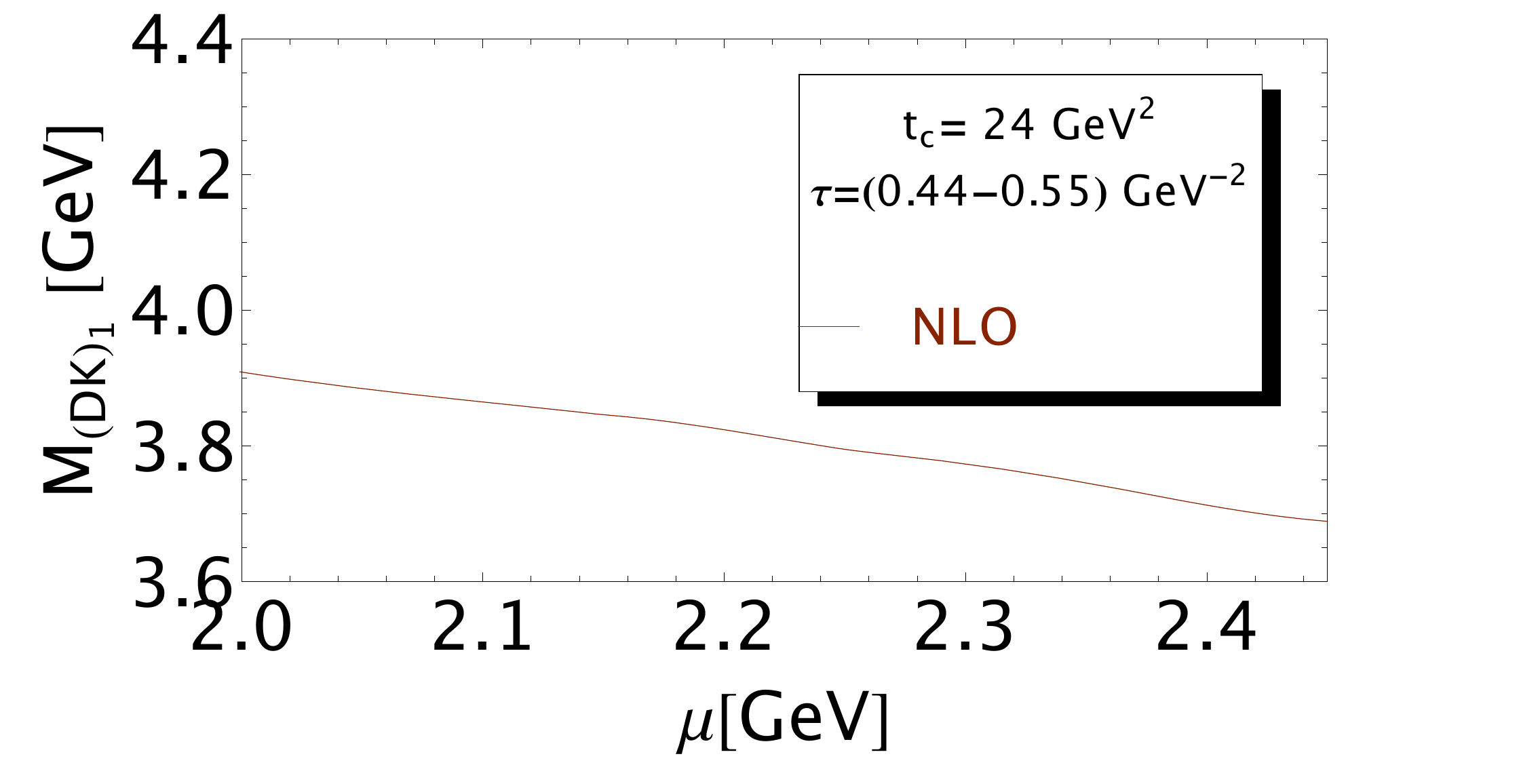}
\includegraphics[width=8.cm]{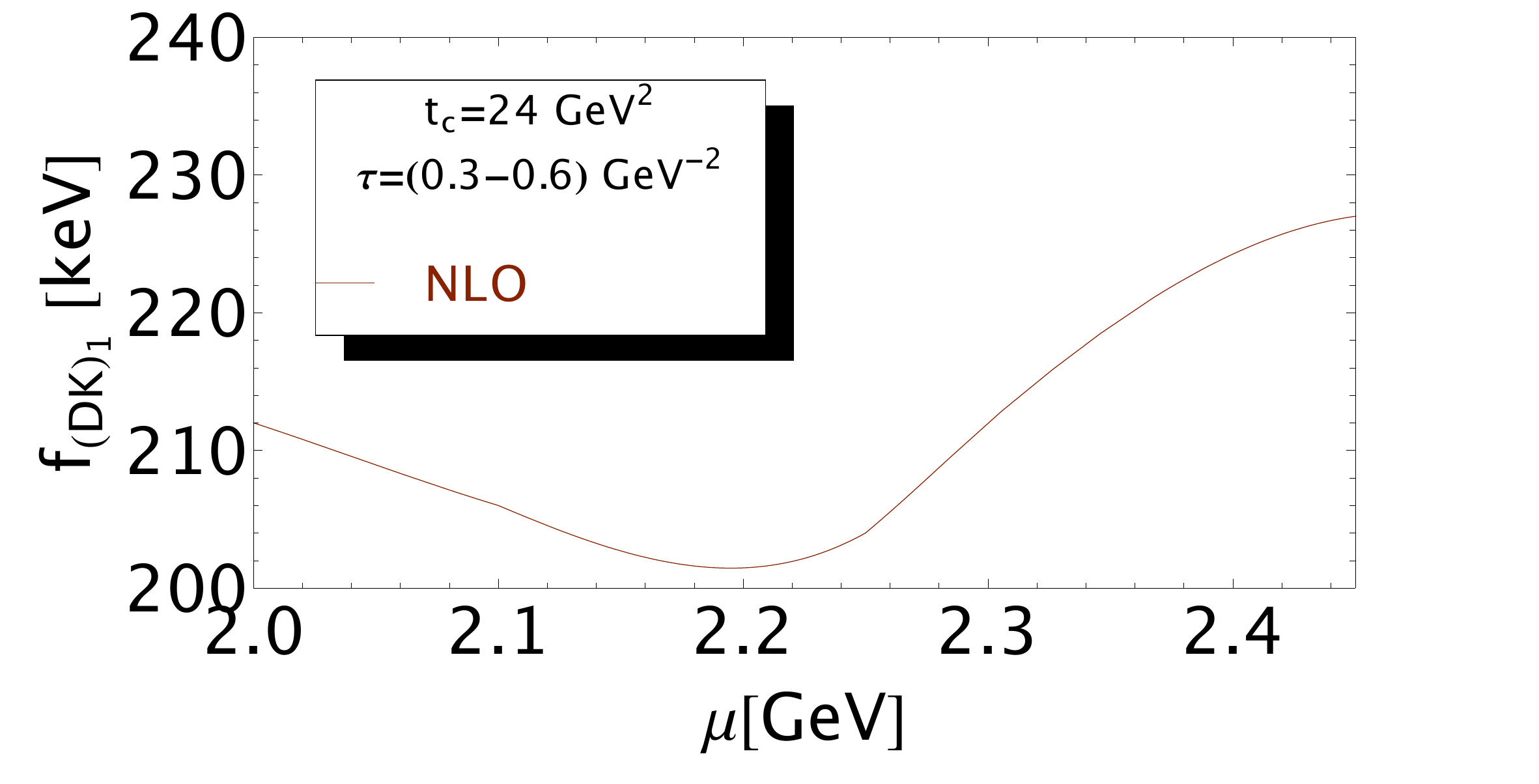}
\vspace*{-0.5cm}
\caption{\footnotesize  $M_{(DK)_1}$ and $f_{(DK)_1}$  as function of $\mu$ at NLO for fixed values of $t_c$ and $\tau$ and for the values of the QCD parameters given in Table\,\ref{tab:param}.} 
\label{fig:dkmu}
\end{center}
\vspace*{-0.5cm}
\end{figure}

We study in Fig.\,\ref{fig:dkmu} the $\mu$-stability fixing $t_c=24$ GeV$^2$ and for $\tau\approx (0.3\sim 0.6)$ GeV$^{-2}$ depending on the value of $\mu$ where the $\tau$-stability is reached. 
\section{The first radial excitation $(\bar D^*K^*)_1$ of the $0^{++}(\bar D^*K^*)$ molecule}
\subsection*{\b $\tau$- and $t_c$-stabilities}
\begin{figure}[hbt]
\begin{center}
\vspace{0.25cm}
\includegraphics[width=8cm]{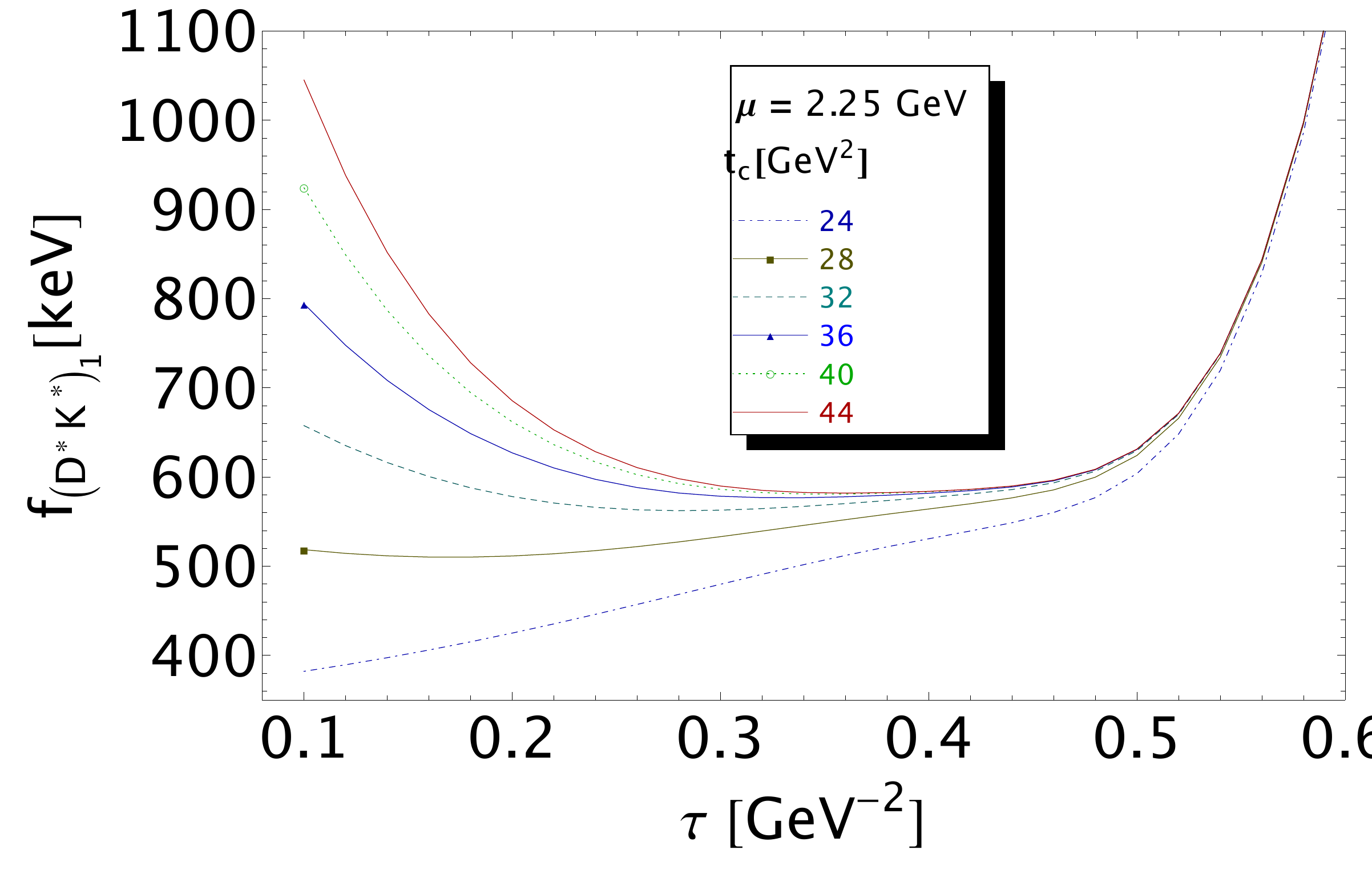}
\includegraphics[width=8cm]{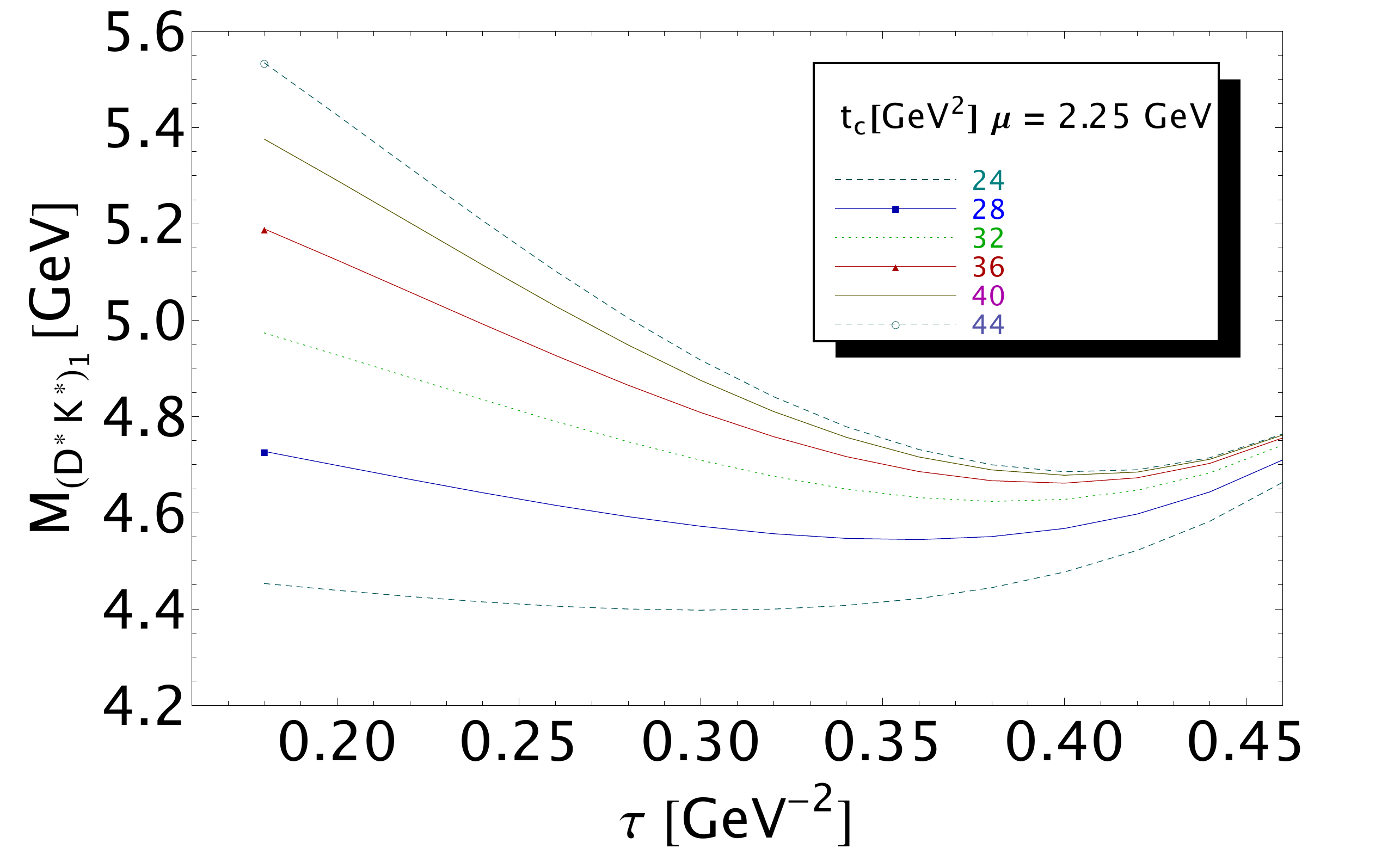}
\vspace*{-0.5cm}
\caption{\footnotesize  $f_{(D^*K^*)_1}$  from the first moment ${\cal L}^c_0$ as function of $\tau$ at NLO for different values of $t_c$, for $\mu$=2.25 GeV and for values of the QCD parameters given in Table\,\ref{tab:param}.} 
\label{fig:f-d*k*1}
\end{center}
\end{figure} 
We show in Fig.\,\ref{fig:f-d*k*1}a)  the $\tau$- and $t_c$-behaviours of the coupling from ${\cal L}^c_0$ and in Fig\,\ref{fig:f-d*k*1}b) the ones of the mass from ${\cal R}_1$ using as input the values of the lowest ground state mass and coupling obtained in Table\,\ref{tab:res0}. The optimal results are obtained for the $(t_c,\tau)$ values given in Table\,\ref{tab:tc-radial}.

\subsection*{\b $\mu$-stability}
\begin{figure}[H]
\begin{center}
\vspace{0.25cm}
\includegraphics[width=7cm]{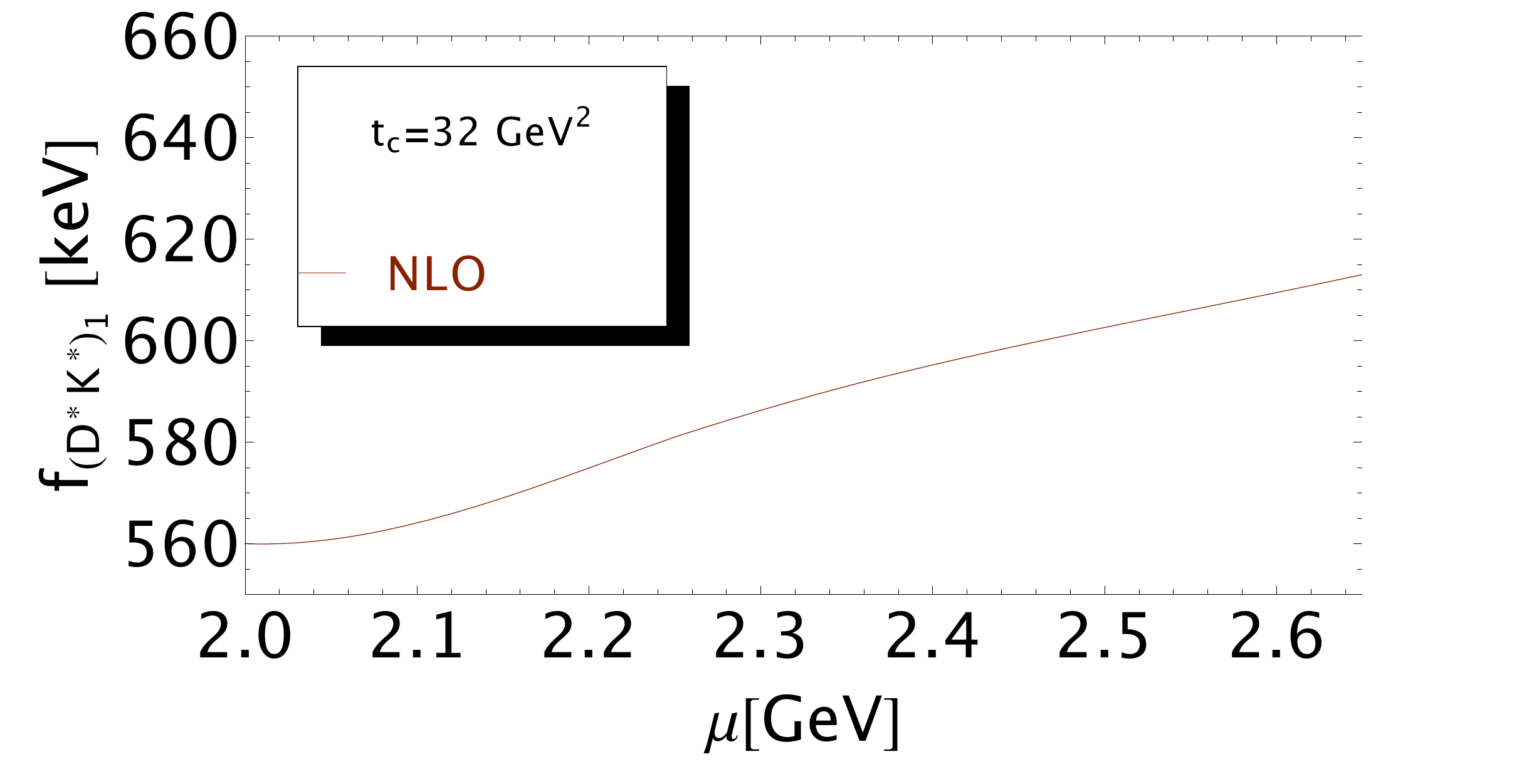}
\includegraphics[width=7cm]{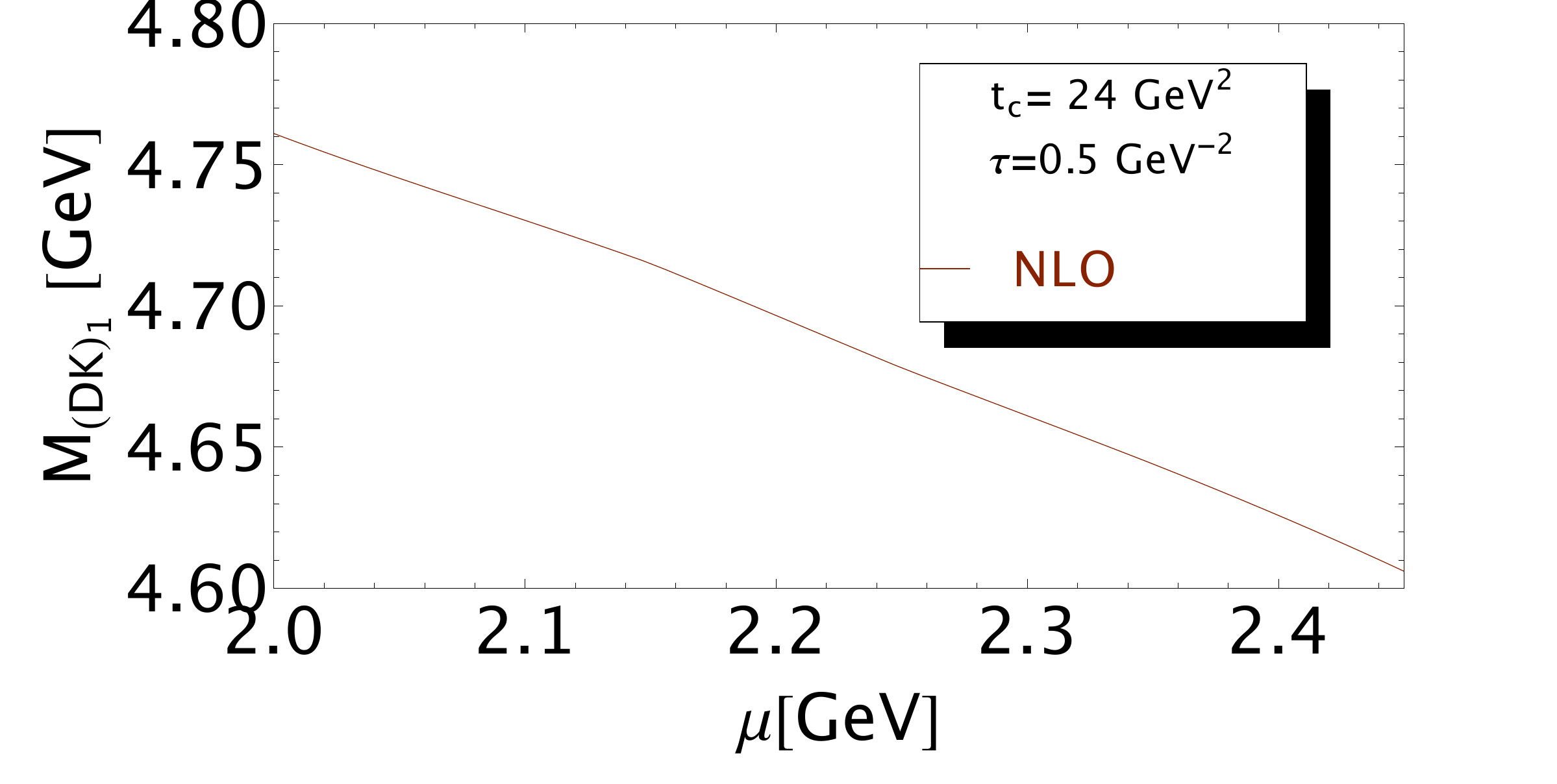}
\vspace*{-0.5cm}
\caption{\footnotesize  $\mu$-behaviour of the $(D^*K^*)_1$  mass and coupling.} 
\label{fig:d*k*1-mu}
\end{center}
\end{figure}

The $\mu$-behaviours of the coupling and mass are shown in Fig.\,\ref{fig:d*k*1-mu}. 
We shall extract  the mass and coupling of the 1st radial excitation $(\bar D^*K^*)_1$.  We shall show the analysis explictily as it may  (a priori) differ from the one of $(DK)_1$ (position of the optimal $\tau$ and value of $t_c$) as the mass of the $\bar D^*K^*$ molecule is higher than that of $\bar DK$. 
\section{The first radial excitation $(\bar PA)_1$ of the $1^{-}(\bar PA)$ tetraquark}
\subsection*{\b $\tau$- and $t_c$-stabilities}
\begin{figure}[hbt]
\begin{center}
\centerline {\hspace*{-7.cm} \bf a)\hspace*{8cm} \bf b) }
\vspace{0.25cm}
\includegraphics[width=8cm]{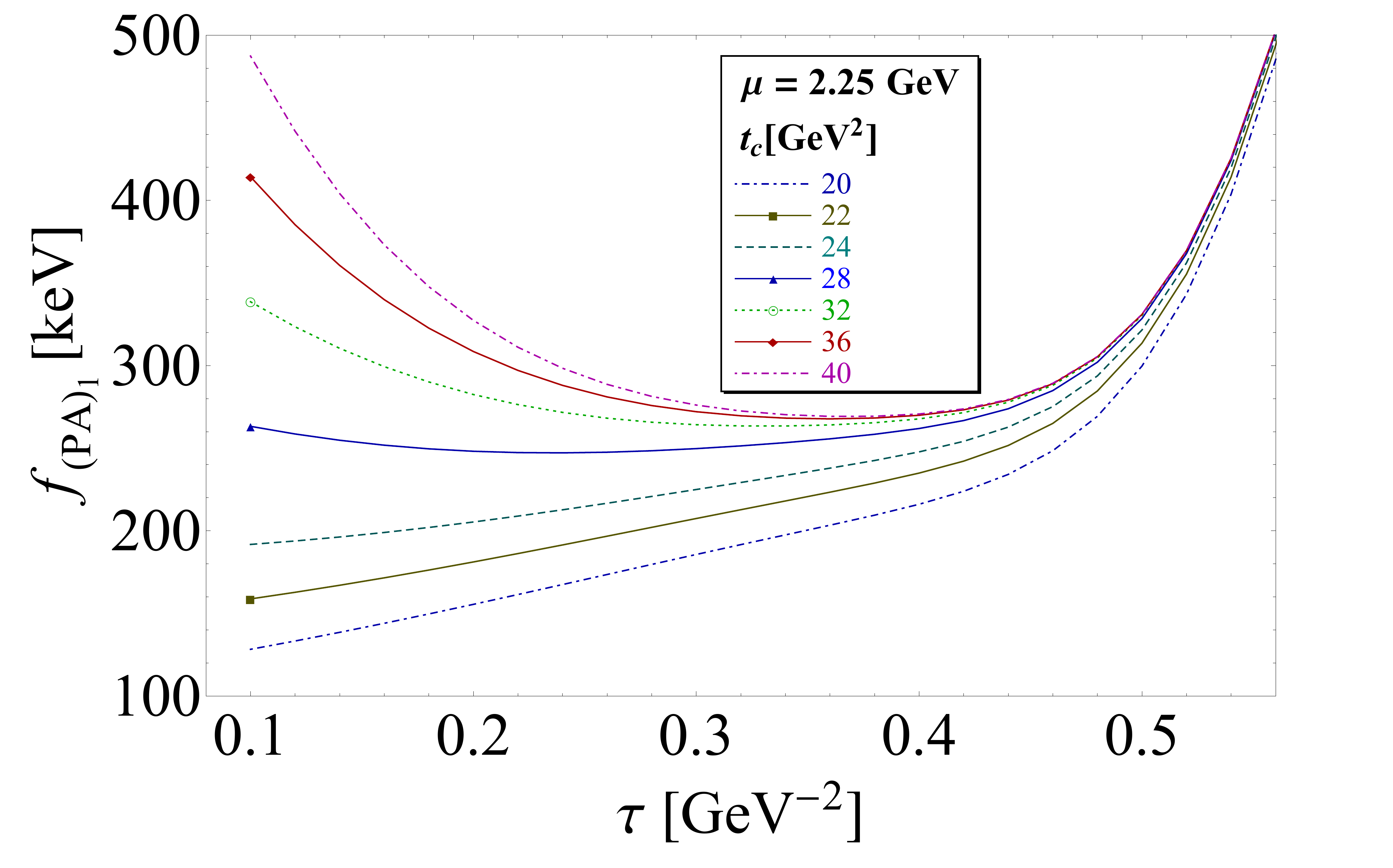}
\includegraphics[width=7.7cm]{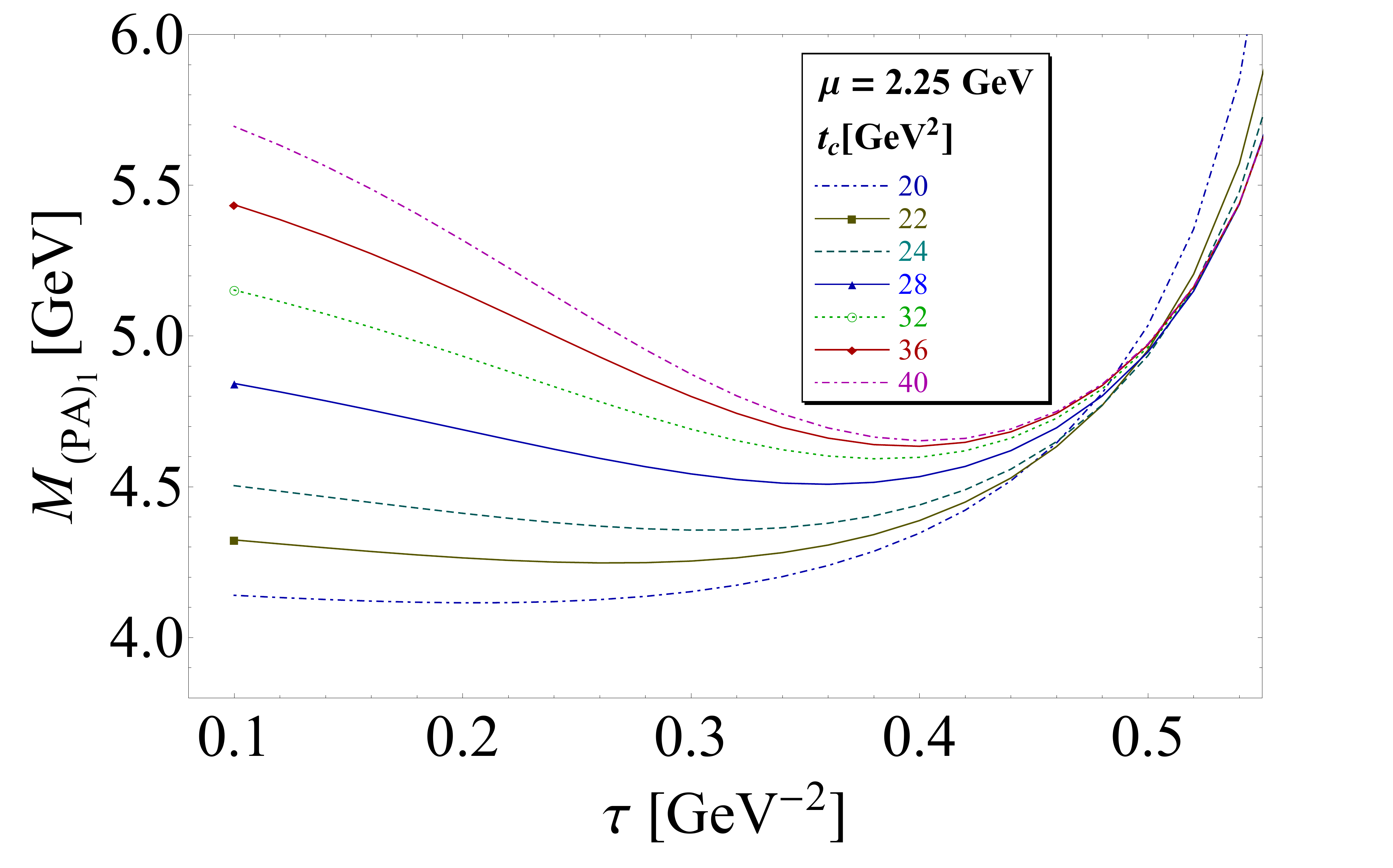}
\vspace*{-0.5cm}
\caption{\footnotesize  a) $f_{(PA)_1}$  from the first moment ${\cal L}^c_0$ and b) $M_{(PA)_1}$  from the 2nd ratio of moments ${\cal R}^c_1$ as function of $\tau$ at NLO for different values of $t_c$, for $\mu$=2.25 GeV and for values of the QCD parameters given in Table\,\ref{tab:param}.} 
\label{fig:f-pa1}
\end{center}
\vspace*{-0.5cm}
\end{figure} 
We show in Fig.\,\ref{fig:f-pa1} the $\tau$- and $t_c$-behaviour of the coupling from ${\cal L}^c_0$ and in Fig\,\ref{fig:pa1-mu} the one of the mass from ${\cal R}_1$ using as input the values of the lowest ground state mass and coupling obtained in Table\,\ref{tab:res0}. The behaviour of the curves for the coupling differs slightly from the previous cases. 
\subsection*{\b $\mu$-stability}
\begin{figure}[hbt]
\begin{center}
\vspace{0.25cm}
\includegraphics[width=8cm]{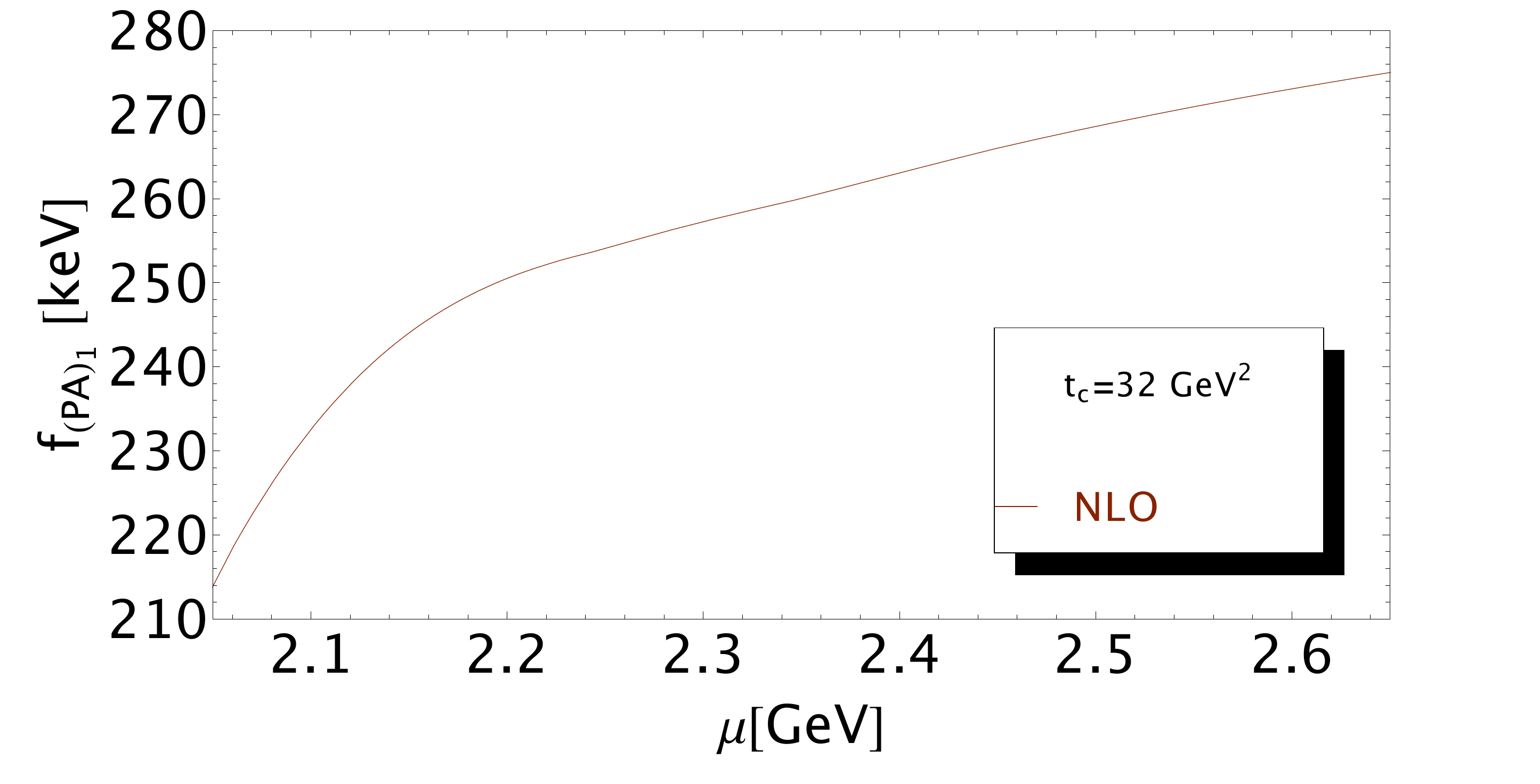}
\includegraphics[width=8cm]{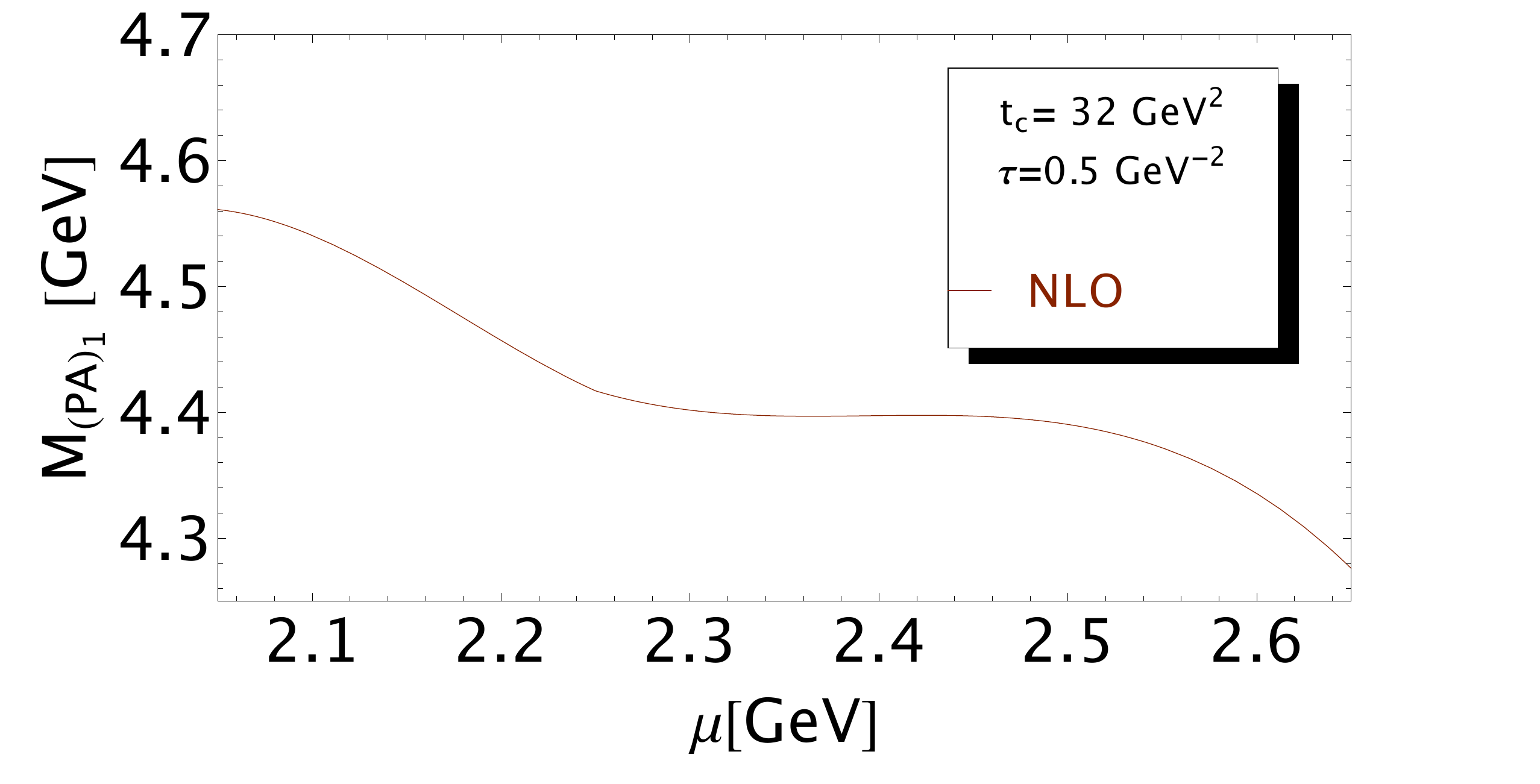}
\vspace*{-0.5cm}
\caption{\footnotesize  $\mu$-behaviour of the $(PA)_1$  mass and coupling.} 
\label{fig:pa1-mu}
\end{center}
\vspace*{-0.5cm}
\end{figure}

The $\mu$-behaviour of the mass and coupling is shown in Fig.\,\ref{fig:pa1-mu}.
\section{The $(D^*_0K^*)_1,~(AA)_1,~(SS)_1$ and $(D_1K)_1,~(SV)_1$ radial excitations}
These radial excitations correspond to the {\it Low Mass} ground states. The analysis of their $\tau,~t_c$ and $\mu$-behaviours shows that they behave like the $(D^*K^*)_1$ and $(DK)_1$ studied explicitly in previous sections. We quote the results of the analysis in Tables\,\ref{tab:resdk0} and\,\ref{tab:resdk1}.  
\begin{table}[hbt]
\setlength{\tabcolsep}{0.13pc}
\catcode`?=\active \def?{\kern\digitwidth}
    {\small
  \begin{tabular*}{\textwidth}{@{}l@{\extracolsep{\fill}}|ccccccccc   c cccccc  l}
\hline
\hline
                      
Observables\,&$ \Delta t_c$&$\Delta\tau$&$ \Delta\mu $ &$\Delta \alpha_s$& $\Delta m_c$ &$\Delta m_s$&$\Delta \bar\psi\psi$&$\Delta \kappa$&$\Delta \alpha_s G^2 $&$\Delta M_0^2$&$\Delta\bar\psi\psi^2$&$\Delta G^3 $&$\Delta M_{G}$&$\Delta f_{G}$&$\Delta M_{(G)_1}$& Values\\
\hline
\bf $0^{+}$ States \\
\cline{0-0} 
$f_{(G)_1}$ [keV] \\
\cline{0-0} 
{\it Molecule}\\
${(DK)_1}$&8.58&0.13&19.0&5.73&2.61&1.20&3.08&17.2&0.7&1.35&21.5&0.05&7.53&25.0&25.0 &211(51)\\
${(D^*K^*)_1}$&3.22&0.11&58.1&14.2&8.82&2.97&6.18&28.9&0.31&8.78&90.8&0.04&14.6&41.2&115&568(167)\\
{\it Tetraquark}\\
$(SS)_1$&6.0&0.28&24.0&9.76&5.60&1.32&3.40&3.91&0.81&2.25&30.0&0.15&11.7&61.8&31.6&359(81)\\
$(AA)_1$& 26.0&0.14&33.0&16.5&8.08&2.05&4.73&5.39&0.54&3.17&43.9&0.20&19.3&45.1&49.9&547(95)\\
\cline{0-0} 
$M_{(G)_1}$ [MeV] \\
\cline{0-0}
{\it Molecule}\\
${(DK)_1}$&43.5&1.30&98.9&42.6&18.9&4.13&9.67&83.1&3.33&14.8&173&0.27&58.7&204&--&3678(310)\\
${(D^*K^*)_1}$&26.0 &8.13&82.7&21.6&12.0&4.31&9.74&44.5&0.51&15.9&109&0.07& 41.1&199&--&4626(252)\\
{\it Tetraquark}\\
$(SS)_1$&64.0 &1.75&88.4&25.1&11.1&2.48&7.84&9.0&1.43&5.92&150&0.38&47.8&188&--&4586(268)\\
$(AA)_1$&68.0 &2.40&83.6&27.1&10.8&3.09&7.13&8.13&0.89&5.58&158&0.35&54.2&208&--&4593(289)\\
\\
\hline\hline

\end{tabular*}
{\scriptsize
 \caption{Predictions from LSR at NLO and sources of errors for the decay constants and masses of the 1st radial excitations of the {\it Low Mass} scalar $(0^+)$ molecules and tetraquarks states. The indices $G$ and $(G)_1$ refer to the lowest ground state and to the 1st radial excitation. The errors from the QCD input parameters are from Table\,\ref{tab:param}. $\Delta \mu=0.10$ GeV and $|\Delta \tau|= 0.02$ GeV$^{-2}$.
 }
 \label{tab:resdk0}
}
}
\end{table}
\begin{table}[hbt]
\setlength{\tabcolsep}{0.13pc}
\catcode`?=\active \def?{\kern\digitwidth}
    {\small
  \begin{tabular*}{\textwidth}{@{}l@{\extracolsep{\fill}}|ccccccccc   c cccccc  l}
\hline
\hline
                      
Observables\,&$ \Delta t_c$&$\Delta\tau$&$ \Delta\mu $ &$\Delta \alpha_s$& $\Delta m_c$ &$\Delta m_s$&$\Delta \bar\psi\psi$&$\Delta \kappa$&$\Delta \alpha_s G^2 $&$\Delta M_0^2$&$\Delta\bar\psi\psi^2$&$\Delta G^3 $&$\Delta M_{G}$&$\Delta f_{G}$&$\Delta M_{(G)_1}$& Values\\
\hline

\hline
\bf  $1^{-}$ States \\
\cline{0-0} 
$f_{(G)_1}$ [keV] \\
\cline{0-0} 
$(D_1K)_1$&6.41&0.24&41.2&2.80&1.54&2.50&5.18&43.0&0.60&4.85&17.3&0.11&11.0&16.1&26.8&157(71)\\
$(D^*_0K^*)_1$&5.99& 0.32&26.4&5.72&3.30&0.09&5.76&26.6&0.27&0.47&19.2&0.27&9.19&17.8&41.3&237(63)\\
{\it Tetraquark}\\
${(PA)_1}$&10.3&0.38&25.8&11.3&4.80&2.19&4.85&5.51&0.28&2.75&50.1&0.18&10.7&47.7&33.2&258(82)\\
${(SV)_1}$&7.9&0.29&15.8&8.21&3.10&1.75&4.52&5.15&0.29&6.13&27.0&0.09&8.85&24.7&52.4&243(68)\\
\cline{0-0} 
$M_{(G)_1}$ [MeV] \\
\cline{0-0} 
{\it Molecule}\\
$(D_1K)_1$&62.6&7.04&124&29.5&12.2&2.96&37.9&145&6.64&51.2&252&1.42&152&198&--&4582(414)\\
$(D^*_0K^*)_1$&73.1&6.78&80.8&21.0&9.73&0.42&21.5&94.0&1.44&1.66&149&1.11&56.1&159&--&4662(269)\\
{\it Tetraquark}\\
${(PA)_1}$&63&7.2&90.1&23.0&8.24&4.17&11.8&13.5&0.65&6.78&153&0.41&47.5&82.1&--&4571(213)\\
${(SV)_1}$&59.1&8.0&130&36.4&13.5&9.55&21.5&24.5&1.70&34.6&229&0.54&72.1&191&--&4541(345)\\
\hline\hline
\end{tabular*}
{\scriptsize
 \caption{The same as in Table\,\ref{tab:resdk0} but for the  $(1^-)$ vector molecules and tetraquarks states.
 }
 \label{tab:resdk1}
}
}
\end{table}
\section{Comments on the results}
The results of the analysis are compiled in Tables\,\ref{tab:res0}, ,\ref{tab:res1}, \ref{tab:resdk0} and \ref{tab:resdk1}. 

\subsection*{\b QCD corrections and the spectral functions}
One can notice that :

-- The NLO corrections are relatively small ($\leq 10\%$) which indicate a good convergence of the PT series. The estimate of uncalculated HO corrections using a geometric growth of the series also shows that these corrections are relatively negligible.

-- To LO of PT there are no non-factorised contributions as one has only a product of two   traces, while in the chiral limit $m_s=0$, the PT expressions of the spectral functions are all the same for the scalar (resp. vector) states. This feature indicates that the use of the convolution of two spectral functions of bilinear quark-antiquark currents for the estimate of the NLO corrections is a good approximantion. The smallness of the non-factorized part of the LO PT expressions including quarks and gluon condensates in some other channels (if any) has been also checked in several examples\,\cite{X5568,SU3,QCD16,MOLE16,4Q}. 

-- The contributions of the gluon condensates $\la \alpha_s G^2\ra$ and $\la g^3 G^3\ra$ are negligible, while the ones of the chiral condensates $\la\bar\psi\psi\ra,~\la\bar\psi G\psi\ra, ~\la\bar\psi\psi\ra^2$ are important in this open-charm channel. This feature is typical for the case of the open-charm and beauty states\,\cite{SNFB15,SNFB,SNFB4,SNFB2}. 

-- In the chiral lmit $m_s=0$ and for $\la\bar ss\ra=\la\bar qq\ra$ and  in the scalar channels,  the coefficients of the chiral condensate contributions for $SS, PP, VV, AA$ tetraquarks are opposite of the ones for (respectively) 
$D^*_0K^*_0,DK, D^*K^*,D_1K_1$ molecules modulo some trivial factors.  The same feature is observed in the vector channels : $SV, VS, AP, PA$ versus  $D^*K^*_0, D^*_0K^*, D_1K, DK_1$, which is due to the $\gamma_5$ property. 

\subsection*{\b Errors induced by $t_c$ and $\tau$}
-- The errors due to $t_c$ have been estimated from the mean obtained from the two extremal values of $t_c$ reported in Table\,\ref{tab:lsr-param}. One can notice that the error due to $t_c$ on the meson masses $(SS, ~AA, ~D^*K^*, ~PA,~SV,~D_1K,$ $~D^*_0K^*)$ from the curves presenting inflexion points are relatively small.  This is due to the fact that the position of $\tau$-minimas from the coupling used to localize with a bettrer precision the inflexion point increases when $t_c$ increases. This change of $\tau$ compensates the one due to $t_c$ which induces a final small error due to $t_c$ in the mass determination. 

-- Moreover, the presence of the inflexion point in $\tau$  explains the slightly larger error (about a facctor 2) due to $\tau$ on the masses from inflexion points compared to the ones from minimas $(PP,~VV,~D_1K_1,~D^*_0K^*_0,~AP,~VS,$ $DK_1,~D^*K^*_0)$. 

-- Notice that in the case of asymmetric errors, we take their mean values.

\subsection*{\b Errors induced by the QCD parameters and role of the four-quark condensates}
To illustrate the analysis, we take randomly the example of the $D_1K$ and $DK_1$ molecules which are representative of the light and heavy meson masses. 

-- By inspecting Tables\,\ref{tab:res0} and\,\ref{tab:res1}, one can remark that the main source of errors due to the QCD inputs come mostly from the chiral condensates and especially from the four-quark condensates $\la\bar\psi\psi\ra^2$  and, in some cases, from the $SU(3)$-breaking parameter $\kappa$ of the $\la\bar ss\ra$ condensate. This latter enter in some cases in the four-quark condensate cotributions $\la\bar ss\ra\la\bar qq\ra$ which we parametrize as $\kappa\la\bar qq\ra^2$ in order to take into account the violation of factorization estimated numerically in Table\,\ref{tab:param}. 

-- Noting  that the errors induced by the four-quark condensates are very asymmetric for the two extremal values of $t_c$ given in Table\,\ref{tab:lsr-param}, we shall fix $t_c$ at its mean value for estimating the errors due to the QCD parameters.  In the case of asymmetric errors, we shall take their mean values.

-- Re-examining the derivation of the  Wilson coefficients of the four-quark condensates $\la\bar\psi\psi\ra^2$ in the two channels, we notice that the difference is due to the trace calculations which induces an extra (1-x) Feynman parameter in the integral of $D_1K$. Qualitatively, taking the large $m_c$ limit of the spectral function, we note that this difference of about $m_c^2/3s$ in the two channel induces a suppression factor of about 0.2 in the $D_1K$ contribution relative to the one in $DK_1$. Taking the opposite limit (asymptotic behaviour $m_c^2/s \ll 1$) is not instructive as the contribution obtained in this limit is only a small part of the complete $\la\bar\psi\psi\ra^2$ one. 

--  In addition to the previous qualitative remark which may partially explain the relative strength of the $\la\bar\psi\psi\ra^2$ contribution in the two channels , we attempt a more quantitative numerical explanation by working  with the complete expressions of the Wilson coefficients and by analyzing its effect on the estimate of mass and decay constant. 

-- Truncating the OPE at $d=4$, we find that the $DK_1$ mass and $D_1K$ coupling decrease quickly when $\tau$-increases indicating that the presence of the $d=6$ condensates is crucial for having an optimal estimate. As a result the prediction is very sensitive to the change of $\la \bar\psi\psi\ra^2$ which induces a large error in the determination of the two observables. However, one should note that at this mimimum value the OPE is still convergent as the contribution of $\la\bar\psi\psi\ra^2$ is only about (10-20)\% of the PT contribution to $M_{DK_1}$. 

-- In the case of the $DK_1$ coupling and $D_1K$ mass, the results stabilize without the need of $\la\bar\psi\psi\ra^2$ (plateau for the coupling and minimum for the mass). Therefore, the error induced by$\la\bar\psi\psi\ra^2$ affects only slightly the determinations. 

\subsection*{\b Comparison of the molecules and tetraquarks states}

-- The flip of signs of the chiral condensate contributions due to $\gamma_5$ in the chiral multiplets explains the large spilttings of masses and couplings given in Tables\,\ref{tab:res0} and\,\ref{tab:res1}.

-- Our results indicate that the molecules and tetraquark states leading to the same final states are almost degenerated in masses which can be understood from the properties of the QCD spectral functions discussed previously. 

-- Therefore, we expect that the ``physical state" is a combination of almost degenerated molecules and tetraquark states with the same quantum numbers $J^{PC}$ which we shall call :  {\it  Tetramole} $({\cal T_M}_J)$. 
\subsection*{\b Mass hierarchies}
From our results, one can notice three classes of spectra : 

-- {\it The Low Mass ground states } \\
These states  are around 2.4 to 2.8 GeV.  They are the $0^{++}$ $DK$ and $D^*K^*$ molecules and the $SS$ and $AA$ tetraquarks.  For the $1^-$ states, we have the $D_1K$ and $D^*_0K^*$ molecules and $PA$ and $SV$ tetraquark states.

-- {\it The  High Mass ground states } \\
These states are in the region above 4.5 GeV. For the $0^{++}$ states, they are the $D_1K_1,~D^*_0K^*_0$ molecules and $PP$ and $VV$ tetraquark states, while for the $1^{-}$ states, they are the $DK_1,~D^*K^*_0$ molecules and the $AP,~VS$ tetraquarks.  We have noticed that the shift of the results to higher masses is due to the positivity of the spectral function which is violated  by working at lower energy scale due to the large negative contributions of chiral $\la\bar\psi\psi\ra$ and $\la\bar\psi\psi\ra^2$ in the OPE. 
 
 -- {\it The  First Radial excitations } \\
 The masses of the 1st radial excitations are compiled in Tables\,\ref{tab:resdk0} and \ref{tab:resdk1}, where the large errors in their determinations have been induced by the ones of the ground state couplings. 
 
\section{Comparison with existing results}
\begin{figure}[hbt]
\begin{center}
\centerline {\hspace*{-4.cm} \bf a)\hspace*{3.5cm} \bf b) \hspace*{5.5cm} \bf c)}
\vspace{0.25cm}
\includegraphics[width=7cm]{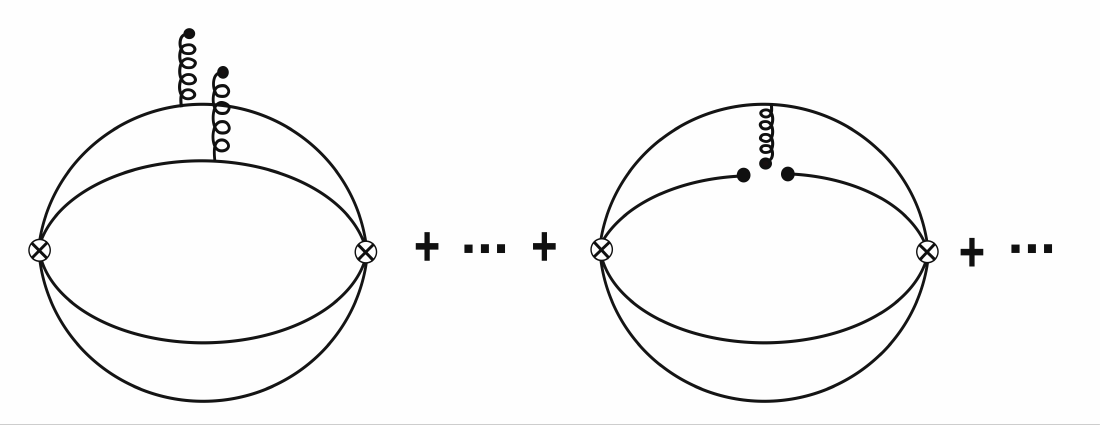}
\includegraphics[width=9cm]{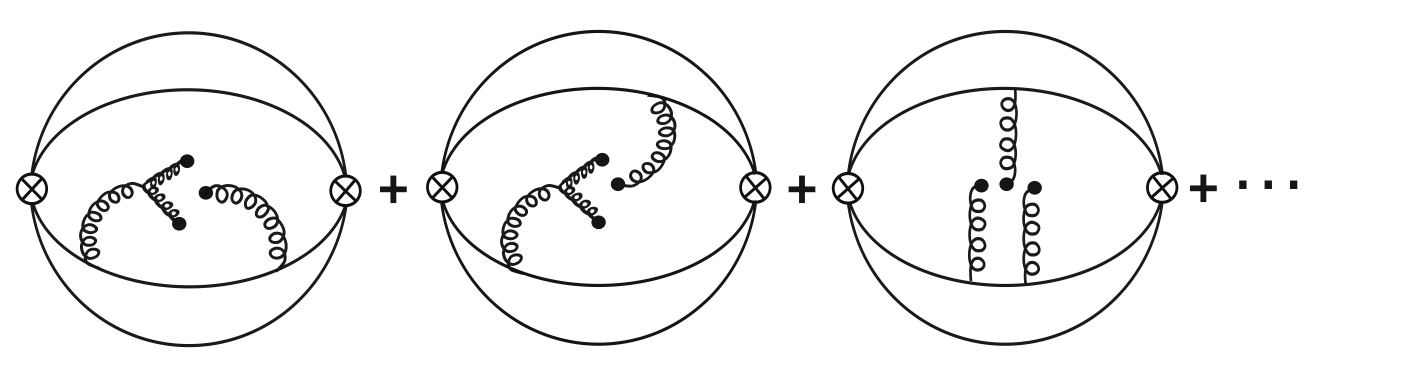}
\vspace*{-0.25cm}
\caption{\footnotesize  a) $d=4~\la \alpha_s G^2\ra$ gluon condensate; b) $d=5~\la \bar\psi G\psi\ra $ mixed quark-gluon condensates; c) $d=6~\la g^3G^3\ra$ triple gluon condensates.} 
\label{fig:zhang}
\end{center}
\vspace*{-0.75cm}
\end{figure} 
\subsection*{\b QCD expressions of the spectral functions}
-- Among the six papers mentioned previously\,\cite{ZHANG,CHEN,WANG,STEELE,AGAEV,TURC}, only the one in\,\cite{ZHANG} gives explict QCD expressions of the $0^{++}$ $SS$ and $AA$ configurations of the tetraquarks while\,\cite{STEELE} gives some explicit expressions of the tetraquark states but with a different choice of the interpolating currents. 

-- Comparing the results step by step with Ref.\,\cite{ZHANG}, we realize that the contributions from the gluon in external fields are systematically missing (see Fig.\,\ref{fig:zhang}). Hopefully, the contributions of these diagrams as well as of the total $\la\alpha_s G^2\ra $ and $\la G^3\ra$ condensates are small which do not affect the numerical results within the precision of the approach. 

-- We notice that the QCD expressions of the spectral functions given by\,\cite{STEELE} correspond to a different choice of interpolating currents  and to the $(\bar c\bar d)(us)$ studied in our previous work\,\cite{MOLE16} but not to $(\bar c\bar s)(ud)$ tetraquark states discussed in this paper.  In the case of the $0^+$ tetraquarks, our optimal choice of current in\,\cite{MOLE16} would correspond to $J_1$  used in\,\cite{STEELE}. Looking at their Table I, one can see that the authors use low values of the set $(t_c,\tau)$ outside the stability region (inflexion point in $\tau$) of our analysis. The complementary study of the coupling which is helpful for fixing with a good precision the stability region (minimum in $\tau$ here) is lacking in the paper. 

\subsection*{\b Results of the analysis}
Though some of our  lowest order (LO) PT results agree within the errors with the recent estimates from QSSR\,\cite{ZHANG,CHEN,WANG,STEELE,AGAEV}, we emphasize that the inclusion of the NLO PT corrections are mandatory for making a sense on the use of the value of the charm quark $\overline{MS}$ mass value in these analyses. 

\section{Confrontation with the LHCb data}
-- From our previous results given in Table\,\ref{tab:res0}, one can notice that {\it High mass states}
corresponding to the  $(0^+)~D_1K_1,~D^*_0K^*_0$ molecules and $PP$ and $VV$ tetraquark states
and to the $(1^-)$ $DK_1,~D^*K^*_0$ molecules and the $AP,~VS$ tetraquarks  states 
are above 5.5 GeV which are too far to contribute to the LHCb $DK$ invariant mass shown in Fig.\,\ref{fig:LHCb}.
\subsection*{\b The 2400 MeV bump around the $DK$ threshold}
This bump coincides with the $D^+K^-$ mass 2400 MeV of the chiral partner of the $D^0K^0$ 
obtained in\,\cite{X5568}. Then, in addition to the $DK$ scattering process which can occur around the $DK$ threshold, we also expect that the $DK$ molecule may participate to this bump.
\subsection*{\b The $X_{0}(2866)$ state and the bump $X_{J}(3150)$}
Taking literally our results in Table\,\ref{tab:res0}, one can see that we have three (almost) degenerate states:
\beq
M_{SS}=2736(21)~{\rm MeV},~~~~~~M_{AA}=2675(65)~{\rm MeV}~~~~{\rm and}~~~~M_{D^*K^*}=2808(41)~{\rm MeV}~,
\label{eq:spin0}
\eeq
and their couplings to the corresponding operators / currents are almost the same:
\beq
 f_{SS}=345(28)~{\rm keV},~~~~~~f_{AA}=498(43)~{\rm keV},~~~~~{\rm and}~~~~~f_{D^*K^*}=405(33)~{\rm keV}~,
\label{eq:fspin0}
\eeq
We assume that the physical state, hereafter called  {\it Tetramole} $({\cal T_M}_J)$,  is a superposition of these nearly degenerated hypothetical states having the same quantum numbers.  Taking its mass and coupling as (quadratic) means of the previous numbers, we obtain  :
\beq
M_{ {\cal T_M}_0}\simeq  2743(18)~{\rm MeV}~,~~~~~~~~~f_{ {\cal T_M}_0}\simeq  395(19)~{\rm keV}~.
\eeq
The $({\cal T_M}_0)$ {\it tetramole} is a good candidate for explaining the $X_{0}(2866)$ though its mass is slightly lighter. 

One can also see from Table\,\ref{tab:resdk0} that the radial excitation $(DK)_1$  mass and coupling are :
\beq
M_{(DK)_1}\simeq 3678(310)~{\rm MeV}~,~~~~~~~~~f_{(DK)_1}\simeq 199(62)~{\rm keV}~,
\eeq
which is the lightest $0^{++}$ first radial excitation. Assuming that the  $X_{J}(3150)$ bump is a scalar state (J=0), we attempt to use a {\it two-component minimal mixing model} between the  {\it Tetramole} $({\cal T_M}_0)$ and the $(DK)_1$ radially excited molecule :
\bea
\vert X_0(2866)\ra&=&\,\,\,\,\,\,\cos \theta_0 \vert {\cal T_M}_0 \ra +\sin\theta_0 \vert (DK)_1\ra\nnb\\
\vert X_{0}(3150)\ra&=&-\sin \theta_0 \vert  {\cal T_M}_0 \ra +\cos\theta_0 \vert (DK)_1\ra~.
\eea
We reproduce the data with a tiny mixing  angle :
\beq
\theta_0\simeq (5.2\pm 1.9)^0~.
\eeq
\subsection*{\b The $X_{1}(2904)$ state and the $X_{J}(3350)$ bump}
\hspace*{0.5cm} -- From our result in Table\,\ref{tab:res0}, one can see that there are four degenerate states with masses :
\bea
M_{PA}&=&2666(47)~{\rm MeV},~~~~~~~~M_{SV}=2593(31)~{\rm MeV}\nnb\\
M_{D_1K}&=&2676(47)~{\rm MeV},~~~~~~~~M_{D^*_0K^*}=2744(41)~{\rm MeV}~,
\eea
and couplings :
\bea
f_{PA}&=&285(29)~{\rm keV},~~~~~~~~f_{SV}=259(25)~{\rm keV}\nnb\\
f_{D_1K}&=&191(21)~{\rm keV},~~~~~~~~f_{D^*_0K^*}=216(22)~{\rm keV}~.
\eea
Like previously, we assume that the (unmixed) physical state is a combination of these hypothetical states.  We
take the mass and coupling of this {\it Tetramole} as the (geometric) means:
\beq
M_{{\cal T_M}_1}=2656(20)~{\rm MeV},~~~~~~~~~f_{ {\cal T_M}_1}\simeq  229(12)~{\rm keV},
\eeq
where one may notice that it can contribute to the $X_{1}(2904)$ state but its mass is slightly lower. 

-- One can also notice from Tables\,\ref{tab:resdk0} and  \,\ref{tab:resdk1} that the radial excitations other than the one of $DK$ are almost degenerated around 4.5 GeV from which one can extract
the masses and couplings (geometric mean) of the spin 0 excluding $(DK)_1$ and spin one Tetramoles:
\bea
M_{({\cal T_M}_0)_1}&\simeq& 4603(155)~{\rm MeV},~~~~~~~M_{({\cal T_M}_1)_1}\simeq 4592(141)~{\rm MeV}~,\nnb\\
f_{({\cal T_M}_0)_1}&\simeq& 454(58)~{\rm keV},~~~~~~~~~~~f_{({\cal T_M}_1)_1}\simeq 223(35)~{\rm keV}~.
\eea
Then, we may consider a {\it minimal two-component mixing } of the spin 1 Tetramole ($ {\cal T_M}_1$)
with its 1st radial excitation $({\cal T_M}_1)_1$ to explain the $X_{1}(2904)$ state and the $X_{J}(3350)$ bump  assuming that the latter is a spin 1 state. The data can be fitted with a tiny mixing angle :
\beq
\theta_1\simeq (9.1\pm 0.6)^0~.
\eeq
A (non)-confirmation of these two  {\it minimal mixing models} requires an experimental identification of the quantum numbers of the bumps at 3150 and 3350 MeV.  

\section{Summary and conclusions}
\b Motivated by the recent LHCb data on the $D^-K^+$ invariant mass from $B\to D^+D^-K^+$
decay (see Fig\,\ref{fig:LHCb}), we have systematically calculated the masses and couplings of some possible configurations of the molecules and tetraquarks states using QCD Laplace sum rules (LSR) within stability criteria where we have added to the LO perturbative term, the NLO radiative corrections which are essential for giving a meaning on the input value of the charm quark which plays an important role in the analysis. We consider our results as improvement and a completion of the results obtained to LO from QCD spectral sum rules\,\cite{ZHANG,CHEN,WANG,STEELE,AGAEV,TURC}. 

\b We have added to the PT contributions the ones of quark and gluon condensates up to dimension-6 in the OPE. We have noted that in some channels, these condensates contributions are large and negative which pushes to work at higher values of energy $s$ for respecting the positivity of the QCD spectral functions. By duality, the resulting values of the corresponding resonances masses are high (see Table\,\ref{tab:res0}) which are outside the region reached by LHCb. 

\b Therefore, we have used the results of the {\it Low Mass} resonances for an attempt to understand the whole range of $DK$ invariant mass found by LHCb:

-- The bumb around the $DK$ threshold can be due to $DK$ scattering amplitude $\oplus$ the $DK(2400)$ lowest mass molecule.

-- The ($0^{++}$) $X_0(2866)$ and $X_J(3150)$ (if it is a $0^{++}$ state)  can e.g result from a mixing of the  {\it Tetramole} ($ {\cal T_M}_0$) with the 1st radial excitation $(DK)_1$ of the molecule state  $(DK)$ with a tiny mixing angle $\theta_0\simeq (5.2\pm 1.9)^0$.

--  The ($1^{-}$) $X_1(2904)$ and $X_J(3350)$ (if it is a $1^{-}$ state)  can result from a mixing of the  {\it Tetramole} ($ {\cal T_M}_1$) with its 1st radial excitation $({\cal T_M})_1$  with a tiny mixing angle $\theta_1\simeq (9.1\pm 0.6)^0$.

\b In addition  to the QSSR approaches, 
some alternative explanations using other models are given in the literature\,\cite{ROSNER,HU,ZHU,HLIU,LU,LIU,HE,HUANG,OSET,SWANSON}. However, to our knowledge, the discussions in the existing papers are limited to the interpretation of the two resonances $X_0(2866)$ and $X_1(2904)$. More data on the precise quantum numbers of the $X_J(3150)$ and $X_J(3350)$ states are nedeed for testing the previous two {\it minimal mixing models} proposal. For completing our study, we plan to estimate the widths of the previous states in a future publication. 



\newpage
\appendix
 \vspace{-0.5cm}
\section{ Scalar Tetraquarks ($0^+$)}
We list below the compact integrated expressions of the spectral functions in different channels to LO of perturbative QCD and including up to dimension-six quark and gluon condensates. These expressions can be useful for further study and check of our numerical analysis. 

Hereafter, we define :
 $\rho_J\equiv \frac{1}{\pi}{\rm Im}\Pi(s)$ where ${\rm Im}\Pi(s)$ is the spectral function defined in Eq.\,\ref{eq:2-pseudo} with :
 \beqn
 \rho_J(s)\simeq \rho_J^{pert}+ \rho_J^{\langle \bar{q}q \rangle}+ \rho_J^{\langle G^2 \rangle}+ \rho_J^{\langle \bar{q}Gq \rangle}+ \rho_J^{\langle \bar{q}q \rangle^2}+ \rho_J^{\langle G^3 \rangle}
 \eeqn
 where : 
\beqn
 \la G^2\ra \equiv \la g^2G^2\ra,~\la \bar qGq\ra\equiv M_0^2\la\bar qq\ra,~\la G^3\ra \equiv \la g^3G^3\ra~~{\rm and}~~x \equiv  m_c^2/s~.
\eeqn
 $m_c \equiv M_c$ (resp. $m_s$) is the on-shell charm (resp. running strange) quark masses. 

 \subsection*{\b Scalar-Scalar configuration (SS)}
 \vspace{-0.5cm}
\begin{eqnarray*}
  \rho_0^{pert}(s) &=& \frac{m_c^8}{5\cdot 3 \cdot 2^{12} \pi^6} 
  \bigg[ 4x \!+\! 155 \!-\! 60 \bigg(1 \!+\! \frac{4}{x} \!+\! 
  \frac{2}{x^2} \bigg) \log(x) \!+\! \frac{80}{x} \!-\! \frac{220}{x^2} 
  \!-\! \frac{20}{x^3} \!+\! \frac{1}{x^4} \bigg] \\&&
  - \frac{m_s m_c^7}{3 \cdot 2^{10} \pi^6} \bigg[ x \!+\! 28 \!-\! 12 
  \bigg(1 \!+\! \frac{3}{x} \!+\! \frac{1}{x^2} \bigg) \log(x) \!-\! 
  \frac{28}{x^2} \!-\! \frac{1}{x^3} \bigg] \\ && \\
  \rho_0^{\langle \bar{q}q \rangle}(s) &=& \frac{m_c^5 
  \langle \bar{s}s \rangle}{3 \cdot 2^6 \pi^4} \bigg[x \!+\! 9 \!-\! 
  6 \bigg(1 \!+\! \frac{1}{x} \bigg) \log(x) \!-\! \frac{9}{x} \!-\! 
  \frac{1}{x^2} \bigg]
  + \frac{m_s m_c^4 \langle \bar{s}s \rangle}
  {3 \cdot 2^7 \pi^4} \bigg[2x \!+\! 3 \!-\! 6 \log(x) \!-\! \frac{6}{x} \!+\! 
  \frac{1}{x^2} \bigg] \\ && \\
  \rho_0^{\langle G^2 \rangle}(s) &=& \frac{m_c^4 \langle G^2 \rangle}
  {3^2 \cdot 2^{12} \pi^6} \bigg[ 5x \!-\! 6\bigg(2 \!-\! \frac{1}{x} \bigg)
  \log(x) \!-\! \frac{9}{x} \!+\! \frac{4}{x^2} \bigg]
  - \frac{m_s m_c^3 \langle G^2 \rangle}{3^2 \cdot 2^{10} \pi^6} 
  \bigg[ 7x \!+\! 15 \!-\! 3\bigg(7 \!+\! \frac{3}{x} \bigg) \log(x) \!-\! 
  \frac{21}{x} \!-\! \frac{1}{x^2} \bigg] \\ && \\
  \rho_0^{\langle \bar{q}Gq \rangle}(s) &=& - \frac{m_c^3 
  \langle \bar{s}G s \rangle}{2^8 \pi^4} \bigg[ 3x \!+\! 4 \!-\! 
  2\bigg(4 \!+\! \frac{1}{x} \bigg) \log(x) \!-\! \frac{7}{x} \bigg] 
  - \frac{m_s m_c^2 \langle \bar{s} Gs \rangle}{3 \cdot 2^8 \pi^4} 
  \bigg[ 5x \!-\! 4 \!-\! 6\log(x) \!-\! \frac{1}{x} \bigg] \\ && \\
  \rho_0^{\langle \bar{q}q \rangle^2}(s) &=& \frac{m_c^2 
  \langle \bar{q}q \rangle^2}{3 \cdot 2^3 \pi^2} \bigg[ x \!-\! 2 \!+\! 
  \frac{1}{x} \bigg] + \frac{m_s m_c \langle \bar{q}q \rangle^2}
  {3 \cdot 2^2 \pi^2} (1 \!-\! x)  \\ && \\
  \rho_0^{\langle G^3 \rangle}(s) &=& \frac{m_c^2 \langle G^3 \rangle}
  {5 \cdot 3^3 \cdot 2^{12} \pi^6} \bigg[ 76x \!-\! 3 \!-\! 150 \log(x) 
  \!-\! \frac{72}{x} \!-\! \frac{1}{x^2} \bigg]
\end{eqnarray*}

\subsection*{\b Pseudoscalar-Pseudoscalar configuration(PP)}
 \vspace{-0.5cm}
\begin{eqnarray*}
  \rho_0^{pert}(s) &=& \frac{m_c^8}{5\cdot 3 \cdot 2^{12} \pi^6} 
  \bigg[ 4x \!+\! 155 \!-\! 60 \bigg(1 \!+\! \frac{4}{x} \!+\! 
  \frac{2}{x^2} \bigg) \log(x) \!+\! \frac{80}{x} \!-\! \frac{220}{x^2} 
  \!-\! \frac{20}{x^3} \!+\! \frac{1}{x^4} \bigg] \\&&
  + \frac{m_s m_c^7}{3 \cdot 2^{10} \pi^6} \bigg[ x \!+\! 28 \!-\! 12 
  \bigg(1 \!+\! \frac{3}{x} \!+\! \frac{1}{x^2} \bigg) \log(x) \!-\! 
  \frac{28}{x^2} \!-\! \frac{1}{x^3} \bigg] \\ && \\
  \rho_0^{\langle \bar{q}q \rangle}(s) &=& - \frac{m_c^5 
  \langle \bar{s}s \rangle}{3 \cdot 2^6 \pi^4} \bigg[x \!+\! 9 \!-\! 
  6 \bigg(1 \!+\! \frac{1}{x} \bigg) \log(x) \!-\! \frac{9}{x} \!-\! 
  \frac{1}{x^2} \bigg]
  + \frac{m_s m_c^4 \langle \bar{s}s \rangle}
  {3 \cdot 2^7 \pi^4} \bigg[2x \!+\! 3 \!-\! 6 \log(x) \!-\! \frac{6}{x} \!+\! 
  \frac{1}{x^2} \bigg] \\ && \\
  \rho_0^{\langle G^2 \rangle}(s) &=& \frac{m_c^4 \langle G^2 \rangle}
  {3^2 \cdot 2^{12} \pi^6} \bigg[ 5x \!-\! 6\bigg(2 \!-\! \frac{1}{x} \bigg)
  \log(x) \!-\! \frac{9}{x} \!+\! \frac{4}{x^2} \bigg]
  + \frac{m_s m_c^3 \langle G^2 \rangle}{3^2 \cdot 2^{10} \pi^6} 
  \bigg[ 7x \!+\! 15 \!-\! 3\bigg(7 \!+\! \frac{3}{x} \bigg) \log(x) \!-\! 
  \frac{21}{x} \!-\! \frac{1}{x^2} \bigg] \\ && \\
  \rho_0^{\langle \bar{q}Gq \rangle}(s) &=& \frac{m_c^3 
  \langle \bar{s}G s \rangle}{2^8 \pi^4} \bigg[ 3x \!+\! 4 \!-\! 
  2\bigg(4 \!+\! \frac{1}{x} \bigg) \log(x) \!-\! \frac{7}{x} \bigg] 
  - \frac{m_s m_c^2 \langle \bar{s} Gs \rangle}{3 \cdot 2^8 \pi^4} 
  \bigg[ 5x \!-\! 4 \!-\! 6\log(x) \!-\! \frac{1}{x} \bigg] \\ && \\
  \rho_0^{\langle \bar{q}q \rangle^2}(s) &=& - \frac{m_c^2 
  \langle \bar{q}q \rangle^2}{3 \cdot 2^3 \pi^2} \bigg[ x \!-\! 2 \!+\! 
  \frac{1}{x} \bigg] + \frac{m_s m_c \langle \bar{q}q \rangle^2}
  {3 \cdot 2^2 \pi^2} (1 \!-\! x)  \\ && \\
  \rho_0^{\langle G^3 \rangle}(s) &=& \frac{m_c^2 \langle G^3 \rangle}
  {5 \cdot 3^3 \cdot 2^{12} \pi^6} \bigg[ 76x \!-\! 3 \!-\! 150 \log(x) 
  \!-\! \frac{72}{x} \!-\! \frac{1}{x^2} \bigg]
\end{eqnarray*}

\subsection*{\b Vector-Vector configuration (VV)}
 \vspace{-0.5cm}
\begin{eqnarray*}
  \rho_0^{pert}(s) &=& \frac{m_c^8}{5\cdot 3 \cdot 2^{10} \pi^6} 
  \bigg[ 4x \!+\! 155 \!-\! 60 \bigg(1 \!+\! \frac{4}{x} \!+\! 
  \frac{2}{x^2} \bigg) \log(x) \!+\! \frac{80}{x} \!-\! \frac{220}{x^2} 
  \!-\! \frac{20}{x^3} \!+\! \frac{1}{x^4} \bigg] \\&&
  + \frac{m_s m_c^7}{3 \cdot 2^{9} \pi^6} \bigg[ x \!+\! 28 \!-\! 12 
  \bigg(1 \!+\! \frac{3}{x} \!+\! \frac{1}{x^2} \bigg) \log(x) \!-\! 
  \frac{28}{x^2} \!-\! \frac{1}{x^3} \bigg] \\ && \\
  \rho_0^{\langle \bar{q}q \rangle}(s) &=& - \frac{m_c^5 
  \langle \bar{s}s \rangle}{3 \cdot 2^5 \pi^4} \bigg[x \!+\! 9 \!-\! 
  6 \bigg(1 \!+\! \frac{1}{x} \bigg) \log(x) \!-\! \frac{9}{x} \!-\! 
  \frac{1}{x^2} \bigg]
  + \frac{m_s m_c^4 \langle \bar{s}s \rangle}
  {3 \cdot 2^5 \pi^4} \bigg[2x \!+\! 3 \!-\! 6 \log(x) \!-\! \frac{6}{x} 
  \!+\! \frac{1}{x^2} \bigg] \\ && \\
  \rho_0^{\langle G^2 \rangle}(s) &=& \frac{m_c^4 \langle G^2 \rangle}
  {3^2 \cdot 2^{11} \pi^6} \bigg[ 7x \!+\! 18 \!-\! 6\bigg(4 \!+\! 
  \frac{1}{x} \bigg) \log(x) \!-\! \frac{27}{x} \!+\! \frac{2}{x^2} \bigg]
  - \frac{m_s m_c^3 \langle G^2 \rangle}{3^2 \cdot 2^{11} \pi^6} 
  \bigg[ 17x \!-\! 24 \!-\! 6\bigg(4 \!-\! \frac{3}{x} \bigg) \log(x) \!+\! 
  \frac{3}{x} \!+\! \frac{4}{x^2} \bigg] \\ && \\
  \rho_0^{\langle \bar{q}Gq \rangle}(s) &=& \frac{m_c^3 
  \langle \bar{s}G s \rangle}{2^7 \pi^4} \bigg[ x \!-\! 2 \log(x) \!-\! \frac{1}{x} \bigg] - \frac{m_s m_c^2 \langle \bar{s} Gs \rangle}
  {3 \cdot 2^7 \pi^4} \bigg[ x \!-\! 2 \!+\! \frac{1}{x} \bigg] \\ && \\
  \rho_0^{\langle \bar{q}q \rangle^2}(s) &=& - \frac{m_c^2 
  \langle \bar{q}q \rangle^2}{3 \cdot 2^2 \pi^2} \bigg[ x \!-\! 2 \!+\! 
  \frac{1}{x} \bigg] + \frac{m_s m_c \langle \bar{q}q \rangle^2}
  {3 \pi^2} (1 \!-\! x)  \\ && \\
  \rho_0^{\langle G^3 \rangle}(s) &=& \frac{m_c^2 \langle G^3 \rangle}
  {5 \cdot 3^3 \cdot 2^{10} \pi^6} \bigg[ 31x \!-\! 3 \!-\! 60 \log(x) 
  \!-\! \frac{27}{x} \!-\! \frac{1}{x^2} \bigg]
\end{eqnarray*}
  
\subsection*{\b  Axial-Axial configuration(AA)}
 \vspace{-0.5cm}
\begin{eqnarray*}
  \rho_0^{pert}(s) &=& \frac{m_c^8}{5\cdot 3 \cdot 2^{10} \pi^6} 
  \bigg[ 4x \!+\! 155 \!-\! 60 \bigg(1 \!+\! \frac{4}{x} \!+\! 
  \frac{2}{x^2} \bigg) \log(x) \!+\! \frac{80}{x} \!-\! \frac{220}{x^2} 
  \!-\! \frac{20}{x^3} \!+\! \frac{1}{x^4} \bigg] \\&&
  - \frac{m_s m_c^7}{3 \cdot 2^{9} \pi^6} \bigg[ x \!+\! 28 \!-\! 12 
  \bigg(1 \!+\! \frac{3}{x} \!+\! \frac{1}{x^2} \bigg) \log(x) \!-\! 
  \frac{28}{x^2} \!-\! \frac{1}{x^3} \bigg] \\ && \\
  \rho_0^{\langle \bar{q}q \rangle}(s) &=& \frac{m_c^5 
  \langle \bar{s}s \rangle}{3 \cdot 2^5 \pi^4} \bigg[x \!+\! 9 \!-\! 
  6 \bigg(1 \!+\! \frac{1}{x} \bigg) \log(x) \!-\! \frac{9}{x} \!-\! 
  \frac{1}{x^2} \bigg]
  + \frac{m_s m_c^4 \langle \bar{s}s \rangle}
  {3 \cdot 2^5 \pi^4} \bigg[2x \!+\! 3 \!-\! 6 \log(x) \!-\! \frac{6}{x} 
  \!+\! \frac{1}{x^2} \bigg] \\ && \\
  \rho_0^{\langle G^2 \rangle}(s) &=& \frac{m_c^4 \langle G^2 \rangle}
  {3^2 \cdot 2^{11} \pi^6} \bigg[ 7x \!+\! 18 \!-\! 6\bigg(4 \!+\! 
  \frac{1}{x} \bigg) \log(x) \!-\! \frac{27}{x} \!+\! \frac{2}{x^2} \bigg]
  + \frac{m_s m_c^3 \langle G^2 \rangle}{3^2 \cdot 2^{11} \pi^6} 
  \bigg[ 17x \!-\! 24 \!-\! 6\bigg(4 \!-\! \frac{3}{x} \bigg) \log(x) \!+\! 
  \frac{3}{x} \!+\! \frac{4}{x^2} \bigg] \\ && \\
  \rho_0^{\langle \bar{q}Gq \rangle}(s) &=& - \frac{m_c^3 
  \langle \bar{s}G s \rangle}{2^7 \pi^4} \bigg[ x \!-\! 2 \log(x) \!-\! \frac{1}{x} \bigg] - \frac{m_s m_c^2 \langle \bar{s} Gs \rangle}
  {3 \cdot 2^7 \pi^4} \bigg[ x \!-\! 2 \!+\! \frac{1}{x} \bigg] \\ && \\
  \rho_0^{\langle \bar{q}q \rangle^2}(s) &=& \frac{m_c^2 
  \langle \bar{q}q \rangle^2}{3 \cdot 2^2 \pi^2} \bigg[ x \!-\! 2 \!+\! 
  \frac{1}{x} \bigg] + \frac{m_s m_c \langle \bar{q}q \rangle^2}
  {3 \pi^2} (1 \!-\! x)  \\ && \\
  \rho_0^{\langle G^3 \rangle}(s) &=& \frac{m_c^2 \langle G^3 \rangle}
  {5 \cdot 3^3 \cdot 2^{10} \pi^6} \bigg[ 31x \!-\! 3 \!-\! 60 \log(x) 
  \!-\! \frac{27}{x} \!-\! \frac{1}{x^2} \bigg]
\end{eqnarray*}
\vspace{1cm}

\subsection*{\b ${D_0^\ast K_0^\ast}$ molecule configuration}
 \vspace{-0.5cm}
\begin{eqnarray*}
  \rho_0^{pert}(s) &=& \frac{m_c^8}{5 \cdot 2^{14} \pi^6} 
  \bigg[ 4x \!+\! 155 \!-\! 60 \bigg(1 \!+\! \frac{4}{x} \!+\! 
  \frac{2}{x^2} \bigg) \log(x) \!+\! \frac{80}{x} \!-\! \frac{220}{x^2} 
  \!-\! \frac{20}{x^3} \!+\! \frac{1}{x^4} \bigg] \\ && \\
  \rho_0^{\langle \bar{q}q \rangle}(s) &=& - \frac{m_c^5 
  \langle \bar{q}q \rangle}{2^8 \pi^4} \bigg[x \!+\! 9 \!-\! 
  6 \bigg(1 \!+\! \frac{1}{x} \bigg) \log(x) \!-\! \frac{9}{x} \!-\! 
  \frac{1}{x^2} \bigg]
  + \frac{m_s m_c^4}{2^9 \pi^4} 
  \Big( 2\langle \bar{q}q \rangle \!+\! \langle \bar{s}s \rangle \Big)
  \bigg[2x \!+\! 3 \!-\! 6 \log(x) \!-\! \frac{6}{x} 
  \!+\! \frac{1}{x^2} \bigg] \\ && \\
  \rho_0^{\langle G^2 \rangle}(s) &=& \frac{m_c^4 \langle G^2 \rangle}
  {3 \cdot 2^{13} \pi^6} \bigg[ 4x \!-\! 9 \!-\! 6\bigg(1 \!-\! 
  \frac{2}{x} \bigg) \log(x) \!+\! \frac{5}{x^2} \bigg] 
  \\ && \\
  \rho_0^{\langle \bar{q}Gq \rangle}(s) &=& \frac{3m_c^3 
  \langle \bar{q}G q \rangle}{2^8 \pi^4} \bigg[ x \!+\! 2 \!-\! 
  \Big( 3 + \frac{1}{x} \Big) \log(x) \!-\! \frac{3}{x} \bigg] - 
  \frac{m_s m_c^2}{2^9 \pi^4} 
  \Big( 3\langle \bar{q} Gq \rangle\!-\!2\langle \bar{s} Gs \rangle \Big)
  \bigg[ x \!-\! 2 \!+\! \frac{1}{x} \bigg] \\ && \\
  \rho_0^{\langle \bar{q}q \rangle^2}(s) &=& - \frac{m_c^2 
  \langle \bar{q}q \rangle \langle \bar{s}s \rangle}
  {2^5 \pi^2} \bigg[ x \!-\! 2 \!+\! 
  \frac{1}{x} \bigg] + \frac{m_s m_c}{2^5 \pi^2} 
  \Big( 2\langle \bar{q}q \rangle^2 \!+\! 
  \langle \bar{q}q \rangle \langle \bar{s}s \rangle \Big)(1 \!-\! x)  
  \\ && \\
  \rho_0^{\langle G^3 \rangle}(s) &=& \frac{m_c^2 \langle G^3 \rangle}
  {5 \cdot 3^2 \cdot 2^{14} \pi^6} \bigg[ 166x \!-\! 3 \!-\! 330 \log(x) 
  \!-\! \frac{162}{x} \!-\! \frac{1}{x^2} \bigg]
\end{eqnarray*}

\subsection*{\b ${D K}$ molecule configuration (see Re.\,\cite{X5568})}
 \vspace{-0.5cm}
\begin{eqnarray*}
  \rho_0^{pert}(s) &=& \frac{m_c^8}{5 \cdot 2^{14} \pi^6} 
  \bigg[ 4x \!+\! 155 \!-\! 60 \bigg(1 \!+\! \frac{4}{x} \!+\! 
  \frac{2}{x^2} \bigg) \log(x) \!+\! \frac{80}{x} \!-\! \frac{220}{x^2} 
  \!-\! \frac{20}{x^3} \!+\! \frac{1}{x^4} \bigg] \\ && \\
  \rho_0^{\langle \bar{q}q \rangle}(s) &=& \frac{m_c^5 
  \langle \bar{q}q \rangle}{2^8 \pi^4} \bigg[x \!+\! 9 \!-\! 
  6 \bigg(1 \!+\! \frac{1}{x} \bigg) \log(x) \!-\! \frac{9}{x} \!-\! 
  \frac{1}{x^2} \bigg]
  - \frac{m_s m_c^4}{2^9 \pi^4} 
  \Big( 2\langle \bar{q}q \rangle \!-\! \langle \bar{s}s \rangle \Big)
  \bigg[2x \!+\! 3 \!-\! 6 \log(x) \!-\! \frac{6}{x} 
  \!+\! \frac{1}{x^2} \bigg] \\ && \\
  \rho_0^{\langle G^2 \rangle}(s) &=& \frac{m_c^4 \langle G^2 \rangle}
  {3 \cdot 2^{13} \pi^6} \bigg[ 4x \!-\! 9 \!-\! 6\bigg(1 \!-\! 
  \frac{2}{x} \bigg) \log(x) \!+\! \frac{5}{x^2} \bigg] 
  \\ && \\
  \rho_0^{\langle \bar{q}Gq \rangle}(s) &=& - \frac{3m_c^3 
  \langle \bar{q}G q \rangle}{2^8 \pi^4} \bigg[ x \!+\! 2 \!-\! 
  \Big( 3 + \frac{1}{x} \Big) \log(x) \!-\! \frac{3}{x} \bigg] + 
  \frac{m_s m_c^2}{2^9 \pi^4} 
  \Big( 3\langle \bar{q} Gq \rangle\!+\!2\langle \bar{s} Gs \rangle \Big)
  \bigg[ x \!-\! 2 \!+\! \frac{1}{x} \bigg] \\ && \\
  \rho_0^{\langle \bar{q}q \rangle^2}(s) &=& \frac{m_c^2 
  \langle \bar{q}q \rangle \langle \bar{s}s \rangle}
  {2^5 \pi^2} \bigg[ x \!-\! 2 \!+\! 
  \frac{1}{x} \bigg] + \frac{m_s m_c}{2^5 \pi^2} 
  \Big( 2\langle \bar{q}q \rangle^2 \!-\! 
  \langle \bar{q}q \rangle \langle \bar{s}s \rangle \Big)(1 \!-\! x)  
  \\ && \\
  \rho_0^{\langle G^3 \rangle}(s) &=& \frac{m_c^2 \langle G^3 \rangle}
  {5 \cdot 3^2 \cdot 2^{14} \pi^6} \bigg[ 166x \!-\! 3 \!-\! 330 \log(x) 
  \!-\! \frac{162}{x} \!-\! \frac{1}{x^2} \bigg]
\end{eqnarray*}

\subsection*{\b ${D^\ast K^\ast}$ molecule configuration}
 \vspace{-0.5cm}
\begin{eqnarray*}
  \rho_0^{pert}(s) &=& \frac{m_c^8}{5 \cdot 2^{12} \pi^6} 
  \bigg[ 4x \!+\! 155 \!-\! 60 \bigg(1 \!+\! \frac{4}{x} \!+\! 
  \frac{2}{x^2} \bigg) \log(x) \!+\! \frac{80}{x} \!-\! \frac{220}{x^2} 
  \!-\! \frac{20}{x^3} \!+\! \frac{1}{x^4} \bigg] \\ && \\
  \rho_0^{\langle \bar{q}q \rangle}(s) &=& \frac{m_c^5 
  \langle \bar{q}q \rangle}{2^7 \pi^4} \bigg[x \!+\! 9 \!-\! 
  6 \bigg(1 \!+\! \frac{1}{x} \bigg) \log(x) \!-\! \frac{9}{x} \!-\! 
  \frac{1}{x^2} \bigg]
  - \frac{m_s m_c^4}{2^7 \pi^4} 
  \Big( \langle \bar{q}q \rangle \!-\! \langle \bar{s}s \rangle \Big)
  \bigg[2x \!+\! 3 \!-\! 6 \log(x) \!-\! \frac{6}{x} 
  \!+\! \frac{1}{x^2} \bigg] \\ && \\
  \rho_0^{\langle G^2 \rangle}(s) &=& \frac{m_c^4 \langle G^2 \rangle}
  {3 \cdot 2^{11} \pi^6} \bigg[ x \!+\! 9 \!-\! 6\bigg(1 \!+\! 
  \frac{1}{x} \bigg) \log(x) \!-\! \frac{9}{x} \!-\! \frac{1}{x^2} \bigg] 
  \\ && \\
  \rho_0^{\langle \bar{q}Gq \rangle}(s) &=& - \frac{3m_c^3 
  \langle \bar{q}G q \rangle}{2^8 \pi^4} \bigg[ x \!-\! 2 \log(x) \!-\! \frac{1}{x} \bigg] + \frac{m_s m_c^2}{2^8 \pi^4} 
  \Big( 3\langle \bar{q} Gq \rangle\!-\!2\langle \bar{s} Gs \rangle \Big)
  \bigg[ x \!-\! 2 \!+\! \frac{1}{x} \bigg] \\ && \\
  \rho_0^{\langle \bar{q}q \rangle^2}(s) &=& \frac{m_c^2 
  \langle \bar{q}q \rangle \langle \bar{s}s \rangle}
  {2^4 \pi^2} \bigg[ x \!-\! 2 \!+\! 
  \frac{1}{x} \bigg] + \frac{m_s m_c}{2^4 \pi^2} 
  \Big( 4\langle \bar{q}q \rangle^2 \!-\! 
  \langle \bar{q}q \rangle \langle \bar{s}s \rangle \Big)(1 \!-\! x)  
  \\ && \\
  \rho_0^{\langle G^3 \rangle}(s) &=& - \frac{m_c^2 \langle G^3 \rangle}
  {5 \cdot 3^2 \cdot 2^{12} \pi^6} \bigg[ 14x \!+\! 3 \!-\! 30 \log(x) 
  \!-\! \frac{18}{x} \!+\! \frac{1}{x^2} \bigg]
\end{eqnarray*}

\subsection*{\b ${D_1 K_1}$ molecule configuration}
 \vspace{-0.5cm}
\begin{eqnarray*}
  \rho_0^{pert}(s) &=& \frac{m_c^8}{5 \cdot 2^{12} \pi^6} 
  \bigg[ 4x \!+\! 155 \!-\! 60 \bigg(1 \!+\! \frac{4}{x} \!+\! 
  \frac{2}{x^2} \bigg) \log(x) \!+\! \frac{80}{x} \!-\! \frac{220}{x^2} 
  \!-\! \frac{20}{x^3} \!+\! \frac{1}{x^4} \bigg] \\ && \\
  \rho_0^{\langle \bar{q}q \rangle}(s) &=& - \frac{m_c^5 
  \langle \bar{q}q \rangle}{2^7 \pi^4} \bigg[x \!+\! 9 \!-\! 
  6 \bigg(1 \!+\! \frac{1}{x} \bigg) \log(x) \!-\! \frac{9}{x} \!-\! 
  \frac{1}{x^2} \bigg]
  + \frac{m_s m_c^4}{2^7 \pi^4} 
  \Big( \langle \bar{q}q \rangle \!+\! \langle \bar{s}s \rangle \Big)
  \bigg[2x \!+\! 3 \!-\! 6 \log(x) \!-\! \frac{6}{x} 
  \!+\! \frac{1}{x^2} \bigg] \\ && \\
  \rho_0^{\langle G^2 \rangle}(s) &=& \frac{m_c^4 \langle G^2 \rangle}
  {3 \cdot 2^{11} \pi^6} \bigg[ x \!+\! 9 \!-\! 6\bigg(1 \!+\! 
  \frac{1}{x} \bigg) \log(x) \!-\! \frac{9}{x} \!-\! \frac{1}{x^2} \bigg] 
  \\ && \\
  \rho_0^{\langle \bar{q}Gq \rangle}(s) &=& \frac{3m_c^3 
  \langle \bar{q}G q \rangle}{2^8 \pi^4} \bigg[ x \!-\! 2 \log(x) \!-\! \frac{1}{x} \bigg] - \frac{m_s m_c^2}{2^8 \pi^4} 
  \Big( 3\langle \bar{q} Gq \rangle\!+\!2\langle \bar{s} Gs \rangle \Big)
  \bigg[ x \!-\! 2 \!+\! \frac{1}{x} \bigg] \\ && \\
  \rho_0^{\langle \bar{q}q \rangle^2}(s) &=& -\frac{m_c^2 
  \langle \bar{q}q \rangle \langle \bar{s}s \rangle}
  {2^4 \pi^2} \bigg[ x \!-\! 2 \!+\! 
  \frac{1}{x} \bigg] + \frac{m_s m_c}{2^4 \pi^2} 
  \Big( 4\langle \bar{q}q \rangle^2 \!+\! 
  \langle \bar{q}q \rangle \langle \bar{s}s \rangle \Big)(1 \!-\! x)  
  \\ && \\
  \rho_0^{\langle G^3 \rangle}(s) &=& - \frac{m_c^2 \langle G^3 \rangle}
  {5 \cdot 3^2 \cdot 2^{12} \pi^6} \bigg[ 14x \!+\! 3 \!-\! 30 \log(x) 
  \!-\! \frac{18}{x} \!+\! \frac{1}{x^2} \bigg]
\end{eqnarray*}
\vspace{-1cm}
\section{Vector Tetraquark and Molecule States $(1^-)$}
\subsection*{\b Axial-Pseudoscalar tetraquark configuration (AP)}
 \vspace{-0.5cm}
\begin{eqnarray*}
  \rho_1^{pert}(s) &=& \frac{m_c^8}{5\cdot 3^2 \cdot 2^{13} \pi^6} 
  \bigg[ x^2 \!+\! 555 \!-\! 60 \bigg(3 \!+\! \frac{16}{x} \!+\! 
  \frac{9}{x^2} \bigg) \log(x) \!+\! \frac{480}{x} \!-\! \frac{945}{x^2} 
  \!-\! \frac{96}{x^3} \!+\! \frac{5}{x^4} \bigg] \\&&
  - \frac{m_s m_c^7}{3 \cdot 2^{10} \pi^6} \bigg[ x \!+\! 
  28 \!-\! 12\bigg(1 \!+\! \frac{3}{x} \!+\! \frac{1}{x^2} \bigg) 
  \log(x) \!-\! \frac{28}{x^2} \!-\! 
  \frac{1}{x^3} \bigg] \\ && \\
  \rho_1^{\langle \bar{q}q \rangle}(s) &=& \frac{m_c^5 
  \langle \bar{s}s \rangle}{3 \cdot 2^6 \pi^4} \bigg[x \!+\! 9 \!-\! 
  6 \bigg(1 \!+\! \frac{1}{x} \bigg) \log(x) \!-\! \frac{9}{x} \!-\! 
  \frac{1}{x^2} \bigg]
  + \frac{m_s m_c^4 \langle \bar{s}s \rangle}
  {3 \cdot 2^9 \pi^4} \bigg[x^2 \!+\! 12 \!-\! 12 \log(x) \!-\! 
  \frac{16}{x} \!+\! \frac{3}{x^2} \bigg] \\ && \\
  \rho_1^{\langle G^2 \rangle}(s) &=& \frac{m_c^4 \langle G^2 \rangle}
  {3^3 \cdot 2^{13} \pi^6} \bigg[ 5x^2 \!+\! 12x \!+\! 198 \!-\! 48\bigg(
  3 \!+\! \frac{2}{x} \bigg) \log(x) \!-\! \frac{212}{x} \!-\! 
  \frac{3}{x^2} \bigg] \\&&
  - \frac{m_s m_c^3 \langle G^2 \rangle}
  {3^2 \cdot 2^{10} \pi^6} \bigg[ 7x \!+\! 15  \!-\! 
  3\bigg(7 \!+\! \frac{3}{x} \bigg) \log(x) \!-\! \frac{21}{x} \!-\!
  \frac{1}{x^2} \bigg] \\ && \\
  \rho_1^{\langle \bar{q}Gq \rangle}(s) &=& - \frac{m_c^3 
  \langle \bar{s}G s \rangle}{2^7 \pi^4} \bigg[ x \!-\! 
  2 \log(x) \!-\! \frac{1}{x} \bigg] 
  - \frac{m_s m_c^2 \langle \bar{s} Gs \rangle}
  {3^2 \cdot 2^9 \pi^4} \bigg[ 5x^2 \!-\! 12x \!-\! 9 \!+\! 
  18 \log(x) \!+\! \frac{16}{x} \bigg] \\ && \\
  \rho_1^{\langle \bar{q}q \rangle^2}(s) &=& - \frac{m_c^2 
  \langle \bar{q}q \rangle^2}{3^2 \cdot 2^3 \pi^2} \bigg[ x^2 \!-\! 3 
  \!+\! \frac{2}{x} \bigg] - \frac{m_s m_c \langle \bar{q}q \rangle^2}
  {3 \cdot 2^2 \pi^2} (1 \!-\! x)  \\ && \\
  \rho_1^{\langle G^3 \rangle}(s) &=& \frac{m_c^2 \langle G^3 \rangle}
  {5 \cdot 3^3 \cdot 2^{14} \pi^6} \bigg[ 11x^2 \!-\! 240x \!+\! 
  24 \!+\! 420 \log(x) \!+\! \frac{208}{x} \!-\! \frac{3}{x^2} \bigg]
\end{eqnarray*}

\subsection*{\b Pseudoscalar-Axial tetraquark configuration(PA)}
 \vspace{-0.5cm}
\begin{eqnarray*}
  \rho_1^{pert}(s) &=& \frac{m_c^8}{5\cdot 3^2 \cdot 2^{13} \pi^6} 
  \bigg[ x^2 \!+\! 555 \!-\! 60 \bigg(3 \!+\! \frac{16}{x} \!+\! 
  \frac{9}{x^2} \bigg) \log(x) \!+\! \frac{480}{x} \!-\! \frac{945}{x^2} 
  \!-\! \frac{96}{x^3} \!+\! \frac{5}{x^4} \bigg] \\&&
  + \frac{m_s m_c^7}{5 \cdot 3 \cdot 2^{12} \pi^6} \bigg[ x^2 \!+\! 
  340 \!-\! 60\bigg(2 \!+\! \frac{8}{x} \!+\! \frac{3}{x^2} \bigg) 
  \log(x) \!+\! \frac{80}{x} \!-\! \frac{405}{x^2} \!-\! 
  \frac{16}{x^3} \bigg] \\ && \\
  \rho_1^{\langle \bar{q}q \rangle}(s) &=& - \frac{m_c^5 
  \langle \bar{s}s \rangle}{3^2 \cdot 2^8 \pi^4} \bigg[x^2 \!+\! 72 \!-\! 
  12 \bigg(3 \!+\! \frac{4}{x} \bigg) \log(x) \!-\! \frac{64}{x} \!-\! 
  \frac{9}{x^2} \bigg]
  + \frac{m_s m_c^4 \langle \bar{s}s \rangle}
  {3 \cdot 2^9 \pi^4} \bigg[x^2 \!+\! 12 \!-\! 12 \log(x) \!-\! 
  \frac{16}{x} \!+\! \frac{3}{x^2} \bigg] \\ && \\
  \rho_1^{\langle G^2 \rangle}(s) &=& \frac{m_c^4 \langle G^2 \rangle}
  {3^3 \cdot 2^{13} \pi^6} \bigg[ x^2 \!-\! 24x \!-\! 54 \!+\! 24\bigg(
  3 \!+\! \frac{1}{x} \bigg) \log(x) \!+\! \frac{80}{x} \!-\! 
  \frac{3}{x^2} \bigg] \\&&
  + \frac{m_s m_c^3 \langle G^2 \rangle}
  {3^2 \cdot 2^{12} \pi^6} \bigg[ 2x^2 \!-\! 12x \!+\! 45 \!-\! 
  6\bigg(1 \!+\! \frac{4}{x} \bigg) \log(x) \!-\! \frac{32}{x} \!-\!
  \frac{3}{x^2} \bigg] \\ && \\
  \rho_1^{\langle \bar{q}Gq \rangle}(s) &=& \frac{m_c^3 
  \langle \bar{s}G s \rangle}{3^2 \cdot 2^6 \pi^4} \bigg[ x^2 \!+\! 
  9 \!-\! 3\bigg(3 + \frac{1}{x} \bigg) \log(x) \!-\! \frac{10}{x} \bigg] 
  - \frac{m_s m_c^2 \langle \bar{s} Gs \rangle}
  {3^2 \cdot 2^9 \pi^4} \bigg[ 7x^2 \!-\! 3 \!-\! 
  18 \log(x) \!-\! \frac{4}{x} \bigg] \\ && \\
  \rho_1^{\langle \bar{q}q \rangle^2}(s) &=& \frac{m_c^2 
  \langle \bar{q}q \rangle^2}{3 \cdot 2^3 \pi^2} \bigg[ x \!-\! 2 
  \!+\! \frac{1}{x} \bigg] - \frac{m_s m_c \langle \bar{q}q \rangle^2}
  {3 \cdot 2^2 \pi^2} (1 \!-\! x)  \\ && \\
  \rho_1^{\langle G^3 \rangle}(s) &=& \frac{m_c^2 \langle G^3 \rangle}
  {5^2 \cdot 3^3 \cdot 2^{14} \pi^6} \bigg[ 8x^3 \!+\! 235x^2 \!+\! 
  760 \!-\! 1500 \log(x) \!-\! \frac{1000}{x} \!-\! \frac{3}{x^2} 
  - \frac{4m_c^2 \tau}{x^3} (1-x)^5 \bigg]
\end{eqnarray*}

\subsection*{\b Scalar-Vector tetraquark configuration (SV)}
 \vspace{-0.5cm}
\begin{eqnarray*}
  \rho_1^{pert}(s) &=& \frac{m_c^8}{5\cdot 3^2 \cdot 2^{13} \pi^6} 
  \bigg[ x^2 \!+\! 555 \!-\! 60 \bigg(3 \!+\! \frac{16}{x} \!+\! 
  \frac{9}{x^2} \bigg) \log(x) \!+\! \frac{480}{x} \!-\! \frac{945}{x^2} 
  \!-\! \frac{96}{x^3} \!+\! \frac{5}{x^4} \bigg] \\&&
  + \frac{m_s m_c^7}{3 \cdot 2^{10} \pi^6} \bigg[ x \!+\! 
  28 \!-\! 12\bigg(1 \!+\! \frac{3}{x} \!+\! \frac{1}{x^2} \bigg) 
  \log(x) \!-\! \frac{28}{x^2} \!-\! 
  \frac{1}{x^3} \bigg] \\ && \\
  \rho_1^{\langle \bar{q}q \rangle}(s) &=& - \frac{m_c^5 
  \langle \bar{s}s \rangle}{3 \cdot 2^6 \pi^4} \bigg[x \!+\! 9 \!-\! 
  6 \bigg(1 \!+\! \frac{1}{x} \bigg) \log(x) \!-\! \frac{9}{x} \!-\! 
  \frac{1}{x^2} \bigg]
  + \frac{m_s m_c^4 \langle \bar{s}s \rangle}
  {3 \cdot 2^9 \pi^4} \bigg[x^2 \!+\! 12 \!-\! 12 \log(x) \!-\! 
  \frac{16}{x} \!+\! \frac{3}{x^2} \bigg] \\ && \\
  \rho_1^{\langle G^2 \rangle}(s) &=& \frac{m_c^4 \langle G^2 \rangle}
  {3^3 \cdot 2^{13} \pi^6} \bigg[ 5x^2 \!+\! 12x \!+\! 198 \!-\! 48\bigg(
  3 \!+\! \frac{2}{x} \bigg) \log(x) \!-\! \frac{212}{x} \!-\! 
  \frac{3}{x^2} \bigg] \\&&
  + \frac{m_s m_c^3 \langle G^2 \rangle}
  {3^2 \cdot 2^{10} \pi^6} \bigg[ 7x \!+\! 15  \!-\! 
  3\bigg(7 \!+\! \frac{3}{x} \bigg) \log(x) \!-\! \frac{21}{x} \!-\!
  \frac{1}{x^2} \bigg] \\ && \\
  \rho_1^{\langle \bar{q}Gq \rangle}(s) &=& \frac{m_c^3 
  \langle \bar{s}G s \rangle}{2^7 \pi^4} \bigg[ x \!-\! 
  2 \log(x) \!-\! \frac{1}{x} \bigg] 
  - \frac{m_s m_c^2 \langle \bar{s} Gs \rangle}
  {3^2 \cdot 2^9 \pi^4} \bigg[ 5x^2 \!-\! 12x \!-\! 9 \!+\! 
  18 \log(x) \!+\! \frac{16}{x} \bigg] \\ && \\
  \rho_1^{\langle \bar{q}q \rangle^2}(s) &=& \frac{m_c^2 
  \langle \bar{q}q \rangle^2}{3^2 \cdot 2^3 \pi^2} \bigg[ x^2 \!-\! 3 
  \!+\! \frac{2}{x} \bigg] - \frac{m_s m_c \langle \bar{q}q \rangle^2}
  {3 \cdot 2^2 \pi^2} (1 \!-\! x)  \\ && \\
  \rho_1^{\langle G^3 \rangle}(s) &=& \frac{m_c^2 \langle G^3 \rangle}
  {5 \cdot 3^3 \cdot 2^{14} \pi^6} \bigg[ 11x^2 \!-\! 240x \!+\! 
  24 \!+\! 420 \log(x) \!+\! \frac{208}{x} \!-\! \frac{3}{x^2} \bigg]
\end{eqnarray*}
\subsection*{\b Scalar-Vector tetraquark configuration (VS)}
 \vspace{-0.5cm}
\begin{eqnarray*}
  \rho_1^{pert}(s) &=& \frac{m_c^8}{5\cdot 3^2 \cdot 2^{13} \pi^6} 
  \bigg[ x^2 \!+\! 555 \!-\! 60 \bigg(3 \!+\! \frac{16}{x} \!+\! 
  \frac{9}{x^2} \bigg) \log(x) \!+\! \frac{480}{x} \!-\! \frac{945}{x^2} 
  \!-\! \frac{96}{x^3} \!+\! \frac{5}{x^4} \bigg] \\&&
  - \frac{m_s m_c^7}{5 \cdot 3 \cdot 2^{12} \pi^6} \bigg[ x^2 \!+\! 
  340 \!-\! 60\bigg(2 \!+\! \frac{8}{x} \!+\! \frac{3}{x^2} \bigg) 
  \log(x) \!+\! \frac{80}{x} - \frac{405}{x^2} \!-\! 
  \frac{16}{x^3} \bigg] \\ && \\
  \rho_1^{\langle \bar{q}q \rangle}(s) &=& \frac{m_c^5 
  \langle \bar{s}s \rangle}{3^2 \cdot 2^8 \pi^4} \bigg[x^2 \!+\! 72 \!-\! 
  12 \bigg(3 \!+\! \frac{4}{x} \bigg) \log(x) \!-\! \frac{64}{x} \!-\! 
  \frac{9}{x^2} \bigg]
  + \frac{m_s m_c^4 \langle \bar{s}s \rangle}
  {3 \cdot 2^9 \pi^4} \bigg[x^2 \!+\! 12 \!-\! 12 \log(x) \!-\! 
  \frac{16}{x} \!+\! \frac{3}{x^2} \bigg] \\ && \\
  \rho_1^{\langle G^2 \rangle}(s) &=& \frac{m_c^4 \langle G^2 \rangle}
  {3^3 \cdot 2^{13} \pi^6} \bigg[ x^2 \!-\! 24x \!-\! 54 \!+\! 24\bigg(
  3 \!+\! \frac{1}{x} \bigg) \log(x) \!+\! \frac{80}{x} \!-\! 
  \frac{3}{x^2} \bigg] \\&&
  + \frac{m_s m_c^3 \langle G^2 \rangle}
  {3^2 \cdot 2^{10} \pi^6} \bigg[ 7x \!+\! 15  \!-\! 
  3\bigg(7 \!+\! \frac{3}{x} \bigg) \log(x) \!-\! \frac{21}{x} \!-\!
  \frac{1}{x^2} \bigg] \\ && \\
  \rho_1^{\langle \bar{q}Gq \rangle}(s) &=& -\frac{m_c^3 
  \langle \bar{s}G s \rangle}{3^2 \cdot 2^9 \pi^4} \bigg[ 7x^2 \!-\! 
  23 \!-\! 18 \log(x) \!-\! \frac{4}{x} \bigg] 
  - \frac{m_s m_c^2 \langle \bar{s} Gs \rangle}
  {3^2 \cdot 2^9 \pi^4} \bigg[ 7x^2 \!-\! 3 \!-\! 18 \log(x) \!-\! \frac{4}{x} \bigg] \\ && \\
  \rho_1^{\langle \bar{q}q \rangle^2}(s) &=& - \frac{m_c^2 
  \langle \bar{q}q \rangle^2}{3 \cdot 2^3 \pi^2} \bigg[ x \!-\! 2 
  \!+\! \frac{1}{x} \bigg] - \frac{m_s m_c \langle \bar{q}q \rangle^2}
  {3 \cdot 2^2 \pi^2} (1 \!-\! x)  \\ && \\
  \rho_1^{\langle G^3 \rangle}(s) &=& \frac{m_c^2 \langle G^3 \rangle}
  {5^2 \cdot 3^3 \cdot 2^{14} \pi^6} \bigg[ 8x^3 \!+\! 235x^2 \!+\! 
  760 \!-\! 1500 \log(x) \!-\! \frac{1000}{x} \!-\! \frac{3}{x^2} 
  \!-\! \frac{4m_c^2 \tau}{x^3} (1-x)^5 \bigg]
\end{eqnarray*}
\subsection*{\b $D_1 K$ vector molecule configuration}
 \vspace{-0.5cm}
\begin{eqnarray*}
  \rho_1^{pert}(s) &=& \frac{m_c^8}{5\cdot 3 \cdot 2^{15} \pi^6} 
  \bigg[ x^2 \!+\! 555 \!-\! 60 \bigg(3 \!+\! \frac{16}{x} \!+\! 
  \frac{9}{x^2} \bigg) \log(x) \!+\! \frac{480}{x} \!-\! \frac{945}{x^2} 
  \!-\! \frac{96}{x^3} \!+\! \frac{5}{x^4} \bigg]  \\ && \\
  \rho_1^{\langle \bar{q}q \rangle}(s) &=& -\frac{m_c^5 
  \langle \bar{q}q \rangle}{2^{8} \pi^4} \bigg[x \!+\! 9 \!-\! 
  6 \bigg(1 \!+\! \frac{1}{x} \bigg) \log(x) \!-\! \frac{9}{x} \!-\! 
  \frac{1}{x^2} \bigg] 
  - \frac{m_s m_c^4 \langle \bar{q}q \rangle}{2^{11} \pi^4} 
  (2 \langle \bar{q}q \rangle \!-\! \langle \bar{s}s \rangle) \bigg[
  x^2 \!+\! 12 \!-\! 12 \log(x) \!-\! \frac{16}{x} \!+\! 
  \frac{3}{x^2} \bigg] \\ && \\
  \rho_1^{\langle G^2 \rangle}(s) &=& \frac{m_c^4 \langle G^2 \rangle}
  {3 \cdot 2^{15} \pi^6} \bigg[ 3x^2 \!+\! 8x \!+\! 108 \!-\! 12\bigg(
  7 \!+\! \frac{4}{x} \bigg) \log(x) \!-\! \frac{120}{x} \!+\! 
  \frac{1}{x^2} \bigg]  \\ && \\
  \rho_1^{\langle \bar{q}Gq \rangle}(s) &=& \frac{3m_c^3 
  \langle \bar{q}G q \rangle}{2^{9} \pi^4} \bigg[ 
  x \!-\! 2 \log(x) \!-\! \frac{1}{x} \bigg] + 
  \frac{m_s m_c^2}{3 \cdot 2^9 \pi^4}
  \Big(3\langle \bar{q} Gq \rangle + 2\langle \bar{s} Gs \rangle \Big) 
  \bigg[ x^2 \!-\! 3 \!+\! \frac{2}{x} \bigg] \\ && \\
  \rho_1^{\langle \bar{q}q \rangle^2}(s) &=& \frac{m_c^2 
  \langle \bar{q}q \rangle \langle \bar{s}s \rangle}{3 \cdot 2^5 \pi^2} 
  \bigg[ x^2 \!-\! 3 \!+\! \frac{2}{x} \bigg] - 
  \frac{m_s m_c}{2^5 \pi^2} \Big( 2\langle \bar{q}q \rangle^2 
  - \langle \bar{q}q \rangle\langle \bar{s}s \rangle \Big) (1-x) 
  \\ && \\
  \rho_1^{\langle G^3 \rangle}(s) &=& \frac{m_c^2 \langle G^3 \rangle}
  {5 \cdot 3^2 \cdot 2^{16} \pi^6} \bigg[ 31x^2 \!-\! 480x \!+\! 84 
  \!+\! 780 \log(x) \!+\! \frac{368}{x} \!-\! \frac{3}{x^2} \bigg]
\end{eqnarray*}
\subsection*{\b $ D K_1$ vector molecule configuration}
 \vspace{-0.5cm}
\begin{eqnarray*}
  \rho_1^{pert}(s) &=& \frac{m_c^8}{5\cdot 3 \cdot 2^{15} \pi^6} 
  \bigg[ x^2 \!+\! 555 \!-\! 60 \bigg(3 \!+\! \frac{16}{x} \!+\! 
  \frac{9}{x^2} \bigg) \log(x) \!+\! \frac{480}{x} \!-\! \frac{945}{x^2} 
  \!-\! \frac{96}{x^3} \!+\! \frac{5}{x^4} \bigg]  \\ && \\
  \rho_1^{\langle \bar{q}q \rangle}(s) &=& \frac{m_c^5 
  \langle \bar{q}q \rangle}{3 \cdot 2^{10} \pi^4} \bigg[x^2 \!+\! 72 \!-\! 
  12 \bigg(3 \!+\! \frac{4}{x} \bigg) \log(x) \!-\! \frac{64}{x} \!-\! 
  \frac{9}{x^2} \bigg] \\&& 
  + \frac{m_s m_c^4 \langle \bar{q}q \rangle}{2^8 \pi^4} \bigg[
  2x \!+\! 3 \!-\! 6 \log(x) \!-\! \frac{6}{x} \!+\! \frac{1}{x^2} \bigg] 
  + \frac{m_s m_c^4 \langle \bar{s}s \rangle}{2^{11} \pi^4} \bigg[
  x^2 \!+\! 12 \!-\! 12 \log(x) \!-\! \frac{16}{x} \!+\! \frac{3}{x^2} \bigg] 
  \\ && \\
  \rho_1^{\langle G^2 \rangle}(s) &=& \frac{m_c^4 \langle G^2 \rangle}
  {3^2 \cdot 2^{15} \pi^6} \bigg[ x^2 \!-\! 48x \!-\! 180 \!+\! 12\bigg(
  15 \!+\! \frac{8}{x} \bigg) \log(x) \!+\! \frac{224}{x} \!+\! 
  \frac{3}{x^2} \bigg]  \\ && \\
  \rho_1^{\langle \bar{q}Gq \rangle}(s) &=& - \frac{m_c^3 
  \langle \bar{q}G q \rangle}{3 \cdot 2^{10} \pi^4} \bigg[ 
  5x^2 \!+\! 63 \!-\! 6 \bigg( 9 \!+\! \frac{4}{x} \bigg) \log(x) 
  \!-\! \frac{68}{x} \bigg] - \frac{m_s m_c^2}{2^9 \pi^4}
  (3\langle \bar{q} Gq \rangle + 2\langle \bar{s} Gs \rangle) \bigg[ 
  x \!-\! 2 \!+\! \frac{1}{x} \bigg] \\ && \\
  \rho_1^{\langle \bar{q}q \rangle^2}(s) &=& - \frac{m_c^2 
  \langle \bar{q}q \rangle \langle \bar{s}s \rangle}{2^5 \pi^2} 
  \bigg[ x \!-\! 2 \!+\! \frac{1}{x} \bigg] - \frac{m_s m_c 
  \langle \bar{q}q \rangle^2}{2^4 \pi^2} (1-x) -
  \frac{m_s m_c \langle \bar{q}q \rangle \langle \bar{s}s \rangle}
  {2^6 \pi^2} ( 1 \!-\! x^2 )  \\ && \\
  \rho_1^{\langle G^3 \rangle}(s) &=& \frac{m_c^2 \langle G^3 \rangle}
  {5^2 \cdot 3^2 \cdot 2^{16} \pi^6} \bigg[ 8x^3 \!+\! 535x^2 \!+\! 1660 
  \!-\! 3300 \log(x) \!-\! \frac{2200}{x} \!-\! \frac{3}{x^2} \!+\! 
  4m_c^2 \tau \bigg(x^2 \!-\! 5x \!+\! 10 \!-\! \frac{10}{x} \!+\! 
  \frac{5}{x^2} \!-\! \frac{1}{x^3} \bigg) \bigg]
\end{eqnarray*}
\subsection*{\b $ D_0^\ast K^\ast$ vector molecule configuration}
 \vspace{-0.5cm}
\begin{eqnarray*}
  \rho_1^{pert}(s) &=& \frac{m_c^8}{5\cdot 3 \cdot 2^{15} \pi^6} 
  \bigg[ x^2 \!+\! 555 \!-\! 60 \bigg(3 \!+\! \frac{16}{x} \!+\! 
  \frac{9}{x^2} \bigg) \log(x) \!+\! \frac{480}{x} \!-\! \frac{945}{x^2} 
  \!-\! \frac{96}{x^3} \!+\! \frac{5}{x^4} \bigg]  \\ && \\
  \rho_1^{\langle \bar{q}q \rangle}(s) &=& - \frac{m_c^5 
  \langle \bar{q}q \rangle}{3 \cdot 2^{10} \pi^4} \bigg[x^2 \!+\! 72 \!-\! 
  12 \bigg(3 \!+\! \frac{4}{x} \bigg) \log(x) \!-\! \frac{64}{x} \!-\! 
  \frac{9}{x^2} \bigg] \\&& 
  - \frac{m_s m_c^4 \langle \bar{q}q \rangle}{2^8 \pi^4} \bigg[
  2x \!+\! 3 \!-\! 6 \log(x) \!-\! \frac{6}{x} \!+\! \frac{1}{x^2} \bigg] 
  + \frac{m_s m_c^4 \langle \bar{s}s \rangle}{2^{11} \pi^4} \bigg[
  x^2 \!+\! 12 \!-\! 12 \log(x) \!-\! \frac{16}{x} \!+\! \frac{3}{x^2} \bigg] 
  \\ && \\
  \rho_1^{\langle G^2 \rangle}(s) &=& \frac{m_c^4 \langle G^2 \rangle}
  {3^2 \cdot 2^{15} \pi^6} \bigg[ x^2 \!-\! 48x \!-\! 180 \!+\! 12\bigg(
  15 \!+\! \frac{8}{x} \bigg) \log(x) \!+\! \frac{224}{x} \!+\! 
  \frac{3}{x^2} \bigg]  \\ && \\
  \rho_1^{\langle \bar{q}Gq \rangle}(s) &=& \frac{m_c^3 
  \langle \bar{q}G q \rangle}{3 \cdot 2^{10} \pi^4} \bigg[ 
  5x^2 \!+\! 63 \!-\! 6 \bigg( 9 \!+\! \frac{4}{x} \bigg) \log(x) 
  \!-\! \frac{68}{x} \bigg] + \frac{m_s m_c^2}{2^9 \pi^4}
  (3\langle \bar{q} Gq \rangle - 2\langle \bar{s} Gs \rangle) \bigg[ 
  x \!-\! 2 \!+\! \frac{1}{x} \bigg] \\ && \\
  \rho_1^{\langle \bar{q}q \rangle^2}(s) &=& \frac{m_c^2 
  \langle \bar{q}q \rangle \langle \bar{s}s \rangle}{2^5 \pi^2} 
  \bigg[ x \!-\! 2 \!+\! \frac{1}{x} \bigg] - \frac{m_s m_c 
  \langle \bar{q}q \rangle^2}{2^4 \pi^2} (1-x) +
  \frac{m_s m_c \langle \bar{q}q \rangle \langle \bar{s}s \rangle}
  {2^6 \pi^2} ( 1 \!-\! x^2 )  \\ && \\
  \rho_1^{\langle G^3 \rangle}(s) &=& \frac{m_c^2 \langle G^3 \rangle}
  {5^2 \cdot 3^2 \cdot 2^{16} \pi^6} \bigg[ 8x^3 \!+\! 535x^2 \!+\! 1660 
  \!-\! 3300 \log(x) \!-\! \frac{2200}{x} \!-\! \frac{3}{x^2} \!+\! 
  4m_c^2 \tau \bigg(x^2 \!-\! 5x \!+\! 10 \!-\! \frac{10}{x} \!+\! 
  \frac{5}{x^2} \!-\! \frac{1}{x^3} \bigg) \bigg]
\end{eqnarray*}

\subsection*{\b $ D^\ast K^\ast_0$ vector molecule configuration}
 \vspace{-0.5cm}
\begin{eqnarray*}
  \rho_1^{pert}(s) &=& \frac{m_c^8}{5\cdot 3 \cdot 2^{15} \pi^6} 
  \bigg[ x^2 \!+\! 555 \!-\! 60 \bigg(3 \!+\! \frac{16}{x} \!+\! 
  \frac{9}{x^2} \bigg) \log(x) \!+\! \frac{480}{x} \!-\! \frac{945}{x^2} 
  \!-\! \frac{96}{x^3} \!+\! \frac{5}{x^4} \bigg]  \\ && \\
  \rho_1^{\langle \bar{q}q \rangle}(s) &=& \frac{m_c^5 
  \langle \bar{q}q \rangle}{2^8 \pi^4} \bigg[x \!+\! 9 \!-\! 
  6 \bigg(1 \!+\! \frac{1}{x} \bigg) \log(x) \!-\! \frac{9}{x} \!-\! 
  \frac{1}{x^2} \bigg]
  + \frac{m_s m_c^4}{2^{11} \pi^4}
  (2 \langle \bar{q}q \rangle \!+\! \langle \bar{s}s \rangle) 
  \bigg[x^2 \!+\! 12 \!-\! 12 \log(x) \!-\! 
  \frac{16}{x} \!+\! \frac{3}{x^2} \bigg] \\ && \\
  \rho_1^{\langle G^2 \rangle}(s) &=& \frac{m_c^4 \langle G^2 \rangle}
  {3 \cdot 2^{15} \pi^6} \bigg[ 3x^2 \!+\! 8x \!+\! 108 \!-\! 12\bigg(
  7 \!+\! \frac{4}{x} \bigg) \log(x) \!-\! \frac{120}{x} \!+\! 
  \frac{1}{x^2} \bigg]  \\ && \\
  \rho_1^{\langle \bar{q}Gq \rangle}(s) &=& - \frac{3m_c^3 
  \langle \bar{q}G q \rangle}{2^9 \pi^4} \bigg[ x \!-\! 2 \log(x) \!-\! \frac{1}{x} \bigg] - \frac{m_s m_c^2}{3 \cdot 2^9 \pi^4}
  (3\langle \bar{q} Gq \rangle - 2\langle \bar{s} Gs \rangle) \bigg[ 
  x^2 \!-\! 3 \!+\! \frac{2}{x} \bigg] \\ && \\
  \rho_1^{\langle \bar{q}q \rangle^2}(s) &=& - \frac{m_c^2 
  \langle \bar{q}q \rangle \langle \bar{s}s \rangle}{3 \cdot 2^5 \pi^2} 
  \bigg[ x^2 \!-\! 3 \!+\! \frac{2}{x} \bigg] - \frac{m_s m_c}
  {2^5 \pi^2} (2\langle \bar{q}q \rangle^2 + \langle \bar{q}q \rangle 
  \langle \bar{s}s \rangle) \bigg[ 1 \!-\! x \bigg]  \\ && \\
  \rho_1^{\langle G^3 \rangle}(s) &=& \frac{m_c^2 \langle G^3 \rangle}
  {5 \cdot 3^2 \cdot 2^{16} \pi^6} \bigg[ 31x^2 \!-\! 480x \!+\! 84 
  \!+\! 780 \log(x) \!+\! \frac{368}{x} \!-\! \frac{3}{x^2} \bigg]
\end{eqnarray*}


\end{document}